\documentclass[aip,jcp,reprint,superscriptaddress,amsmath,amssymb]{revtex4-2}

\usepackage[utf8]{inputenc}
\usepackage{graphicx}
\usepackage{subcaption}
\usepackage{physics}
\usepackage{float}
\usepackage{comment}
\usepackage{booktabs}
\usepackage{gensymb}
\usepackage{array}
\usepackage{xcolor}
\usepackage{xr}
\usepackage{hyperref}
\usepackage{ragged2e}
\usepackage{subcaption}

\hypersetup{hidelinks}

\begin{document}

\title{
Active Hydrodynamics Couples Polymer Organization, Shape Fluctuations, and Motility in Deformable Droplets}
\author{Ritu Raj}
\affiliation{Department of Physics, Indian Institute of Technology Madras, Chennai, 600036, India.}
\affiliation{Center for Soft and Biological Matter, Indian Institute of Technology Madras, Chennai, 600036, India.}
\author{P. B. Sunil Kumar}
\email{sunil@physics.iitm.ac.in}
\affiliation{Department of Physics, Indian Institute of Technology Madras, Chennai, 600036, India.}
\affiliation{Center for Soft and Biological Matter, Indian Institute of Technology Madras, Chennai, 600036, India.}

\date{\today}

\begin{abstract}

We study semiflexible active polymers confined within a soft, deformable droplet suspended in a fluid using dissipative particle dynamics (DPD) simulations. Extensile and contractile force dipoles along the polymer backbone generate distinct hydrodynamic flow fields that mediate effective interactions between polymer segments. Extensile activity promotes parallel alignment and lateral attraction, whereas contractile activity favors predominantly perpendicular organization, leading to qualitatively different collective behavior. The activity-generated flows couple strongly to the deformable interface, selectively enhancing low-order spherical-harmonic modes associated with long-wavelength droplet deformations. Activity also drives the interfacial relaxation away from passive capillary behavior, with extensile and contractile droplets exhibiting distinct mode-dependent dynamics. These differences in internal organization and interfacial fluctuations strongly influence droplet motility. Extensile activity produces increasingly persistent droplet motion with increasing polymer number, whereas contractile activity can sustain long-lived near-ballistic motion whose duration depends sensitively on polymer number and activity strength. Our results demonstrate how active hydrodynamics governs the interplay between internal structure, interfacial fluctuations, and emergent motility in confined active matter systems.


\end{abstract}

\pacs{}

\maketitle

\section{Introduction}

Fluid--fluid interfaces are ubiquitous in natural and engineered systems, ranging from emulsions and multiphase flows~\cite{Stone1994DropDynamics} to membraneless organelles formed through liquid--liquid phase separation within cells~\cite{Banani2017Condensates,Brangwynne2009PGranules}. These interfaces are intrinsically deformable, with their shape fluctuations and dynamics governed by the interplay among interfacial tension, viscous stresses, and thermal fluctuations~\cite{MillerScriven1968DropletOscillations,Aarts2004ThermalCapillaryWaves}. Although passive liquid droplets have been extensively studied~\cite{MillerScriven1968DropletOscillations,Aarts2004ThermalCapillaryWaves}, recent work has increasingly focused on active droplets, in which continuous energy injection by active constituents can strongly modify interfacial fluctuations and dynamics~\cite{GiomiDeSimone2014ActiveNematicDroplets,RuskeYeomans2021Morphology,Kokot2022ActiveDroplet,Adkins2022ActiveInterfaces}.

Experimental studies have revealed clear signatures of nonequilibrium dynamics in active droplets. In quasi-two-dimensional droplets containing Quincke rollers, activity enhances interfacial shape fluctuations, producing a nonequilibrium fluctuation spectrum and circulating probability currents in shape-mode space that signal broken detailed balance; at higher activity, collective roller motion drives large-scale deformations and intermittent self-propulsion~\cite{Kokot2022ActiveDroplet}. Similarly, in three-dimensional phase-separated droplets containing motile bacteria, collective active flows generate scale-dependent interfacial fluctuations with propagating wave-like modes and, at higher activity, drive pronounced deformations and filament-like protrusions, revealing strong coupling between internal flows and interfacial dynamics~\cite{Chang2026BacterialTurbulence}.

The behavior of droplet interfaces becomes considerably richer when the encapsulated constituents are polymers rather than point-like particles. Owing to their extended conformations and internal degrees of freedom, polymers couple to the confining interface through their conformational entropy, bending elasticity, orientational organization, and excluded-volume interactions. Under strong confinement, this coupling can modify both the equilibrium droplet shape and its spectrum of interfacial fluctuations~\cite{Goulian1996PolymerDropletFluctuations}. Recent experiments on semiflexible filaments embedded within biomolecular condensates have further shown that competition among filament bending, interfacial tension, and wetting can generate pronounced nonspherical droplet morphologies~\cite{Wolf2026CondensateFilaments,Mansour2026ActinDroplets}. While these studies have established the important role of filament organization in shaping passive droplets, much less is known about droplets containing active polymers, where internally generated stresses may couple polymer conformations, hydrodynamic flows, and interfacial fluctuations in fundamentally different ways.

Active emulsions based on microtubule--kinesin active fluids provided a prominent experimental platform for investigating the interplay between internally generated activity and deformable interfaces. In these systems, the filamentous active network couples strongly to the droplet boundary, giving rise to a wide range of emergent phenomena. In droplets coated with three-dimensional microtubule networks, adsorption of the filaments onto the interface leads to the formation of active nematic layers whose internally generated flows and topological defect dynamics drive autonomous droplet motion~\cite{Sanchez2012SpontaneousMotion}. In water-in-oil droplets containing bulk microtubule--kinesin active fluids, coherent intradroplet circulation hydrodynamically couples to the surrounding passive oil, while the external flow field, in turn, influences the organization of the active network~\cite{Chen2021FlowCoupling}. Similarly, microtubule--kinesin networks confined within aqueous phase-separated droplets accumulate near the interface, generate persistent vortical flows, and propel the droplets through self-organized active stresses~\cite{Sakuta2023VortexFlow}.

The coupling between active stresses and deformable boundaries has been extensively investigated through continuum theories and numerical simulations. Continuum analyses have shown that active slip velocities or surface tractions excite distinct spherical-harmonic deformation modes, with their amplitudes and dynamics determined by the competition between active forcing and interfacial tension~\cite{KreeZippelius2023Deformations}. Continuum models of active nematic droplets further demonstrate that active stresses can spontaneously break symmetry, driving droplet motility through the emergence of polar order~\cite{Tjhung2012SpontaneousSymmetry}, while the interplay between active stresses and topological defects gives rise to droplet deformation, self-propulsion, and even spontaneous division~\cite{GiomiDeSimone2014ActiveNematicDroplets}. Complementing these continuum approaches, particle-resolved simulations have provided microscopic insights into the feedback between active constituents and deformable interfaces. In two-dimensional systems, active particles confined within deformable droplets enhance interfacial fluctuations, induce large shape deformations, and can ultimately trigger droplet breakup~\cite{DiazPagonabarraga2024ActiveParticles}. Similarly, active particles enclosed by flexible boundaries exhibit persistent motion, boundary polarization, and transitions between vortical and collectively polarized states~\cite{Paoluzzi2016SoftVesicles,Wen2026StressBoundaryMemory}. While these studies have significantly advanced our understanding of active matter in deformable confinement, they have largely been restricted to two-dimensional geometries and systems composed of active particles rather than active polymers.

Studies of confined active polymers have, to date, focused almost entirely on rigid confinement. While these systems exhibit a rich spectrum of nonequilibrium structures and dynamics~\cite{MannaKumar2019ActivePolymericFluids,Saintillan2018Chromatin}, the absence of interfacial deformability eliminates the two-way coupling between polymer-generated hydrodynamic flows and the confining boundary. Consequently, the dynamics of active polymers under soft confinement remain largely unexplored. Understanding how polymer activity drives interfacial fluctuations, droplet deformation, and emergent droplet motility is the primary objective of the present work.



Related questions have been addressed in membrane-bound compartments, where activity drives nonequilibrium flickering, enhanced fluctuations, and shape transitions~\cite{Turlier2016RBCFlickering,Sciortino2025ActiveMembrane,Iyer2022ActiveVesicles,Peterson2021ActiveFilaments,KrishnanSunilKumar2022ActiveRecycling}. However, in these systems, shape relaxation is primarily governed by bending elasticity, whereas fluid--fluid interfaces of interest here  are primarily tension-dominated, with a correspondingly different mode spectrum and relaxation hierarchy~\cite{Loubet2012EffectiveTension,MilnerSafran1987DynamicalFluctuations,Faucon1989BendingElasticity}.

Here, we address these questions by studying active polymers confined within a soft, deformable droplet bounded by a nearly spherical fluid--fluid interface (Fig.~\ref{fig:schematic}a). Using dissipative particle dynamics simulations with an explicit solvent, we introduce dipolar activity along the polymer backbone (Fig.~\ref{fig:schematic}b), allowing us to resolve how polymer-generated forces are transmitted through the solvent to the interface and how the resulting hydrodynamic coupling influences polymer organization, interfacial fluctuations, and droplet dynamics. We show that the sign of the active dipole fundamentally alters the collective organization of the confined polymers, leading to qualitatively different interfacial fluctuation spectra, relaxation dynamics, and droplet motility. In particular, extensile and contractile activity produce distinct modes of transport and shape fluctuations, demonstrating how active hydrodynamic interactions couple the internal dynamics of confined polymers to the deformation and motion of the enclosing fluid compartment.

\section{Model and simulation details}
\label{sec:model_and_methods}

\begin{figure*}[!ht]
    \centering
    \begin{subfigure}[t]{0.98\textwidth}
        \captionsetup{position=top, justification=raggedright, singlelinecheck=false}
        \includegraphics[width=\linewidth]{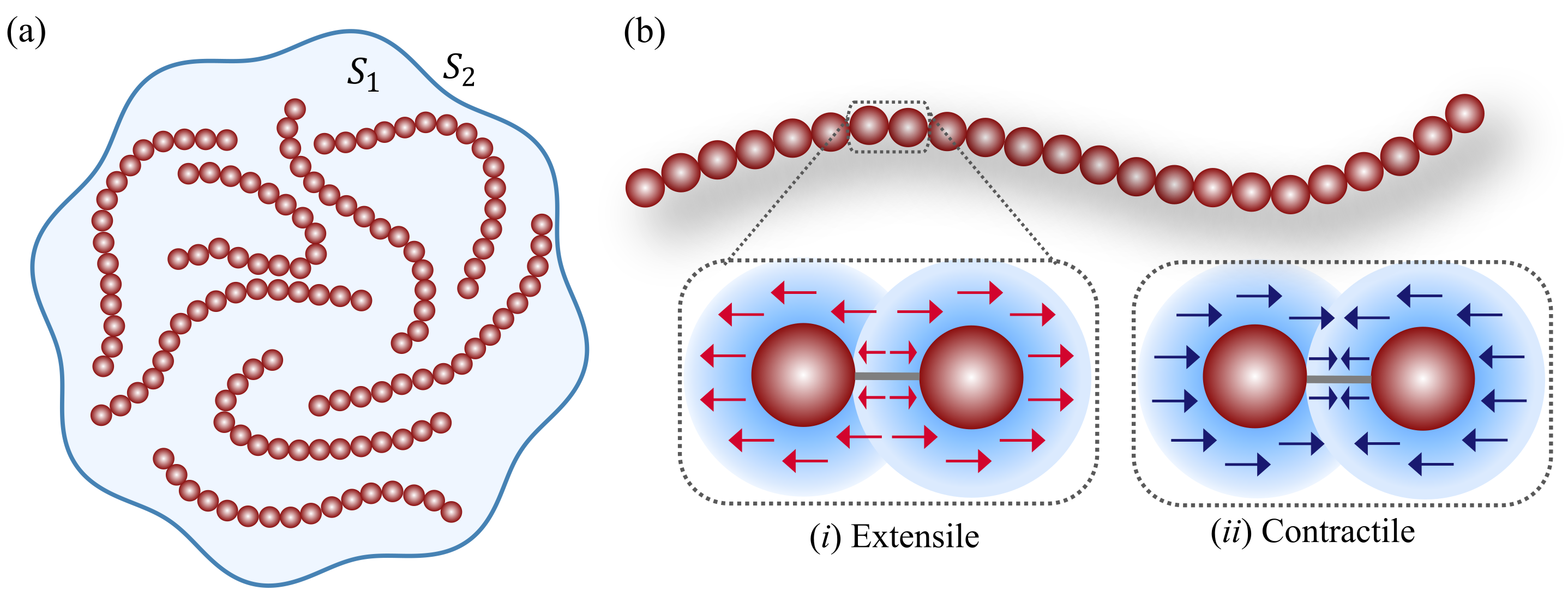}
    \end{subfigure}
    \captionsetup{position=top, justification=raggedright, singlelinecheck=false}
    \caption{\justifying
    \textbf{Model schematic.} 
    \textbf{(a)} Schematic of bead--spring semiflexible polymers confined within a soft, deformable droplet. The irregular boundary represents the fluctuating droplet interface. The solvent inside the droplet and the surrounding solvent are denoted by $S_1$ and $S_2$, respectively.
\textbf{(b)} Local implementation of active force dipoles along the polymer backbone. For each bonded monomer pair $(i,j)$, spherical force-distribution regions of radius $R_m=1.3$ are defined around the two monomers, as indicated by the blue shaded regions. (i) For extensile activity, equal and opposite forces of magnitude $F_0$ are distributed among the neighboring beads within the two regions and directed away from the bond center along the local bond axis. (ii) For contractile activity, the force directions are reversed, driving the neighboring beads towards the bond center. Further details are provided in Sec.~\ref{sec:model_and_methods}. 
 }
    \label{fig:schematic}
\end{figure*}
We consider a system of $N_p$ polymers confined within a droplet of solvent $S_1$, which is suspended in an immiscible solvent $S_2$. The coarse-grained model comprises three types of beads: polymer beads, solvent beads representing the $S_1$ phase, and solvent beads representing the $S_2$ phase. The interactions between the solvent beads are chosen to render the two solvents immiscible, leading to the spontaneous formation of an approximately spherical $S_1$ droplet containing the polymers and bounded by a fluctuating interface, as illustrated in Fig.~\ref{fig:schematic}a. The droplet thus provides a soft, deformable confining environment for the polymers. This model may be regarded as a coarse-grained representation of polymers confined within an oil-like droplet dispersed in a water-like medium. The dynamics of all beads are evolved using dissipative particle dynamics (DPD)~\cite{Espanol1995DPD, Groot1997DPD}, with their motion governed by Newton’s equations of motion,
\begin{equation}
m\frac{d^2\vec{r}_i}{dt^2} = \vec{F}_i ,
\end{equation}
where $\vec{r}_i$ is the position of bead $i$, $m$ is the bead mass, and $\vec{F}_i$ is the total force acting on bead $i$, given by
\begin{equation}
\vec{F}_i =
\vec{F}^{C}_i
+
\sum_{j\neq i}
\left(
\vec{F}^{D}_{ij}
+
\vec{F}^{R}_{ij}
\right)
+
\vec{F}^{\mathrm{act}}_i ,
\end{equation}
where $\vec{F}^{C}_i$ is the total conservative force, $\vec{F}^{D}_{ij}$ and $\vec{F}^{R}_{ij}$ are the dissipative and random DPD forces between beads $i$ and $j$, respectively, and $\vec{F}^{\mathrm{act}}_i$ is the contribution from active force dipoles applied along the polymer backbone.

The conservative force $\vec{F}^{C}_i$ consists of non-bonded interactions and, for polymer beads,  intrapolymer forces arising from bond-stretching and bending potentials:
\begin{equation}
\vec{F}^{C}_i
=
\sum_{j\neq i}\vec{F}^{C,\mathrm{nb}}_{ij}
-
\nabla_i U_{\mathrm{bond}}
-
\nabla_i U_{\mathrm{bend}},
\end{equation}
where $\nabla_i$ denotes the gradient with respect to the position $\vec{r}_i$ of bead $i$.

The pairwise non-bonded conservative force on any bead $i$ due to bead $j$ is taken to be a soft repulsive force:
\begin{equation}
\vec{F}^{C,\mathrm{nb}}_{ij}
=
-a_{ij}
\left(1-\frac{r_{ij}}{r_c}\right)
\hat{r}_{ij},
\qquad r_{ij}<r_c,
\end{equation}
and is zero otherwise. Here, $a_{ij}>0$ sets the repulsion strength between beads $i$ and $j$, $r_c$ is the cutoff distance, $\vec{r}_{ij}=\vec{r}_j-\vec{r}_i$, $r_{ij}=|\vec{r}_{ij}|$, and $\hat{r}_{ij}=\vec{r}_{ij}/r_{ij}$. The interaction parameter $a_{ij}$ is chosen such that the two solvents are immiscible and the polymer is soluble only in solvent $S_1$. The chosen $a_{ij}$ values are given in Table~\ref{tab:pair_coefficients}. Unless otherwise stated, lengths are measured in units of the bead diameter $r_c$, energies in units of $k_{\mathrm{B}} T$, and times in units of $\tau=r_{c}\sqrt{m/(k_B T)}$.

\begin{table}[htbp]
\centering
\caption{\justifying
Matrix of non-bonded DPD interaction parameters $a_{ij}$ between polymer beads, droplet-solvent beads $S_1$, and surrounding-solvent beads $S_2$.}
\label{tab:pair_coefficients}
\begin{tabular}{cccc}
\toprule
$a_{ij}$ & Polymer & $S_1$ & $S_2$ \\
\midrule
Polymer & 80  & 50  & 180 \\
$S_1$   & 50  & 50  & 220 \\
$S_2$   & 180 & 220 & 50  \\
\bottomrule
\end{tabular}
\end{table}

Consecutive beads along each polymer are connected by harmonic springs. For a bonded pair of beads $i$ and $j$, the bond potential energy is
\begin{equation}
U_{\mathrm{bond}}
=
\frac{k_b}{2}
\left(
r_{ij}-r_0
\right)^2 ,
\end{equation}
where $r_{ij}$ is the bond length, $k_b$ is the spring constant, and $r_0$ is the equilibrium bond length. We set $k_b=200$ and $r_0=0.87$. For three consecutive beads $i$, $j$, and $k$ along the polymer backbone, the bending potential is
\begin{equation}
U_{\mathrm{bend}}
=
\kappa
\left(
1-\cos\theta_{ijk}
\right),
\end{equation}
where $\kappa$ is the bending rigidity and $\theta_{ijk}$ is the angle between the two consecutive bond vectors $\vec{r}_{ij}$ and $\vec{r}_{jk}$. We set $\kappa=20$. The pairwise dissipative and random forces are given by~\cite{Espanol1995DPD,Groot1997DPD}
\begin{equation}
\vec{F}^{D}_{ij}
=
-\gamma_{ij}\omega^2(r_{ij})
\left(
\hat{r}_{ij}\cdot\vec{v}_{ij}
\right)
\hat{r}_{ij},
\end{equation}
and
\begin{equation}
\vec{F}^{R}_{ij}
=
\sigma_{ij}\omega(r_{ij})
\frac{\xi_{ij}}{\sqrt{\delta t}}
\hat{r}_{ij},
\end{equation}
where $\gamma_{ij}$ is the friction coefficient, $\sigma_{ij}$ is the noise amplitude, $\vec{v}_{ij}=\vec{v}_i-\vec{v}_j$ is the relative velocity between beads $i$ and $j$, and $\delta t$ is the simulation time step. The quantity $\xi_{ij}$ is a Gaussian random variable with zero mean
and unit variance. The dissipative and random forces are pairwise equal and opposite, ensuring conservation of linear momentum. The fluctuation--dissipation relation is imposed through
\begin{equation}
\sigma_{ij}^{2}=2\gamma_{ij}k_{\mathrm B}T.
\end{equation}
The friction coefficient is set to $\gamma_{ij}=4.5$ for all bead pairs. The weight function is defined as
\begin{equation}
\omega(r_{ij})
=
1-\frac{r_{ij}}{r_c},
\qquad r_{ij}<r_c,
\end{equation}
and is zero otherwise. For this choice of parameters, the solvent Schmidt number is $\mathrm{Sc}\approx 2.58$, indicating that momentum diffuses faster than mass and that hydrodynamic interactions are effectively transmitted through the solvent~\cite{Raj2026ChromatinCondensate,Hansen2006TheorySimpleLiquids}.

The term $\vec{F}^{\mathrm{act}}_i$ represents the active force contribution arising from force dipoles along the polymer backbone~\cite{Raj2026ChromatinCondensate}. Each bonded pair of polymer beads $(i,j)$ defines an active force dipole oriented along the unit bond vector $\hat{r}_{ij}$. As illustrated in Fig.~\ref{fig:schematic}b, each bead in the pair exerts a force, weighted by the  distant dependent function $\varphi(r)$,  on the neighboring beads within a spherical region of radius $R_m = 1.3$ centered on that bead. For a bonded pair $(i,j)$, the net active force distributed over all beads in the spherical region around bead $i$ is $-F_0\hat{r}_{ij}$, while that distributed over all beads in the spherical region around bead $j$ is $F_0\hat{r}_{ij}$. Thus, each active bond generates a local force dipole with zero net force. Therefore, for a bead $k$ lying in the spherical neighborhood of bead $i$, such that $k\in\mathcal{N}_i=\{k' \mid r_{ik'}<R_m,\; k'\notin\{i-1,i,i+1\}\}$, the active force on bead $k$ due to bead $i$ is
\begin{equation}
\vec{F}^{\mathrm{act}}_k
=
-F_0
\frac{\varphi(r_{ik})}
{\sum_{k'\in \mathcal{N}_i} \varphi(r_{ik'})}
\hat{r}_{ij}.
\label{eq:dipole_i}
\end{equation}
Similarly, for a bead $l\in\mathcal{N}_j$ lying within the spherical neighborhood of bead $j$, the active force on bead $l$ due to bead $j$ is
\begin{equation}
\vec{F}^{\mathrm{act}}_l
=
F_0
\frac{\varphi(r_{jl})}
{\sum_{l'\in \mathcal{N}_j} \varphi(r_{jl'})}
\hat{r}_{ij}.
\label{eq:dipole_j}
\end{equation}
Here, $\mathcal{N}_i$ and $\mathcal{N}_j$ denote the sets of neighboring beads within spherical regions of radius $R_m$ centered on beads $i$ and $j$, respectively, excluding each central bead and its bonded neighbors. The distance-dependent weight function is defined as
\[
\varphi(r)
=
\left(1-\frac{r}{R_m}\right)^2,
\qquad r<R_m,
\]
and $\varphi(r)=0$ otherwise.
With this sign convention, Eqs.~\ref{eq:dipole_i} and~\ref{eq:dipole_j} correspond to an extensile force dipole. A contractile force dipole is obtained by reversing the signs of the active forces, equivalently by replacing $F_0$ with $-F_0$. The velocity field generated by an extensile dipole shows outward flow along the bond axis and inward flow along the transverse direction (Fig.~\ref{fig:velocity_field_ext}). For the contractile dipole, the flow directions are reversed, with inward flow along the bond axis and outward flow along the transverse direction (Fig.~\ref{fig:velocity_field_cont}). The dipolar radial velocity amplitude scales as $r^{-2}$, consistent with the far-field stresslet flow generated by a force dipole~\cite{LaugaPowers2009Hydrodynamics} (see Fig.~\ref{fig:velocity_scaling}).

All simulations are performed using LAMMPS~\cite{Thompson2022LAMMPS,Brown11} (version 29 August 2024), supplemented with custom fixes developed to implement the active force dipoles along the polymer backbone. The equations of motion are
integrated using the Shardlow splitting scheme~\cite{Gissinger2024LAMMPSFramework,Larentzos2014DPDShardlow,
Lisal2011DPDShardlow} with a time step $\delta t=0.02\tau$.

The system is simulated in a cubic box with periodic boundary conditions to mimic an unbounded fluid. The total number of beads is fixed at $N=384000$, with bead number density $\rho=3$. Unless otherwise stated, each polymer consists of $N_{\mathrm{poly}}=125$ beads, and the number of polymers is denoted by $N_p$. The combined number of polymer beads and $S_1$ solvent beads is fixed at $40000$, while the remaining beads correspond to the surrounding solvent $S_2$.

To distinguish the effects of increasing polymer number from those of increasing activity, the total active forcing is kept constant by imposing $N_p|F_0| =\mathrm{const}$. Accordingly, for $N_p|F_0|=40$, simulations are performed with $N_p=1$, $5$, and $10$, corresponding to $|F_0|=40$, $8$, and $4$, respectively. Likewise, for $N_p|F_0|=240$, we use $N_p=10$, $20$, $30$, and $40$, corresponding
to $|F_0|=24$, $12$, $8$, and $6$, respectively.

\section{Results and Discussion}
\label{sec:results}

\begin{figure*}[!ht]
    \centering
    \begin{subfigure}[t]{0.98\textwidth}
        \captionsetup{position=top, justification=raggedright, singlelinecheck=false}
        \includegraphics[width=\linewidth]{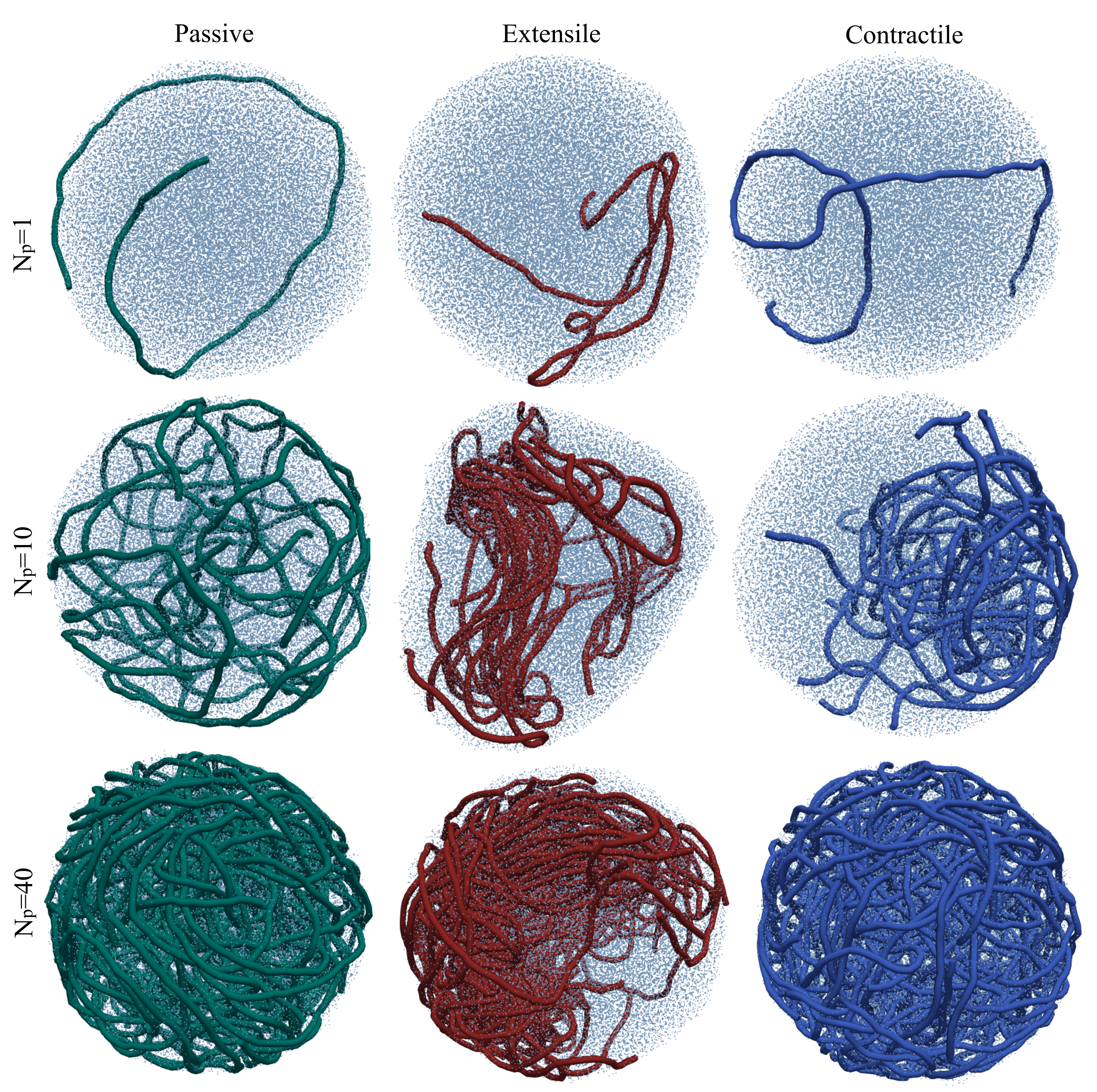}
    \end{subfigure}
    \captionsetup{position=top, justification=raggedright, singlelinecheck=false}
    \caption{\textbf{Representative simulation snapshots.}
    Polymers confined within a fluid droplet for passive, extensile, and contractile activity at $N_p=1$, $10$, and $40$. In the simulation all beads have the same diameter; the droplet-forming $S_1$ solvent beads are rendered smaller, and $S_2$ beads are not shown, for visual clarity.}
    \label{fig:poly_confined_snapshot}
\end{figure*}

\subsection{Organization of confined polymers}

\begin{figure*}[!ht]
    \justifying
    \begin{subfigure}[t]{0.48\textwidth}
        \captionsetup{position=top, justification=raggedright, singlelinecheck=false}
        \caption{}
        \includegraphics[width=\linewidth]{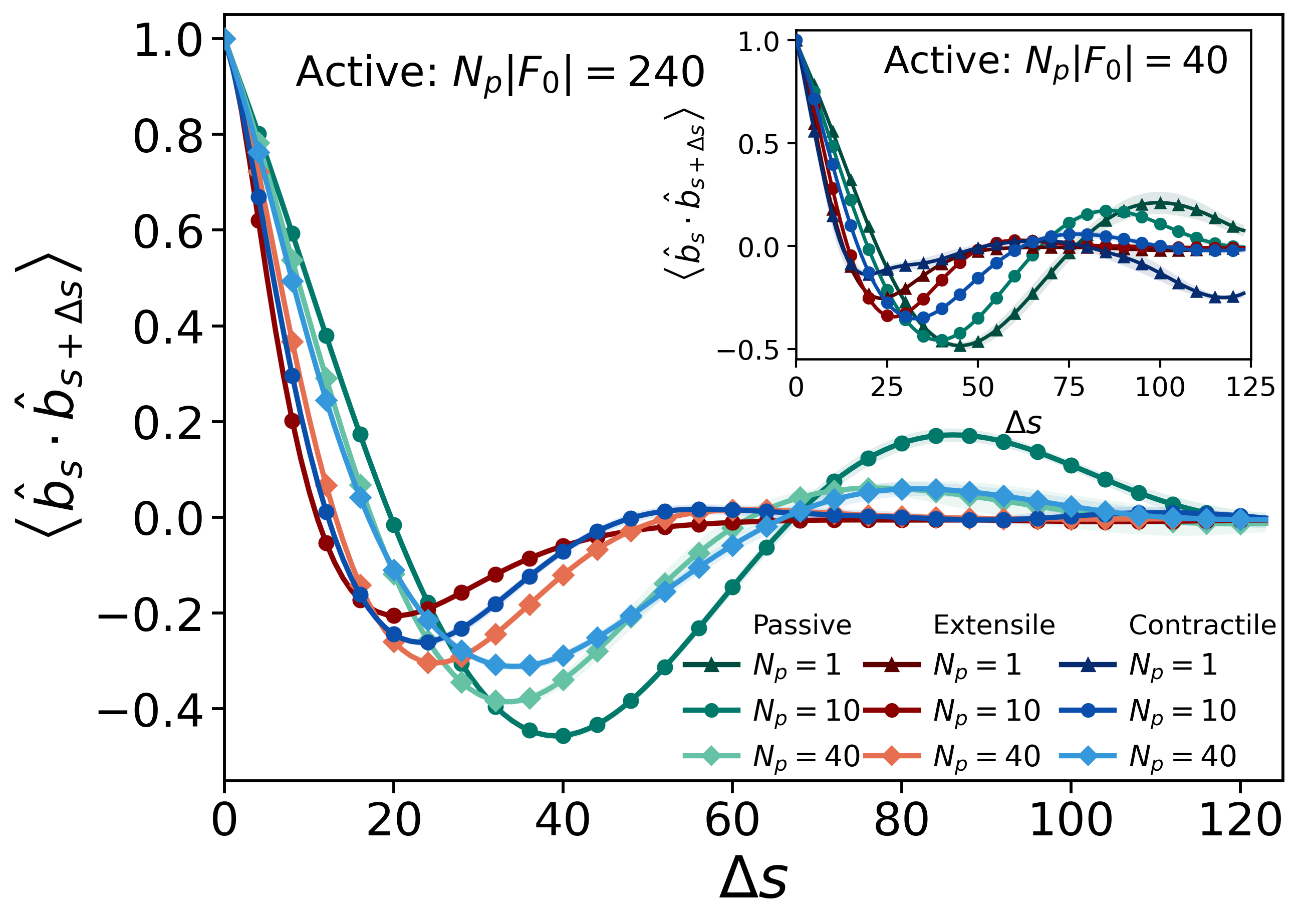}
        \label{fig:tangent_corr_all}
    \end{subfigure}
    \hfill
    \begin{subfigure}[t]{0.48\textwidth}
        \captionsetup{position=top, justification=raggedright, singlelinecheck=false}
        \caption{}
        \includegraphics[width=\linewidth]{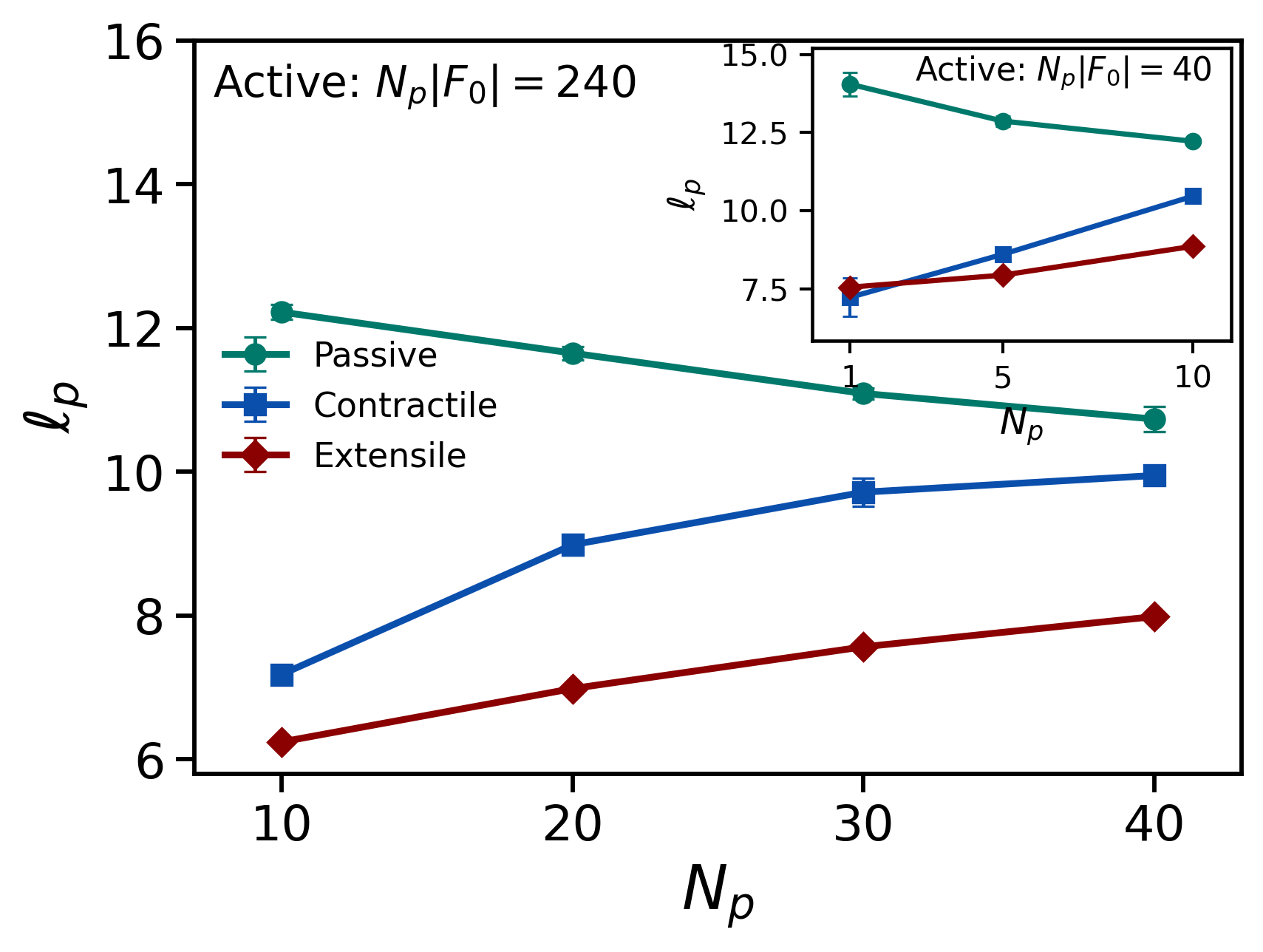}
        \label{fig:effective_persistence_length}
    \end{subfigure}
    \begin{subfigure}[t]{0.48\textwidth}
        \captionsetup{position=top, justification=raggedright, singlelinecheck=false}
        \caption{}
        \includegraphics[width=\linewidth]{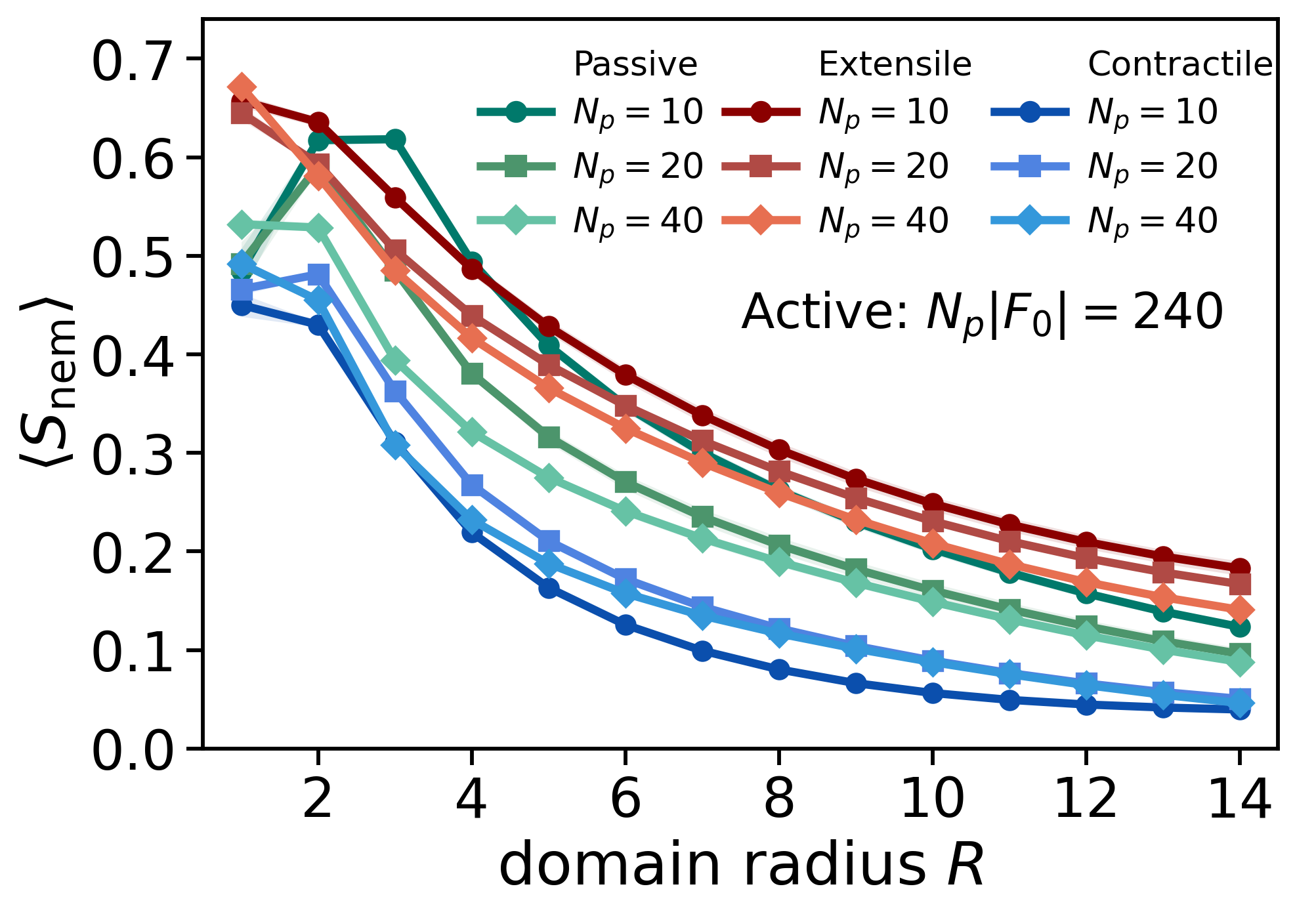}
        \label{fig:nematic_all_comparison}
    \end{subfigure}
    \hfill
    \begin{subfigure}[t]{0.48\textwidth}
        \captionsetup{position=top, justification=raggedright, singlelinecheck=false}
        \caption{}
        \includegraphics[width=\linewidth]{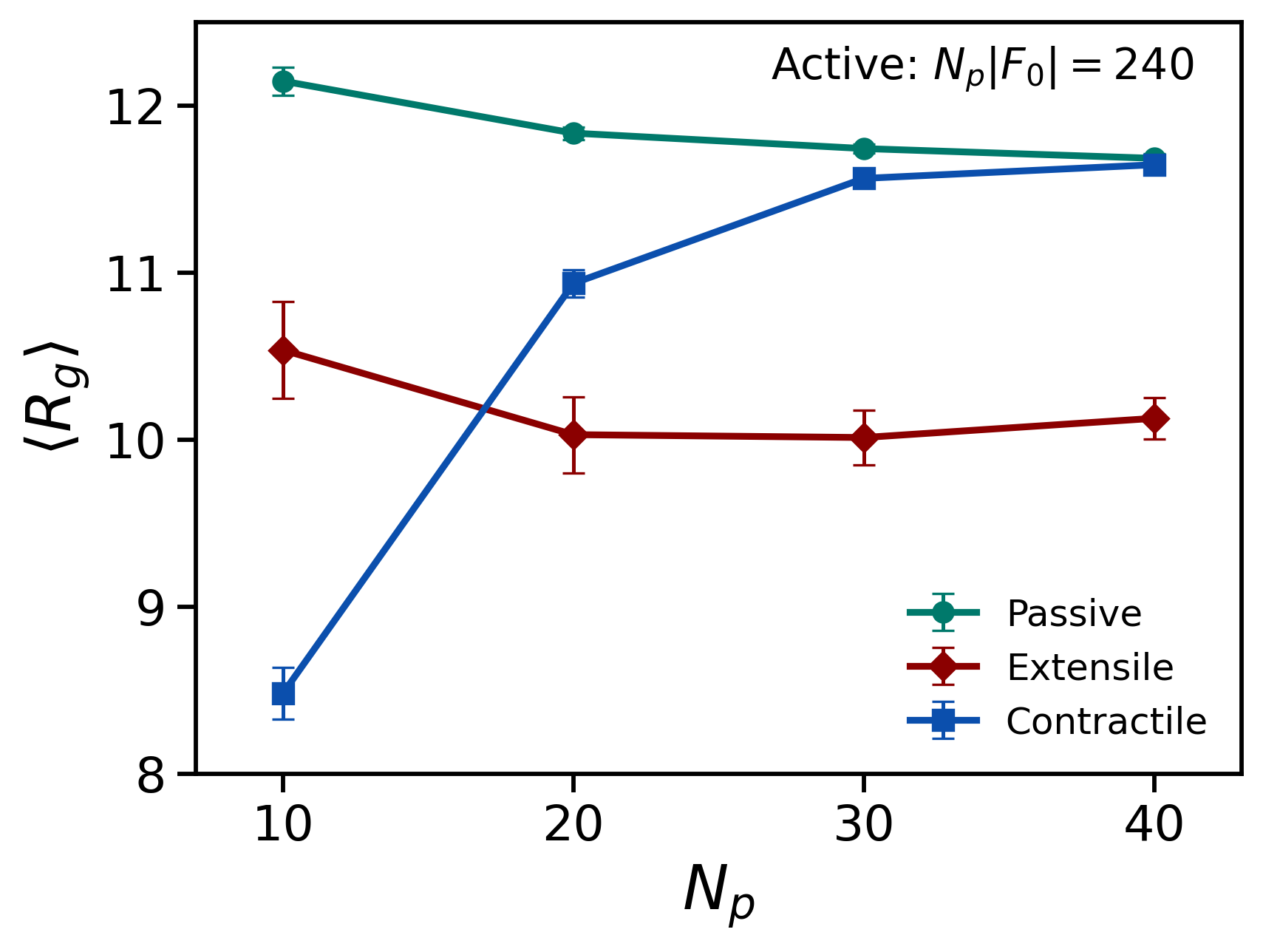}
        \label{fig:Rg_vs_Np}
    \end{subfigure}
    \captionsetup{position=top, justification=raggedright, singlelinecheck=false}
    \caption{\textbf{Internal organization of confined polymers.}
    \textbf{(a)} Bond-orientational correlation function, $\langle \hat{b}_s \cdot \hat{b}_{s+\Delta s}\rangle$, as a function of the contour separation $\Delta s$.
    \textbf{(b)} Effective persistence length, $\ell_p$, extracted from the initial decay of the bond-orientational correlation function, as a function of $N_p$.
    \textbf{(c)} Local nematic order parameter, $\langle S_{\rm nem}(R)\rangle$, as a function of the domain radius $R$ at different values of $N_p$ for passive polymers and for extensile and contractile polymers with $N_p|F_0|=240$.
    \textbf{(d)} Mean radius of gyration of the polymer cluster, $\langle R_g\rangle$, as a function of $N_p$. Error bars represent standard deviations over independent simulations and are shown only when larger than the data symbols.}
    \label{fig:internal_organization}
\end{figure*}
The collective organization of confined polymers is rooted in the activity-dependent conformations of individual chains. To quantify these conformational changes, we compute the bond-orientational correlation function, $\langle \hat{b}_s \cdot \hat{b}_{s+\Delta s} \rangle$, where $\hat{b}_{s}$ is the unit bond vector connecting monomers $s$ and $s+1$, and $\Delta s$ denotes the separation between them along the polymer contour. This correlation measures the persistence of bond orientations along the polymer backbone (Fig.~\ref{fig:tangent_corr_all}, main panel for $N_p|F_0|=240$ and inset for $N_p|F_0|=40$). To quantify the short-range orientational persistence, we extract an effective persistence length, $\ell_p$, from the initial decay of the bond-orientational correlation function (Fig.~\ref{fig:effective_persistence_length}).     For passive polymers, whose contour length exceeds the droplet diameter, confinement causes the chains to wrap around the droplet boundary to minimize bending energy. As a result, the bond orientation reverses along the contour, giving rise to a change in sign of the correlation function at a contour separation corresponding approximately to half the circumference of the droplet (Fig.~\ref{fig:tangent_corr_all}). For extensile filaments, local curvature biases the active hydrodynamic flows, producing filament propulsion in the direction of its curvature. This positive feedback amplifies the bent configuration and stabilizes a curved, motile conformation, leading to a faster decay of bond-orientational correlations and a lower effective persistence length than for passive polymers~\cite{Jayaraman2012AutonomousMotility,MannaKumar2019ActivePolymericFluids}. This behavior is evident in Fig.~\ref{fig:tangent_corr_all}, where the bond-orientational correlations of extensile polymers decay more rapidly than those of contractile and passive polymers, and is reflected in their correspondingly lower effective persistence length, $\ell_p$ (Fig.~\ref{fig:effective_persistence_length}).

Contractile polymers exhibit qualitatively different behavior. It is known that, in the absence of confinement, hydrodynamic interactions generated by contractile activity suppress local bending, rendering isolated polymers stiffer than their passive counterparts~\cite{Jayaraman2012AutonomousMotility,Laskar2013HydrodynamicInstabilities}; see also Fig.~\ref{fig:bond_corr_single_poly}. However, hydrodynamic interactions between contractile segments favor mutually perpendicular configurations~\cite{PandeyAnkita2016}. Under strong confinement, for polymers whose contour length exceeds the droplet diameter, these interactions drive the chains into compact, self-linked conformations. Consequently, the bond-orientational correlation of a confined contractile polymer decays more rapidly than that of its passive counterpart, demonstrating a reversal of the unconfined single-polymer behavior (Fig.~\ref{fig:bond_corr_single_poly_conf}). Consistent with this, the bond-orientational correlations of contractile polymers decay faster than those of passive polymers, as shown in Fig.~\ref{fig:tangent_corr_all}. As the number of confined polymers increases, intrachain perpendicular contacts are progressively replaced by interchain contacts, reducing self-linking and enhancing orientational correlations along individual polymers (Fig.~\ref{fig:tangent_corr_all}). This trend is reflected in the increase of $\ell_p$ with $N_p$ for contractile polymers (Fig.~\ref{fig:effective_persistence_length}).

The distinct single-polymer conformations translate directly into different modes of collective organization within the droplet, as illustrated by the representative snapshots in Fig.~\ref{fig:poly_confined_snapshot} and Supplementary Movie~1.
Extensile activity leads to pronounced polymer bundling, whereas contractile activity results in a more dispersed organization. This behavior originates from solvent-mediated hydrodynamic interactions generated by the active force dipoles. As established in earlier studies~\cite{PandeyAnkita2016,MannaKumar2019ActivePolymericFluids}, extensile dipoles promote parallel alignment and effective attraction between neighboring polymer segments, while contractile dipoles favor predominantly perpendicular configurations. To quantify this orientational organization, we compute the local nematic order (see Sec.~\ref{sec:nematic_order} of the supplementary material). For each bond vector, the nematic order tensor $\mathbf{Q}$ is constructed by averaging over neighboring bond vectors within a spherical region of radius R. The corresponding local nematic order parameter, $S_{\rm nem}(R)$, is defined as the largest eigenvalue of $\mathbf{Q}$~\cite{deGennesProst1993,MilchevEtAl2021}.

Figure~\ref{fig:nematic_all_comparison} shows that extensile polymers exhibit the largest $\langle S_{\rm nem}(R)\rangle$ over a broad range of length scales, indicating strong local alignment that extends over relatively large domains. This observation is consistent with the well-established tendency of extensile active filaments to align and form bundles through hydrodynamic interactions ~\cite{PandeyAnkita2016,MannaKumar2019ActivePolymericFluids}. Passive polymers display an intermediate degree of local nematic order, arising primarily from confinement-induced orientational correlations. In contrast, contractile polymers exhibit the weakest local nematic order, consistent with the tendency of contractile dipoles to orient neighboring segments  perpendicular to one another. In all cases, the ordering is short-ranged, with $\langle S_{\rm nem}(R)\rangle$ decreasing monotonically as $R$ increases. Consequently, the global nematic order parameter remains small for all systems (Fig.~\ref{fig:global_nematic}). 

The activity-induced polymer clustering is quantified through the mean radius of gyration, $\langle R_g\rangle$, defined in Eq.~\eqref{eq:rg}, as shown in Fig.~\ref{fig:Rg_vs_Np}. Extensile polymers remain more compact than passive polymers over the entire range of $N_p$. Moreover, increasing $N_p$ at fixed total activity produces only a weak change in $R_g$, indicating that extensile polymers maintain compact conformations even at higher polymer number.  Contractile polymers show a markedly different trend. For $N_p|F_0|=240$, the contractile cluster is strongly compact at $N_p=10$ and exhibits the smallest $\langle R_g\rangle$ among the three systems. This compactness arises from perpendicular segment--segment interactions, which promote self-linked, network-like conformations, as can be seen in Fig.~\ref{fig:nematic_all_comparison}. As $N_p$ increases, geometrical crowding and the progressive replacement of intrachain contacts by interchain contacts cause the contractile network to spread throughout the droplet. At the same time, the decrease in $|F_0|$ weakens activity-induced compaction. Consequently, $\langle R_g\rangle$ increases and approaches the passive-polymer value at larger $N_p$. To determine whether the increase in $R_g$ arises solely from the decrease in activity per monomer, we performed additional simulations at fixed dipole strength, $|F_0|=6$. Fig.~\ref{fig:Rg_vs_Np_fix_F} shows that $\langle R_g\rangle$ continues to increase with $N_p$, confirming that geometrical constraints and interpolymer interactions also contribute to the expansion of the contractile cluster.

The propensity of extensile polymers to form bundles is reflected in the shape of the resulting polymer clusters. This is quantified using the normalized cluster asphericity, $A$, defined in Eq.~\eqref{eq:asphericity}, which measures the deviation of the cluster from a spherical geometry (Fig.~\ref{fig:Asphericity}). For $N_p|F_0|=240$, extensile clusters exhibit the largest asphericity over the entire range of $N_p$, demonstrating that hydrodynamic alignment produces strongly anisotropic aggregates (Fig.~\ref{fig:Asphericity_a_2}). Although the asphericity decreases with increasing $N_p$, extensile clusters remain more anisotropic than contractile and passive clusters.  Contractile clusters progressively become more isotropic with increasing $N_p$, and at $N_p=40$ their asphericity approaches that of passive polymer clusters. A similar trend is observed for $N_p|F_0|=40$ (Fig.~\ref{fig:Asphericity_a_1}), except at $N_p=1$, where the confined contractile polymer exhibits the largest asphericity because of its compact, self-linked conformation.

\subsection{Interface fluctuations and shape modes}

\begin{figure*}[!ht]
    \justifying
    \begin{subfigure}[t]{0.32\textwidth}
        \captionsetup{position=top, justification=raggedright, singlelinecheck=false}
        \caption{}
        \includegraphics[width=\linewidth]{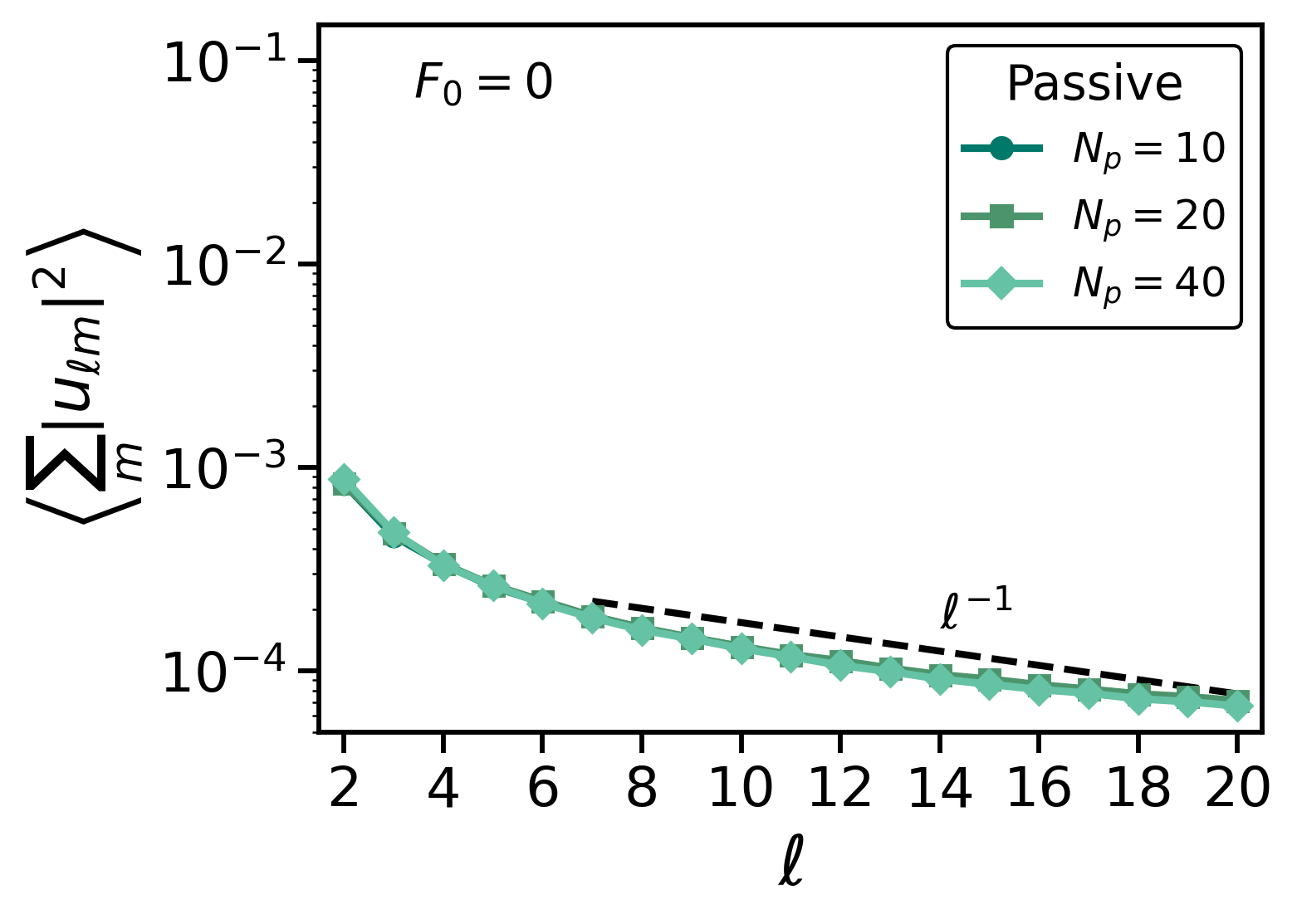}
        \label{fig:spectrum_passive}
    \end{subfigure}
    \hfill
    \begin{subfigure}[t]{0.32\textwidth}
        \captionsetup{position=top, justification=raggedright, singlelinecheck=false}
        \caption{}
        \includegraphics[width=\linewidth]{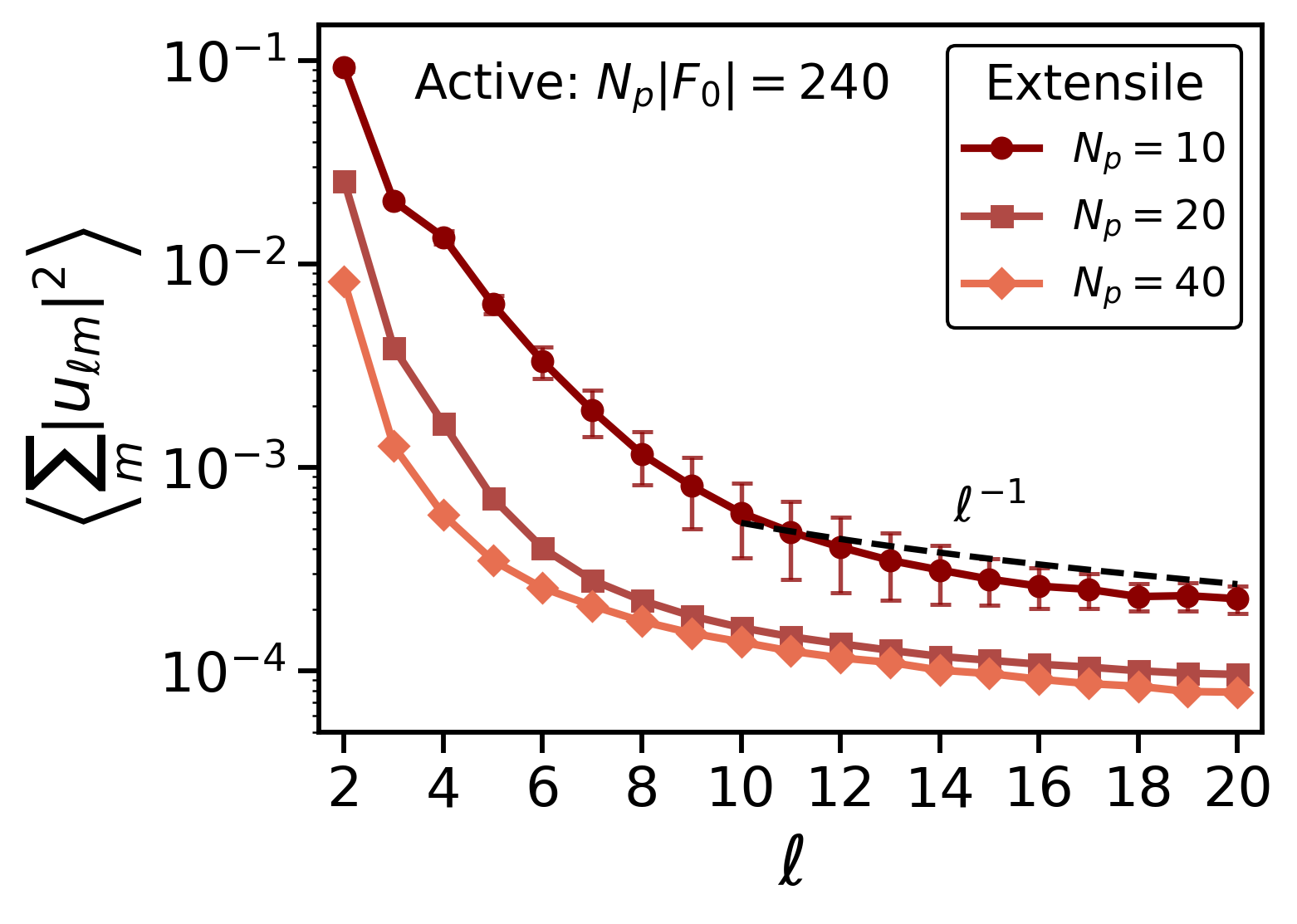}
        \label{fig:spectrum_extensile}
    \end{subfigure}
    \hfill
    \begin{subfigure}[t]{0.32\textwidth}
        \captionsetup{position=top, justification=raggedright, singlelinecheck=false}
        \caption{}
        \includegraphics[width=\linewidth]{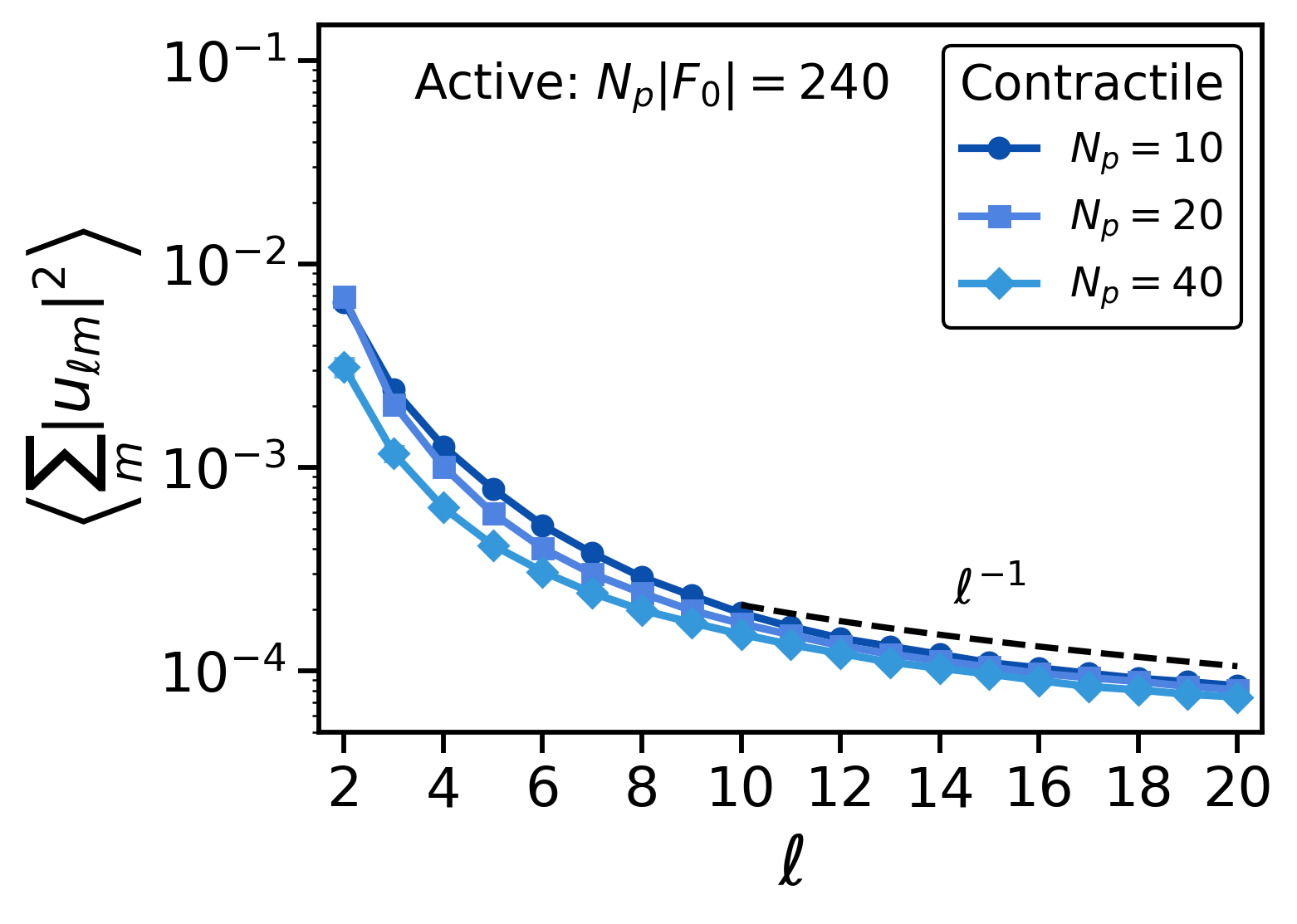}
        \label{fig:spectrum_contractile}
    \end{subfigure}
    \begin{subfigure}[t]{0.32\textwidth}
        \captionsetup{position=top, justification=raggedright, singlelinecheck=false}
        \caption{}
        \includegraphics[width=\linewidth]{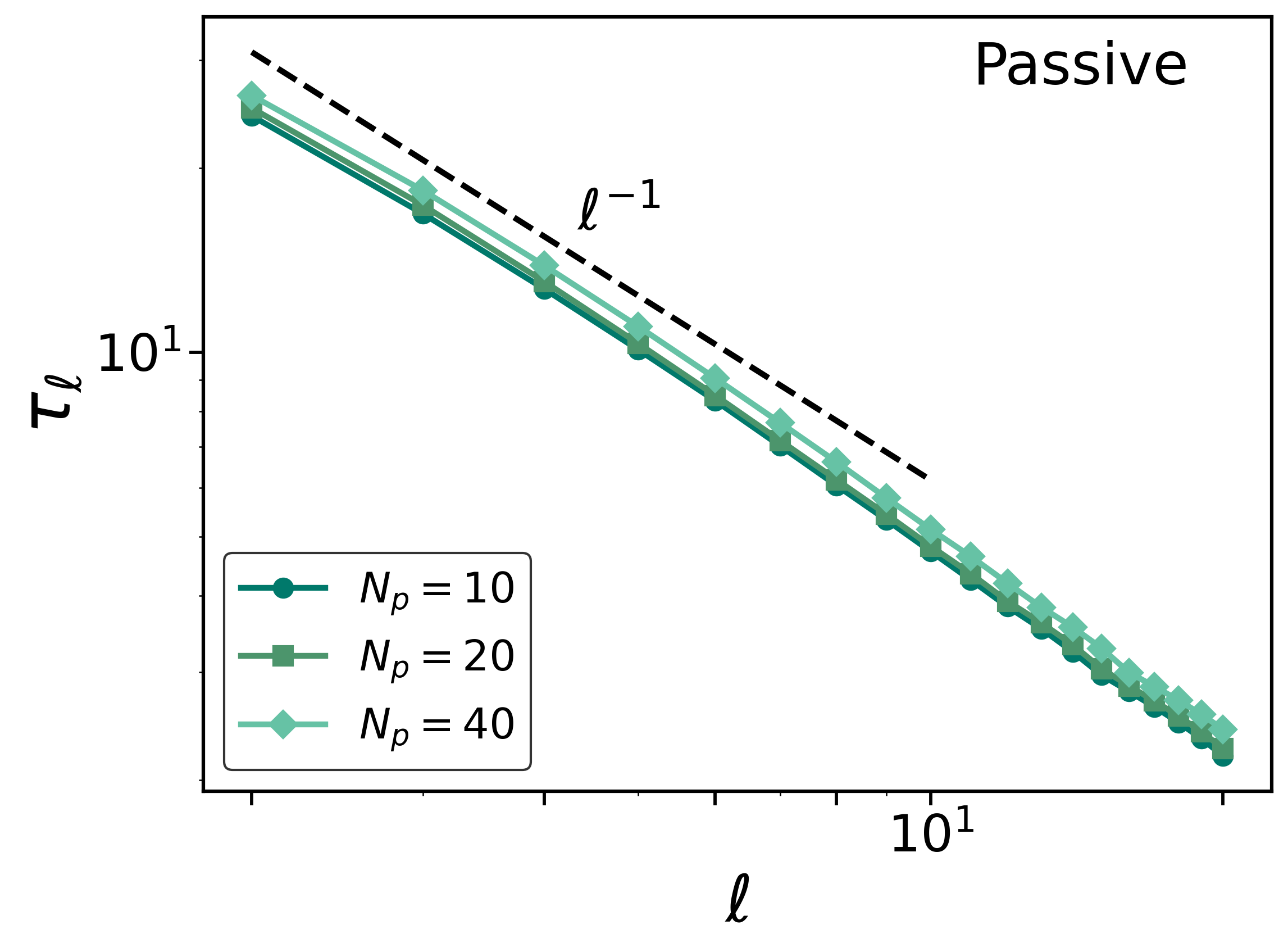}
        \label{fig:corr_time_passive}
    \end{subfigure}
    \hfill
    \begin{subfigure}[t]{0.32\textwidth}
        \captionsetup{position=top, justification=raggedright, singlelinecheck=false}
        \caption{}
        \includegraphics[width=\linewidth]{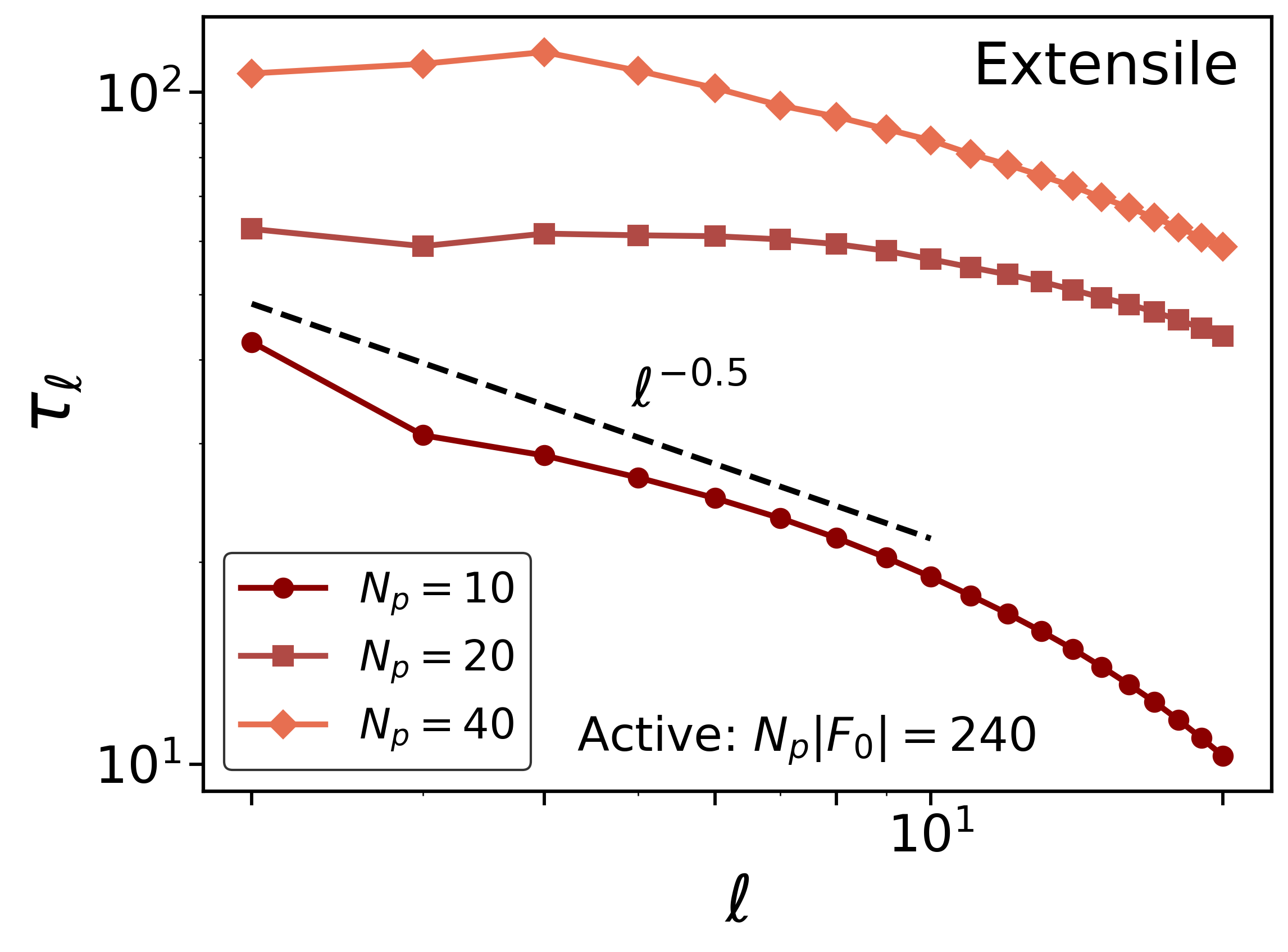}
        \label{fig:corr_time_extensile}
    \end{subfigure}
    \hfill
    \begin{subfigure}[t]{0.32\textwidth}
        \captionsetup{position=top, justification=raggedright, singlelinecheck=false}
        \caption{}
        \includegraphics[width=\linewidth]{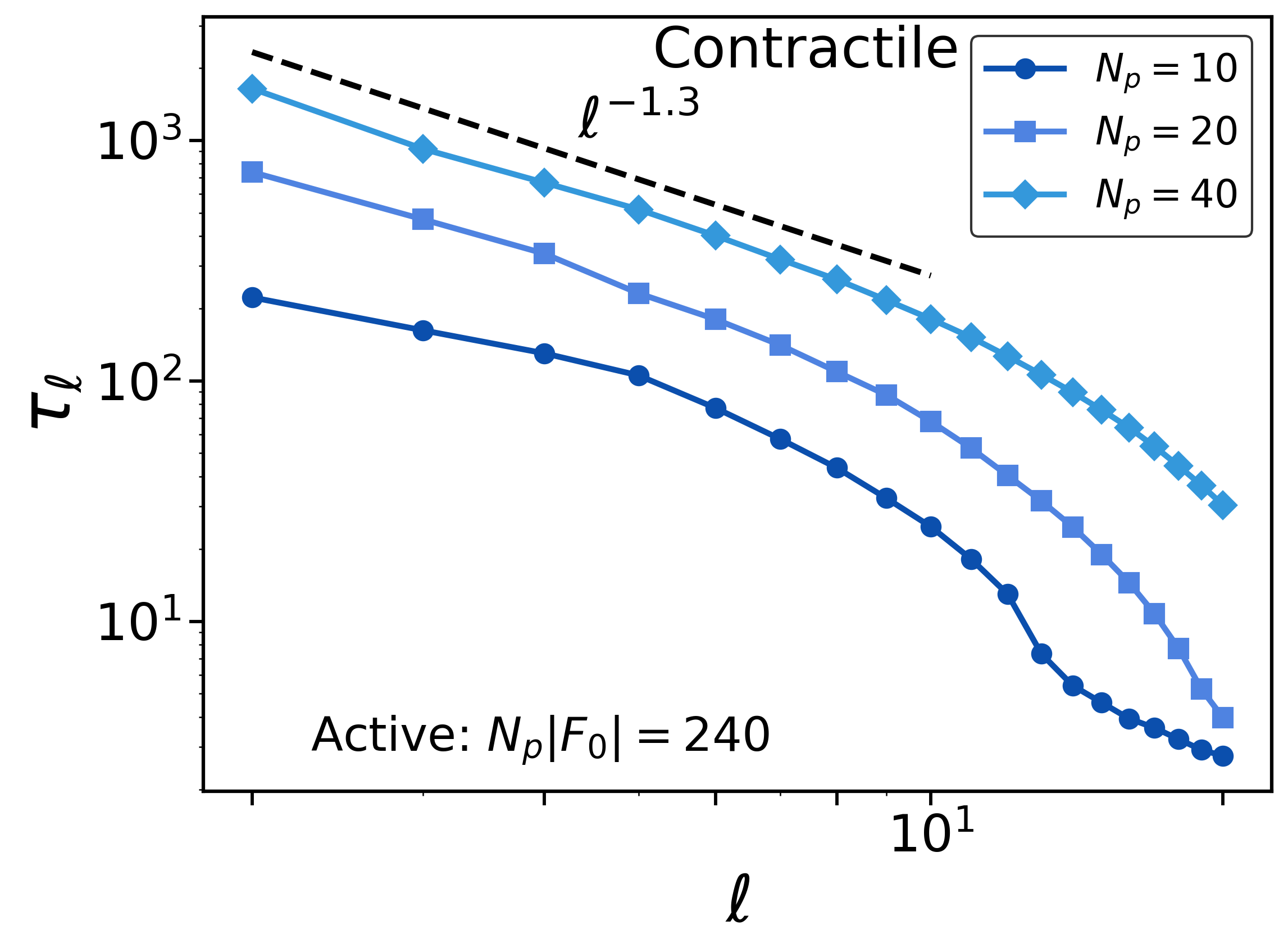}
        \label{fig:corr_time_contractile}
    \end{subfigure}
    \captionsetup{position=top, justification=raggedright, singlelinecheck=false}
    \caption{\justifying 
    \textbf{Interfacial fluctuation spectra and shape-mode relaxation.} Interfacial shape-fluctuation spectrum, $\left\langle \sum_m |u_{\ell m}|^2 \right\rangle$, as a function of the angular mode $\ell$ for droplets containing \textbf{(a)} passive, \textbf{(b)} extensile, and \textbf{(c)} contractile polymers at different $N_p$. Shape-mode relaxation time, $\tau_{\ell}$, obtained from the first crossing of $C_{\ell}(\Delta t)=e^{-1}$, as a function of $\ell$ for droplets containing \textbf{(d)} passive, \textbf{(e)} extensile, and \textbf{(f)} contractile polymers. For the active systems, the total activity is fixed such that $N_p|F_0|=240$. Error bars represent standard deviations over independent simulations and are shown only when larger than the data symbols.}
    \label{fig:shape_fluctuation}
\end{figure*}
 
The droplet--solvent interface enclosing the polymers is influenced by polymer activity. The corresponding activity-dependent interfacial dynamics are shown in Supplementary Movie~2. To characterize its fluctuations, the instantaneous interface is represented by the radial distance of each interfacial bead from the droplet center of mass, $r(\theta,\phi)$, where $\theta$ and $\phi$ are the polar and azimuthal angles, respectively (see supplementary material, Sec.~\ref{sec:droplet_interface}). The interface is written as
\begin{equation}
r(\theta,\phi)=R_0\left[1+u(\theta,\phi)\right],
\label{eq:interface_radius}
\end{equation}
where $R_0$ is the mean droplet radius and $u(\theta,\phi)$ is the dimensionless radial deviation from the mean spherical shape. For the simulated droplets considered here, $R_0 \approx 14.2$.

The dimensionless fluctuation field $u(\theta,\phi)$ is then expanded in spherical harmonics as
\begin{equation}
u(\theta,\phi) = \sum_{\ell=0}^{\infty}\sum_{m=-\ell}^{\ell} 
u_{\ell m}Y_{\ell m}(\theta,\phi).
\label{eq:spherical_harmonic_expansion}
\end{equation}

The shape fluctuation spectrum is quantified by $\left\langle \sum_{m=-\ell}^{\ell} |u_{\ell m}|^2 \right\rangle$, which gives the total orientation-independent fluctuation amplitude for each angular mode $\ell$ (see supplementary material sec.~\ref{sec:spherical_harmonic_analysis}). The angular brackets denote averaging over time and independent trajectories. 
The interfacial fluctuation spectra for $N_p|F_0|=240$ are shown in Fig.~\ref{fig:shape_fluctuation}(a--c), while the corresponding results for $N_p|F_0|=40$ are presented in Supplementary Fig.~\ref{fig:shape_fluctuation_a_1}(a--c).

For passive droplets, the fluctuation spectra are nearly independent of the number of confined polymers, $N_p$, and are consistent with equilibrium thermal capillary fluctuations (Fig.~\ref{fig:spectrum_passive} and supplementary material Fig.~\ref{fig:spectrum_passive_a_1}). In particular, the mode amplitudes, $\langle \sum_m |u_{lm}|^2 \rangle$, decrease as $1/\ell$, as expected for a tension-dominated fluid interface. The spectra are well described by the equilibrium capillary-wave theory, yielding an effective surface tension that is essentially independent of polymer concentration, $\sigma \approx 7.4$ (see supplementary material, Sec.~\ref{sec:passive_fluctuation} and Fig.~\ref{fig:passive_spectrum_fit}). 
These observations indicate that, over the range of polymer concentrations considered here, the passive polymers neither modify the interfacial properties of the droplet nor produce an appreciable change in its effective viscosity.

For droplets containing extensile active polymers, at lower modes,  the interfacial fluctuation spectrum deviates markedly from the equilibrium $1/\ell$ scaling observed for passive droplets, indicating a breakdown of equipartition due to active forcing. As shown in Figs.~\ref{fig:spectrum_extensile} and \ref{fig:spectrum_extensile_a_1}, the extent of this deviation depends strongly on the number of confined polymers. For $N_p|F_0|=240$, the fluctuation amplitudes are largest at $N_p=10$ over the entire range of $\ell$, substantially exceeding those of both passive and contractile systems (Fig.~\ref{fig:spectrum_extensile}). This enhancement reflects the coherent solvent flows generated by extensile polymers, which efficiently transfer active stresses to the fluid--fluid interface and excite large-scale droplet deformations. As $N_p$ increases, the fluctuation amplitudes decrease significantly, with the spectra for $N_p=20$ and $40$ becoming nearly indistinguishable at larger $\ell$. At the lower total activity, $N_p|F_0|=40$, the fluctuation amplitude is likewise largest at the smallest polymer number, $N_p=1$, and decreases with increasing $N_p$, with the spectra for $N_p=5$ and $10$ becoming nearly indistinguishable at larger $\ell$ (Fig.~\ref{fig:spectrum_extensile_a_1}). Since $N_p|F_0|$ is kept constant, the decrease in fluctuation amplitude with increasing $N_p$ cannot be attributed to a reduction in the overall active forcing. Instead, increasing $N_p$ redistributes the activity among a larger number of polymers, thereby reducing $|F_0|$ per monomer and promoting the formation of hydrodynamically coupled bundles. These changes reduce the efficiency with which active stresses excite fluctuations of the droplet interface.
 
For droplets containing contractile polymers, the fluctuation spectra also depend on $N_p$, although this dependence is considerably weaker than for extensile polymers (Figs.~\ref{fig:spectrum_contractile} and \ref{fig:spectrum_contractile_a_1}). At fixed total activity, the fluctuation amplitudes decrease with increasing $N_p$. Contractile activity therefore enhances interfacial fluctuations relative to the passive case, but is less effective than extensile activity in exciting long-wavelength droplet deformations.
Similar to extensile systems, the spectrum deviates significantly from the $1/\ell$ scaling at lower modes.   


To characterize the dynamics of these interfacial deformations, we compute the temporal autocorrelation of the shape modes. The normalized autocorrelation function for mode $\ell$ is defined as
\begin{equation}
C_{\ell}(\Delta t)=
\frac{
\left\langle 
\sum_{m=-\ell}^{\ell}
u_{\ell m}(t+\Delta t)u_{\ell m}^{*}(t)
\right\rangle
}{
\left\langle
\sum_{m=-\ell}^{\ell}
|u_{\ell m}(t)|^2
\right\rangle
},
\label{eq:mode_correlation}
\end{equation}
where the average is taken over time origins and independent trajectories. The characteristic correlation time, $\tau_{\ell}$, is defined as the first time at which $C_{\ell}(\Delta t)$ decays to $e^{-1}$. Since DPD retains finite inertia, this definition provides a robust measure of the mode-relaxation time without assuming a purely exponential decay. For passive droplets, $\tau_{\ell}$ decreases approximately as $\ell^{-1}$, and the curves for different $N_p$ nearly collapse, as shown in Figs.~\ref{fig:corr_time_passive} and \ref{fig:corr_time_pass_a_1}. This indicates that, in the absence of activity, the relaxation dynamics of the interfacial modes are largely insensitive to the number of confined polymers and are governed primarily by capillary relaxation of the droplet interface.

 Activity not only modifies the amplitudes of the interfacial fluctuations but also strongly alters their relaxation dynamics (Figs.~\ref{fig:corr_time_extensile}, \ref{fig:corr_time_contractile}, \ref{fig:corr_time_ext_a_1}, and \ref{fig:corr_time_cont_a_1}). For droplets containing extensile polymers, the mode correlation times depend strongly on $N_p$ and do not follow a single scaling with $\ell$. For $N_p|F_0|=240$, $\tau_{\ell}$ increases with $N_p$, while its dependence on the mode number becomes progressively weaker (Fig.~\ref{fig:corr_time_extensile}). For $N_p\geq20$, where the polymers organize into hydrodynamically coupled bundles, the relaxation times vary only weakly with $\ell$, indicating that collective polymer dynamics dominate interfacial relaxation and suppress the characteristic capillary scaling.  At $N_p=10$, the relaxation retains a weak mode dependence, approximately following $\tau_{\ell}\sim\ell^{-0.5}$ over an intermediate range of modes. As shown in Fig.~\ref{fig:corr_time_ext_a_1}, this mode number dependence of correlation time is stronger at lower total activity, $N_p|F_0|=40$, approximately following $\tau_{\ell}\sim\ell^{-1.3}$ for $N_p=5$ and $10$.


Contractile droplets exhibit a stronger mode dependence than extensile droplets. For both total-activity levels, the correlation times decrease approximately as $\tau_{\ell}\sim\ell^{-1.3}$ over an intermediate range of modes, with deviations becoming apparent at larger $\ell$ (Figs.~\ref{fig:corr_time_contractile} and \ref{fig:corr_time_cont_a_1}). Unlike extensile droplets at large $N_p$, the contractile relaxation retains a pronounced dependence on the mode number, indicating that contractile activity does not suppress the characteristic mode-dependent interfacial relaxation as strongly. These observations highlight the fundamentally different ways in which extensile and contractile activity couple polymer dynamics to interfacial relaxation. A detailed analysis of this coupling will be presented elsewhere.
 

In summary, activity fundamentally alters both the amplitude and the relaxation dynamics of interfacial fluctuations. The breakdown of equilibrium capillary-wave behavior demonstrates that droplet-shape fluctuations are driven by the coupling between active polymer dynamics, the solvent flows they generate, and the deformable interface. 

\subsection{Droplet Motion under Activity}
\label{sec:droplet_motion}

\begin{figure*}[!ht]
    \justifying
    \begin{subfigure}[t]{0.48\textwidth}
        \captionsetup{position=top, justification=raggedright, singlelinecheck=false}
        \caption{}
        \includegraphics[width=\linewidth]{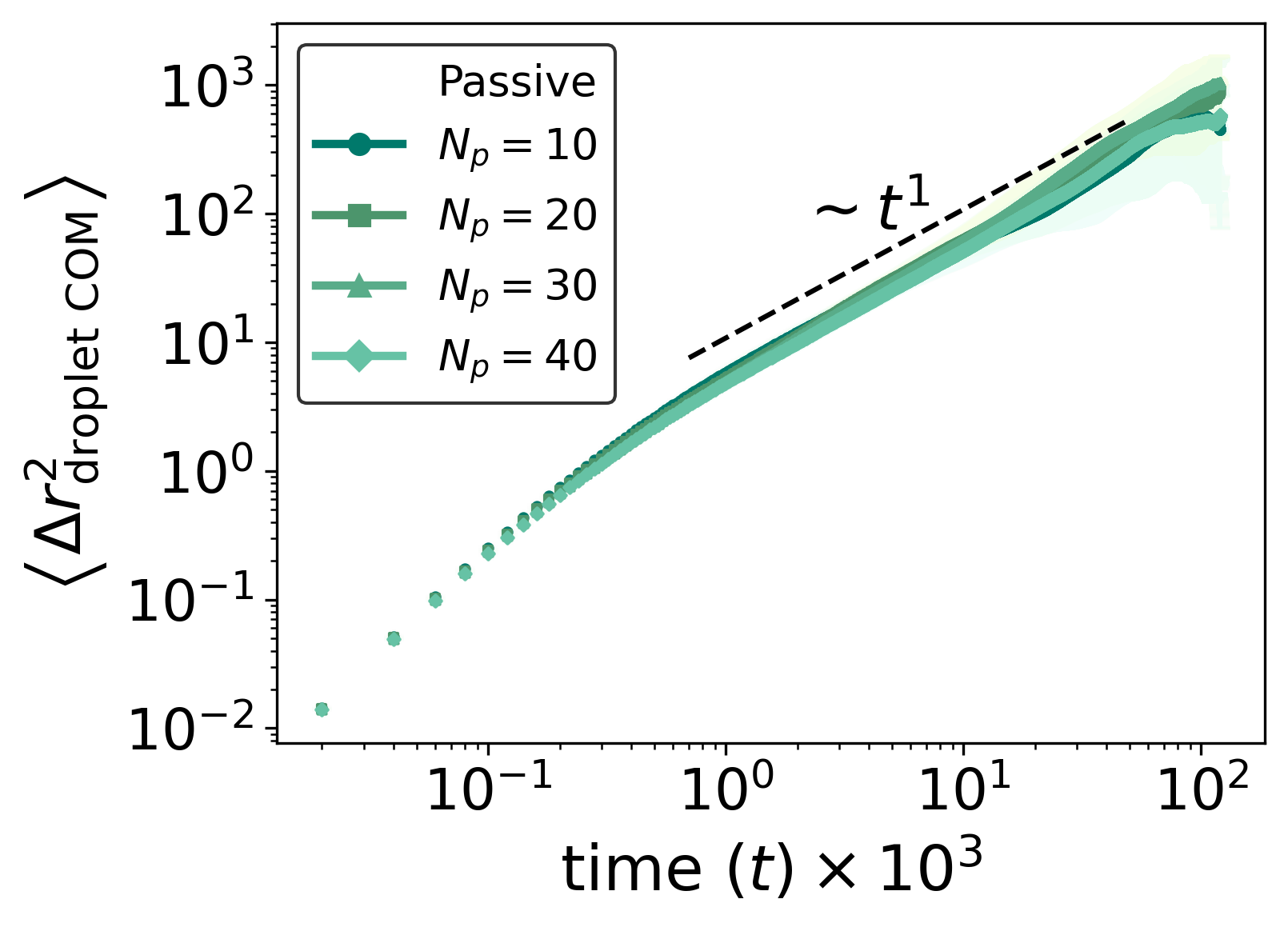}
        \label{fig:MSD_passive}
    \end{subfigure}
    \hfill
    \begin{subfigure}[t]{0.48\textwidth}
        \captionsetup{position=top, justification=raggedright, singlelinecheck=false}
        \caption{}
        \includegraphics[width=\linewidth]{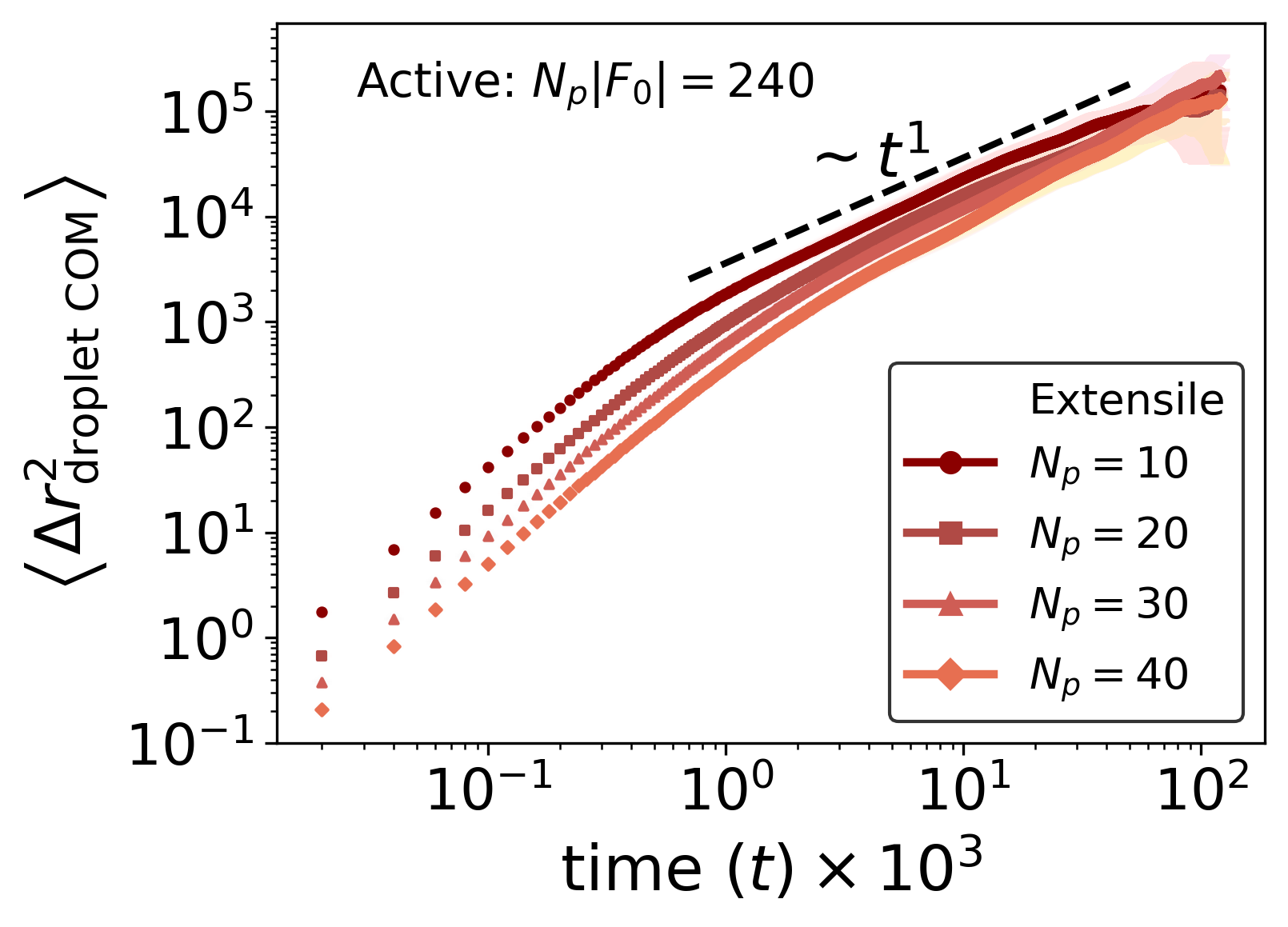}
        \label{fig:MSD_extensile}
    \end{subfigure}
    \begin{subfigure}[t]{0.48\textwidth}
        \captionsetup{position=top, justification=raggedright, singlelinecheck=false}
        \caption{}
        \includegraphics[width=\linewidth]{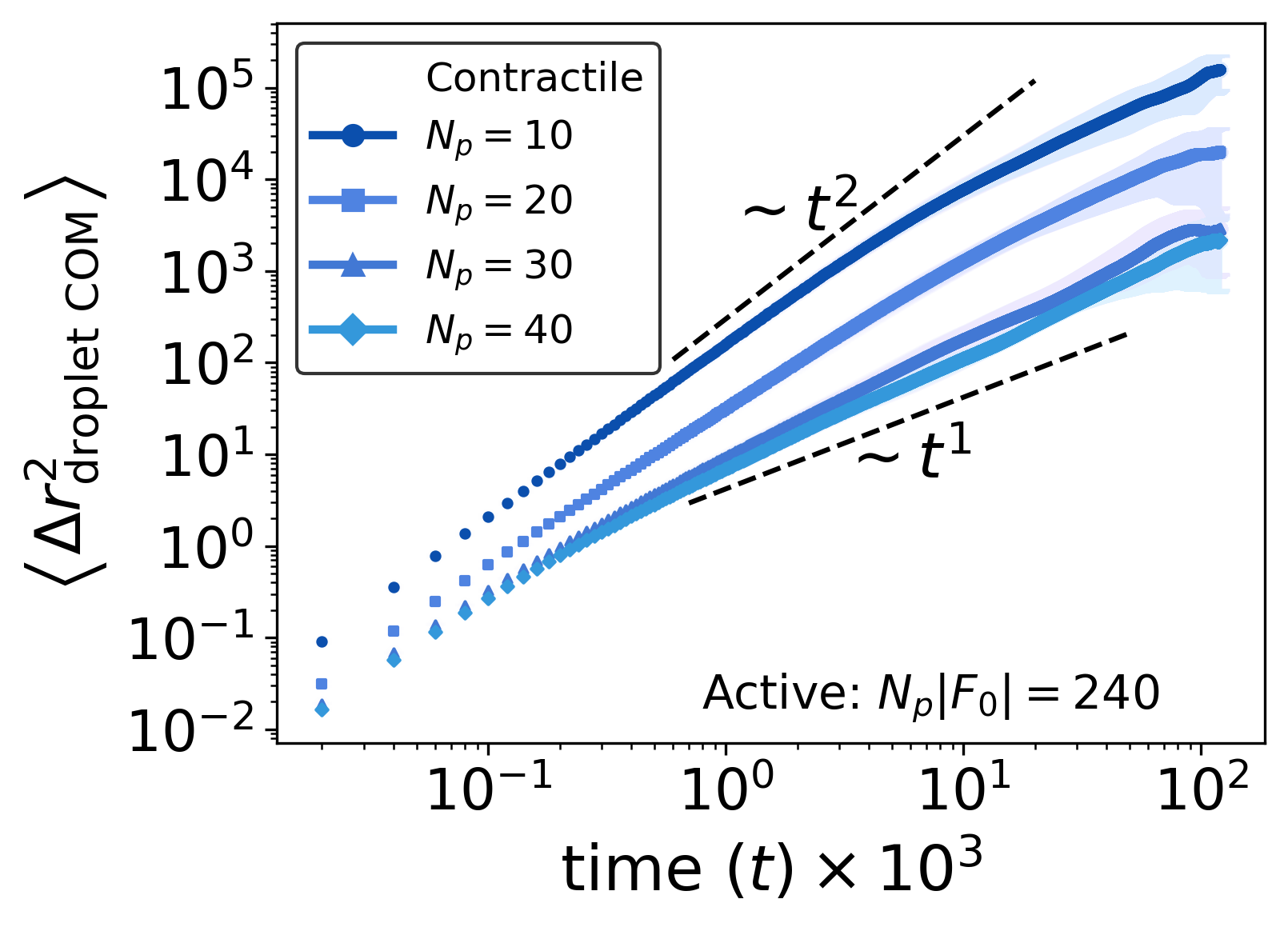}
        \label{fig:MSD_contractile}
    \end{subfigure}
    \hfill
    \begin{subfigure}[t]{0.48\textwidth}
        \captionsetup{position=top, justification=raggedright, singlelinecheck=false}
        \caption{}
        \includegraphics[width=\linewidth]{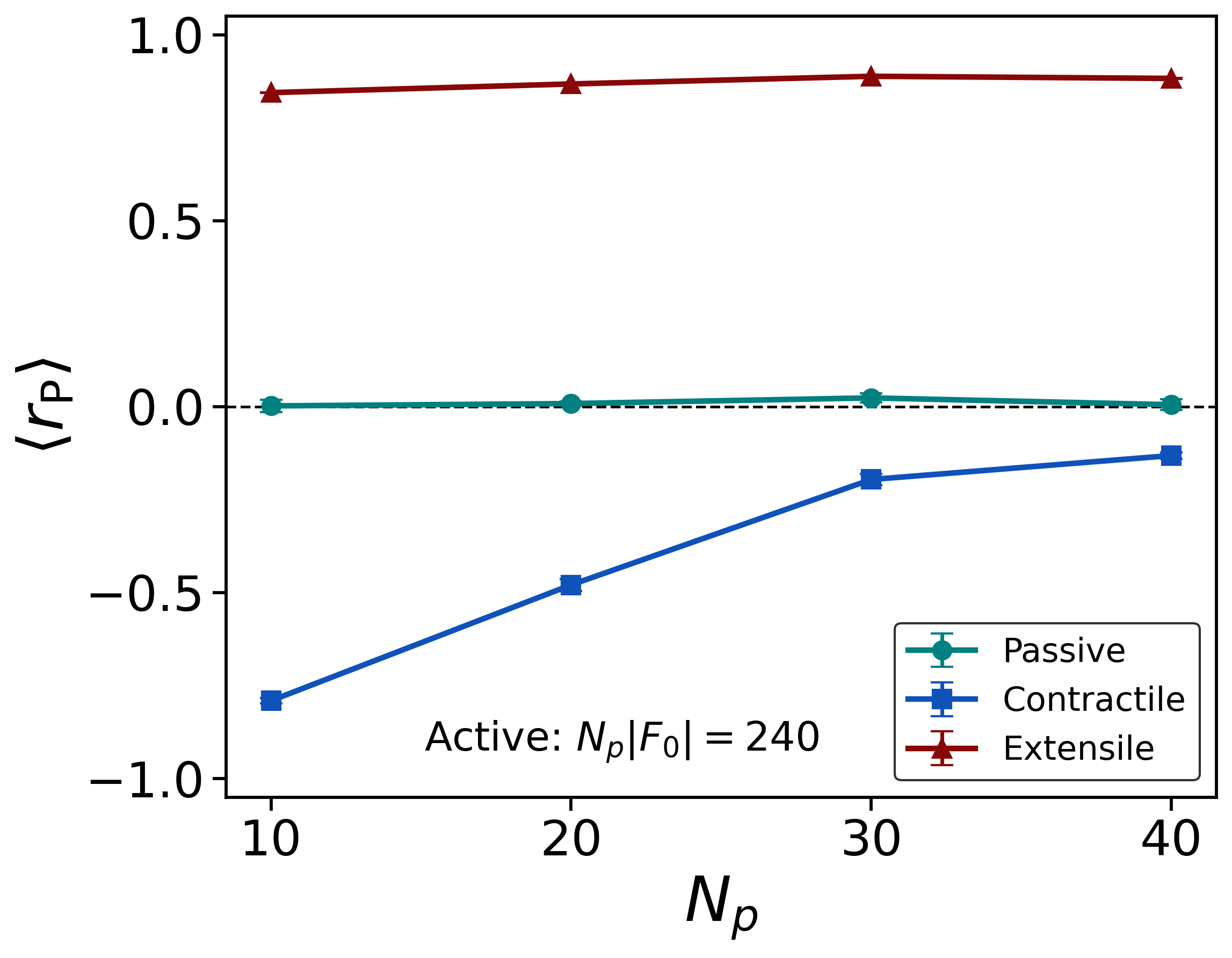}
        \label{fig:pearson_corr}
    \end{subfigure}
    \captionsetup{position=top, justification=raggedright, singlelinecheck=false}
    \caption{\justifying
    \textbf{Activity-dependent droplet motion.}
    Mean-square displacement of the droplet center of mass, $\left\langle \Delta r_{\mathrm{droplet\,COM}}^{2}\right\rangle$, for different polymer numbers, $N_p$, in droplets containing \textbf{(a)} passive, \textbf{(b)} extensile, and \textbf{(c)} contractile polymers.
    \textbf{(d)} Pearson correlation coefficient, $\langle r_p\rangle$, between the
    polymer center-of-mass offset relative to the droplet center, $\Delta\vec{R}_{\mathrm{COM}}
    =\vec{R}_{\mathrm{poly,COM}}-\vec{R}_{\mathrm{droplet,COM}}$, and the instantaneous droplet center-of-mass velocity, $\vec{v}_{\mathrm{droplet,COM}}$, as a function of $N_p$. For the active systems, the total activity is fixed at $N_p|F_0|=240$. Shaded regions in \textbf{(a--c)} denote standard deviations across independent runs.}
    \label{fig:droplet_motion}
\end{figure*}

The active flows that drive interfacial fluctuations also impart translational motion to the droplet, whose nature depends strongly on the organization of the confined polymers. As shown in Figs.~\ref{fig:MSD_passive}, \ref{fig:MSD_extensile}, and \ref{fig:MSD_contractile}, the center-of-mass dynamics of the droplet depend sensitively on the nature of the active force dipole. Passive droplets exhibit diffusive motion, with nearly overlapping droplet COM MSD curves for all values of $N_p$, as expected in the absence of active forcing (Figs.~\ref{fig:MSD_passive} and \ref{fig:MSD_passive_a1}). Representative
activity-dependent droplet motion is shown in Supplementary Movie~3.

Droplets containing extensile polymers exhibit increasingly persistent motion with increasing $N_p$, with the crossover from ballistic to diffusive dynamics shifting to longer times at both total-activity levels, $N_p|F_0|=240$ and $40$ (Figs.~\ref{fig:MSD_extensile} and \ref{fig:MSD_extensile_a1}). Contractile droplets exhibit the opposite $N_p$ dependence for $N_p|F_0|=240$. The droplet COM MSD at $N_p=10$ displays the most extended near-ballistic regime, indicating long-lived persistent motion. As $N_p$ increases, the duration of this regime decreases and the crossover toward diffusive motion shifts to earlier times (Fig.~\ref{fig:MSD_contractile}). However, at $N_p|F_0|=40$, this trend is reversed, the crossover to diffusive motion shifts to later times with increasing $N_p$, and the $N_p=10$ system retains near-ballistic scaling over the longest time interval (Fig.~\ref{fig:MSD_contractile_a1}).


To understand the origin of the directed motion, we next examine how the droplet velocity is coupled to the position of the polymer center of mass within the droplet. To quantify this coupling, we compute the Pearson correlation coefficient, $r_p$, between the polymer center-of-mass offset relative to the droplet center, $\Delta\vec{R}_{\mathrm{COM}}=\vec{R}_{\mathrm{poly,COM}}-\vec{R}_{\mathrm{droplet,COM}}$, and the instantaneous droplet COM velocity, $\vec{v}_{\mathrm{droplet,COM}}$ (see Sec.~\ref{sec:pearson_correlation} in supplementary material). As shown in Figs.~\ref{fig:pearson_corr} and \ref{fig:pearson_corr_a_1}, the Pearson correlation reveals opposite directional couplings between the polymer COM offset and droplet motion for extensile and contractile activity. Extensile systems exhibit a strong positive correlation, $r_p\simeq 0.9$, indicating that the droplet velocity is directed towards the side containing the polymer center of mass. Thus, the polymer-rich side defines the leading edge of the droplet. In contrast, contractile droplets move preferentially away from the polymer-rich side, resulting in a negative correlation between the polymer COM offset and the droplet velocity. The magnitude of this anticorrelation decreases with increasing $N_p$, consistent with the progressive loss of coherent directed motion. Passive systems exhibit near-zero correlations, indicating no preferred directional coupling between the polymer COM offset and droplet motion.

\section{Conclusion}

We studied semiflexible active polymers confined within a soft, deformable, nearly spherical fluid droplet using explicit-solvent dissipative particle dynamics simulations. Activity was introduced through momentum-conserving force dipoles along the polymer backbone. The sign of the force dipole generates distinct hydrodynamic flows, leading to qualitatively different polymer organizations and couplings to fluctuations of the deformable confining interface and to droplet motion.

The two dipole symmetries produce qualitatively different solvent-mediated interactions. Extensile activity promotes effective attraction and parallel alignment between neighboring polymer segments, whereas contractile activity favors attractive interactions in predominantly perpendicular configurations~\cite{PandeyAnkita2016,MannaKumar2019ActivePolymericFluids}. Although earlier results were obtained using stresslet descriptions, the same orientational tendencies emerge in our explicit-solvent, momentum-conserving force-dipole model, highlighting the generality of these hydrodynamically mediated interactions across different implementations of activity.

Confinement qualitatively modifies how these local interactions determine polymer conformations and collective organization. When the polymer contour length greatly exceeds the droplet diameter, contractile activity drives the chains into compact, self-linked conformations, resulting in an effective stiffness lower than that of their passive counterparts. This behavior reverses the activity-induced stiffening reported for isolated contractile filaments in unconfined systems~\cite{Jayaraman2012AutonomousMotility,Laskar2013HydrodynamicInstabilities}. At the collective level, extensile activity produces compact, anisotropic polymer bundles, whereas contractile activity produces compact, self-linked structures at low polymer number. As the polymer number increases, crowding and the progressive replacement of intrachain contacts by interchain contacts cause the contractile polymers to spread more uniformly throughout the droplet, with their organization gradually approaching that of the passive system.

These distinct internal organizations couple strongly to fluctuations of the deformable confining interface. For droplets containing passive polymers, the fluctuation spectrum follows the equilibrium capillary-wave prediction, with a fitted surface tension that remains nearly independent of the number of confined polymers, while the mode-relaxation times scale approximately as $\tau_{\ell}\sim\ell^{-1}$.\cite{MilnerSafran1987DynamicalFluctuations,Goulian1996PolymerDropletFluctuations} Active polymers drive the interface away from this equilibrium behavior, modifying both the amplitudes and relaxation dynamics of the shape fluctuations. Droplets containing extensile polymers exhibit the largest fluctuation amplitudes, particularly in the low-order modes, reflecting the efficient transfer of coherent active solvent flows to large-scale droplet deformations. Droplets containing contractile polymers show qualitatively similar but weaker enhancements. The mode-relaxation times also depart from the passive scaling under activity. For extensile droplets at larger polymer numbers, the relaxation becomes only weakly dependent on $\ell$ and does not follow a single power-law scaling, whereas contractile droplets retain a stronger mode dependence, approximately following $\tau_{\ell}\sim\ell^{-1.3}$ over an intermediate range of modes.

These differences in internal organization and interfacial dynamics translate into distinct modes of droplet transport. While passive droplets exhibit diffusive motion, droplets containing active polymers display persistent motion whose duration depends on the organization of the confined polymers. For extensile polymers, the persistence of droplet motion increases with polymer number as collective polymer structures develop, whereas the persistence of contractile droplets depends more sensitively on polymer number and activity strength. Notably, extensile and contractile polymers generate opposite directional couplings between polymer organization and droplet motion. Extensile droplets move towards the polymer-rich side, which therefore defines the leading edge of the droplet, whereas contractile droplets move preferentially away from it. This sign-dependent propulsion demonstrates that the spatial asymmetry of the confined active polymers, together with the hydrodynamic flows they generate, determines both the direction and persistence of droplet motion.

These results show that the dynamics of droplets containing active polymers arise from feedback among polymer organization, the hydrodynamic flows generated by the polymers, and the response of the deformable interface. The principles identified here may therefore offer insight into active filament networks in deformable droplets and vesicles~\cite{Loiseau2016Blebbing,Litschel2021ActomyosinRings,Sciortino2025ActiveMembrane}, membraneless organelles containing actively remodeled polymers~\cite{Berry2015RNAtranscription,Moore2025ATPChromatin}, and synthetic active emulsions~\cite{Guillamat2018ActiveNematicEmulsions,Adkins2022ActiveInterfaces,Chen2021FlowCoupling} designed for controllable deformation and transport.

Taken together, our work identifies hydrodynamic coupling between confined active polymers and a deformable interface, mediated by their collective organization, as a general mechanism through which microscopic active stresses are converted into large-scale shape deformations and transport of soft compartments.

\section*{SUPPLEMENTARY MATERIAL}

See the supplementary material for details of the nematic-order parameter calculation, the definitions of the radius of gyration and asphericity, the extraction of the droplet interface and the spherical-harmonic analysis of its fluctuations, the equilibrium capillary-wave spectrum, the Pearson correlation analysis, and additional results on polymer organization, interfacial fluctuations
and relaxation, and droplet motion. The supplementary material also includes three movies: Supplementary Movie~1, showing polymer organization and dynamics within the droplet; Supplementary Movie~2, showing interfacial fluctuations; and Supplementary Movie~3, showing droplet motility.

\begin{acknowledgments}
Computational support from the PARAM Shakti supercomputing facility at IIT Madras is gratefully acknowledged. PBSK acknowledges the financial support provided by Anusandhan National Research Foundation (ANRF) under J. C. Bose grant (ANRF/JBG/2025/000187/PS).
\end{acknowledgments}

\section*{AUTHOR DECLARATIONS}

\subsection*{Conflict of Interest}

The authors have no conflicts to disclose.

\subsection*{Author Contributions}

Ritu Raj: Conceptualization, Methodology, Formal analysis, Data curation,
Writing--original draft, and Writing--review and editing.

P. B. Sunil Kumar: Conceptualization, Methodology, Supervision,
Writing--original draft, and Writing--review and editing.

\section*{DATA AVAILABILITY}

The data that support the findings of this study are available within the article and its supplementary material.

\bibliographystyle{apsrev4-2} 
\bibliography{references} 

\end{document}


\title{
Supplementary Material for\\
Active Hydrodynamics Couples Polymer Organization, Shape Fluctuations,
and Motility in Deformable Droplets
}

\author{Ritu Raj}
\affiliation{Department of Physics, Indian Institute of Technology Madras, Chennai, 600036, India.}
\affiliation{Center for Soft and Biological Matter, Indian Institute of Technology Madras, Chennai, 600036, India.}
\author{P. B. Sunil Kumar}
\email{sunil@physics.iitm.ac.in}
\affiliation{Department of Physics, Indian Institute of Technology Madras, Chennai, 600036, India.}
\affiliation{Center for Soft and Biological Matter, Indian Institute of Technology Madras, Chennai, 600036, India.}

\maketitle

\onecolumngrid

\section{Supplementary Methods}
\label{sec:SI}

\setcounter{figure}{0}
\renewcommand{\thefigure}{S\arabic{figure}}

\setcounter{table}{0}
\renewcommand{\thetable}{S\arabic{table}}

\setcounter{equation}{0}
\renewcommand{\theequation}{S\arabic{equation}}

\subsection{Nematic order parameter}
\label{sec:nematic_order}

For each unit bond vector $\hat{b}_i$, the local nematic order tensor is defined as
\begin{equation}
Q_{\alpha\beta}^{(i)}(R)
=
\frac{1}{N_i(R)}
\sum_{j\in\mathcal{N}_i(R)}
\left(
\frac{3}{2}
\hat{b}_{j,\alpha}\hat{b}_{j,\beta}
-
\frac{1}{2}\delta_{\alpha\beta}
\right),
\label{eq:local_nematic_tensor}
\end{equation}
where $\alpha,\beta\in\{x,y,z\}$, $\delta_{\alpha\beta}$ is the Kronecker delta, and $\mathcal{N}_i(R)$ denotes the set of bonds whose midpoints lie within a spherical region of radius $R$ centered at the midpoint of bond $i$. The quantity $N_i(R)$ is the number of bonds contained within this local neighborhood.

The local nematic order associated with bond $i$ is defined as the largest eigenvalue of $\mathbf{Q}^{(i)}(R)$,
\begin{equation}
S_{\mathrm{nem}}^{(i)}(R)
=
\lambda_{\max}
\left[
\mathbf{Q}^{(i)}(R)
\right].
\label{eq:local_nematic_eigenvalue}
\end{equation}
The reported local nematic order parameter for a neighborhood of radius $R$ is obtained by averaging over all bonds,
\begin{equation}
S_{\mathrm{nem}}(R)
=
\left\langle
\frac{1}{N_{\mathrm{b}}}
\sum_{i=1}^{N_{\mathrm{b}}}
S_{\mathrm{nem}}^{(i)}(R)
\right\rangle,
\label{eq:local_nematic_order}
\end{equation}
where $N_{\mathrm{b}}$ is the total number of polymer bonds. A value $S_{\mathrm{nem}}(R)\simeq 0$ corresponds to locally isotropic bond orientations, whereas $S_{\mathrm{nem}}(R)\simeq 1$ indicates strong local nematic alignment~\cite{deGennesProst1993,MilchevEtAl2021}.

To quantify the global orientational order, we construct the nematic order tensor using all polymer bonds within the droplet,
\begin{equation}
Q_{\alpha\beta}^{\mathrm{glob}}
=
\frac{1}{N_{\mathrm{b}}}
\sum_{j=1}^{N_{\mathrm{b}}}
\left(
\frac{3}{2}
\hat{b}_{j,\alpha}\hat{b}_{j,\beta}
-
\frac{1}{2}\delta_{\alpha\beta}
\right).
\label{eq:global_nematic_tensor}
\end{equation}
The global nematic order parameter is then given by
\begin{equation}
\left\langle S\right\rangle
=
\left\langle
\lambda_{\max}
\left[
\mathbf{Q}^{\mathrm{glob}}
\right]
\right\rangle,
\label{eq:global_nematic_order}
\end{equation}
where the angular brackets denote an average over sampled simulation configurations.

\subsection{Radius of gyration and asphericity}

The gyration tensor of the polymer cluster is defined as
\begin{equation}
G_{\alpha\beta}(t)
=
\frac{1}{N_m}
\sum_{i=1}^{N_m}
\left[
r_{i,\alpha}(t)-R_{\mathrm{poly,COM},\alpha}(t)
\right]
\left[
r_{i,\beta}(t)-R_{\mathrm{poly,COM},\beta}(t)
\right],
\end{equation}
where $N_m=N_pN_{\mathrm{poly}}$ is the total number of polymer beads and $\vec{R}_{\mathrm{poly,COM}}$ is the center of mass of the polymer cluster. The radius of gyration is obtained from
\begin{equation}
R_g^2(t)
=
\mathrm{Tr}\,\mathbf{G}(t)
=
\lambda_1(t)+\lambda_2(t)+\lambda_3(t),
\label{eq:rg}
\end{equation}
where $\lambda_1\leq\lambda_2\leq\lambda_3$ are the eigenvalues of the gyration tensor. The normalized asphericity is calculated as
\begin{equation}
A
=
\frac{
\langle\lambda_3\rangle
-\frac{1}{2}
\left(
\langle\lambda_1\rangle+\langle\lambda_2\rangle
\right)
}{
\langle R_g^2\rangle
}.
\label{eq:asphericity}
\end{equation}
Here, $A=0$ corresponds to a spherical configuration, whereas $A=1$ corresponds to a perfectly linear configuration.

\subsection{Extraction of the droplet interface}
\label{sec:droplet_interface}

For each frame, all particle coordinates were shifted relative to the instantaneous center of mass of the droplet-forming solvent, $S_1$. An $S_1$ or polymer particle $i$ was identified as an interface bead when
\begin{equation}
\min_{j\in S_2} d_{\mathrm{PBC}}(\vec r_i,\vec r_j)<r_{\mathrm{int}},
\qquad r_{\mathrm{int}}=1.3,
\end{equation}
where $S_2$ denotes the surrounding-solvent particles and $d_{\mathrm{PBC}}$ is the minimum-image distance. 

The radial position of each retained particle relative to the instantaneous droplet center, $\vec R_{\mathrm d}$, was calculated as
\begin{equation}
r_i=\left|\vec r_i-\vec R_{\mathrm d}\right|.
\end{equation}
The resulting spherical coordinates $(r_i,\theta_i,\phi_i)$ were used to construct the interface profile $r(\theta,\phi)$.

\subsection{Fluctuation spectrum of the droplet interface}
\label{sec:spherical_harmonic_analysis}

The instantaneous droplet interface is represented as a radial deformation about its mean spherical shape,
\begin{equation}
r(\theta,\phi,t)=R_0\left[1+u(\theta,\phi,t)\right],
\end{equation}
where $R_0$ is the mean droplet radius and $u(\theta,\phi,t)$ is the dimensionless radial deviation from the mean spherical shape. The deformation is expanded in spherical harmonics as
\begin{equation}
u(\theta,\phi,t)=
\sum_{\ell=0}^{\infty}\sum_{m=-\ell}^{\ell}
u_{\ell m}(t)Y_{\ell m}(\theta,\phi),
\end{equation}
where the spherical harmonics satisfy
\[
\int d\Omega\,
Y_{\ell m}^{*}(\theta,\phi)
Y_{\ell' m'}(\theta,\phi)
=
\delta_{\ell\ell'}\delta_{mm'}.
\]

For each interfacial bead $i$, the dimensionless radial deformation $u_i(t)$ was calculated as
\begin{equation}
u(\theta_i,\phi_i,t)
=
\frac{r_i(t)}{R_0}-1.
\end{equation}
The spherical-harmonic coefficients were obtained from the discrete projection of the instantaneous interface deformation,
\begin{equation}
u_{\ell m}(t)
=
\Delta\Omega(t)
\sum_{i=1}^{N_{\mathrm{int}}(t)}
u_i(t)Y_{\ell m}^{*}(\theta_i,\phi_i),
\label{eq:ulm_projection}
\end{equation}
where $\Delta\Omega(t)=4\pi/N_{\mathrm{int}}(t)$, $N_{\mathrm{int}}(t)$ is the number of interfacial beads at time $t$, and $Y_{\ell m}^{*}$ denotes the complex conjugate of $Y_{\ell m}$.

The fluctuation spectrum associated with mode $\ell$ was calculated as
\begin{equation}
S_{\ell}
=
\left\langle
\sum_{m=-\ell}^{\ell}
|u_{\ell m}(t)|^2
\right\rangle,
\label{eq:mode_spectrum}
\end{equation}
where $|u_{\ell m}|^2=u_{\ell m}u_{\ell m}^{*}$ and
$\langle\cdots\rangle$ denotes an average over simulation time.

\subsection{Equilibrium fluctuation spectrum of a passive droplet}
\label{sec:passive_fluctuation}
For a droplet whose deformation energy is dominated by the interfacial tension $\sigma$, the free energy is
\begin{equation}
\mathcal{F}=\sigma A,
\end{equation}
where $A$ is the instantaneous surface area. Expanding the surface area to quadratic order in the deformation amplitudes and imposing constant droplet volume gives~\cite{MilnerSafran1987DynamicalFluctuations}
\begin{equation}
\Delta\mathcal{F}
=
\frac{\sigma R_0^2}{2}
\sum_{\ell=2}^{\infty}\sum_{m=-\ell}^{\ell}
(\ell-1)(\ell+2)|u_{\ell m}|^2.
\end{equation}
The $\ell=0$ mode is constrained by volume conservation, while the $\ell=1$ modes correspond to translations and are removed by measuring the interface relative to the instantaneous droplet COM.

Applying equipartition to each independent fluctuation mode gives
\begin{equation}
\left\langle |u_{\ell m}|^2\right\rangle
=
\frac{k_{\mathrm B}T}
{\sigma R_0^2(\ell-1)(\ell+2)}.
\end{equation}
Summing over the $2\ell+1$ azimuthal modes therefore gives
\begin{equation}
\left\langle
\sum_{m=-\ell}^{\ell}|u_{\ell m}|^2
\right\rangle
=
\frac{k_{\mathrm B}T(2\ell+1)}
{\sigma R_0^2(\ell-1)(\ell+2)}.
\label{eq:equilibrium_fluctuation_spectrum}
\end{equation}

\subsection{Pearson correlation between polymer center-of-mass offset and droplet velocity}
\label{sec:pearson_correlation}
The instantaneous polymer center-of-mass offset relative to the droplet center of mass is defined as
\begin{equation}
\Delta\vec{R}_{\mathrm{COM}}(t)
=
\vec{R}_{\mathrm{poly,COM}}(t)
-
\vec{R}_{\mathrm{droplet, COM}}(t).
\label{eq:relative_polymer_com}
\end{equation}

The Pearson correlation coefficient between
$\Delta\vec{R}_{\mathrm{COM}}(t)$ and the instantaneous droplet velocity
$\vec{v}_{\mathrm{droplet,COM}}(t)$ is calculated by:
\begin{equation}
r_p =
\frac{
\displaystyle
\sum_{\alpha=x,y,z}\sum_{t=1}^{N_t}
\left[
\Delta R_{\alpha}(t)
-
\left\langle\Delta R_{\alpha}\right\rangle
\right]
\left[
v_{\alpha}(t)
-
\left\langle v_{\alpha}\right\rangle
\right]
}{
\displaystyle
\sqrt{
\sum_{\alpha=x,y,z}\sum_{t=1}^{N_t}
\left[
\Delta R_{\alpha}(t)
-
\left\langle\Delta R_{\alpha}\right\rangle
\right]^2
}
\sqrt{
\sum_{\alpha=x,y,z}\sum_{t=1}^{N_t}
\left[
v_{\alpha}(t)
-
\left\langle v_{\alpha}\right\rangle
\right]^2
}
}.
\label{eq:pearson_correlation}
\end{equation}

\clearpage

\section{Supplementary Figures}

\begin{figure*}[!ht]
    \justifying
    \begin{subfigure}[t]{0.49\textwidth}
        \captionsetup{position=top, justification=raggedright, singlelinecheck=false}
        \caption{}
        \includegraphics[width=\linewidth]{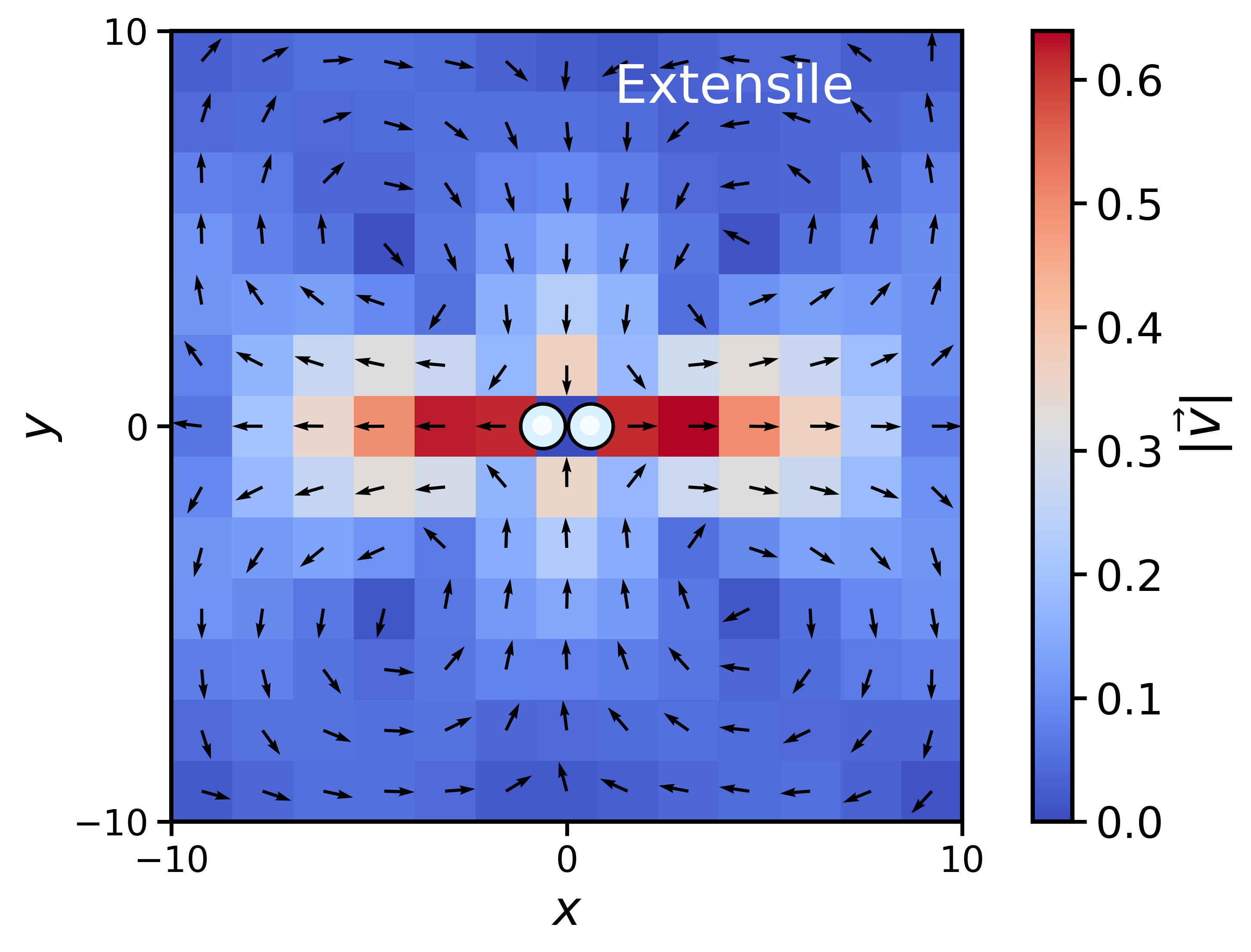}
        \label{fig:velocity_field_ext}
    \end{subfigure}
    \hfill
    \begin{subfigure}[t]{0.49\textwidth}
        \captionsetup{position=top, justification=raggedright, singlelinecheck=false}
        \caption{}
        \includegraphics[width=\linewidth]{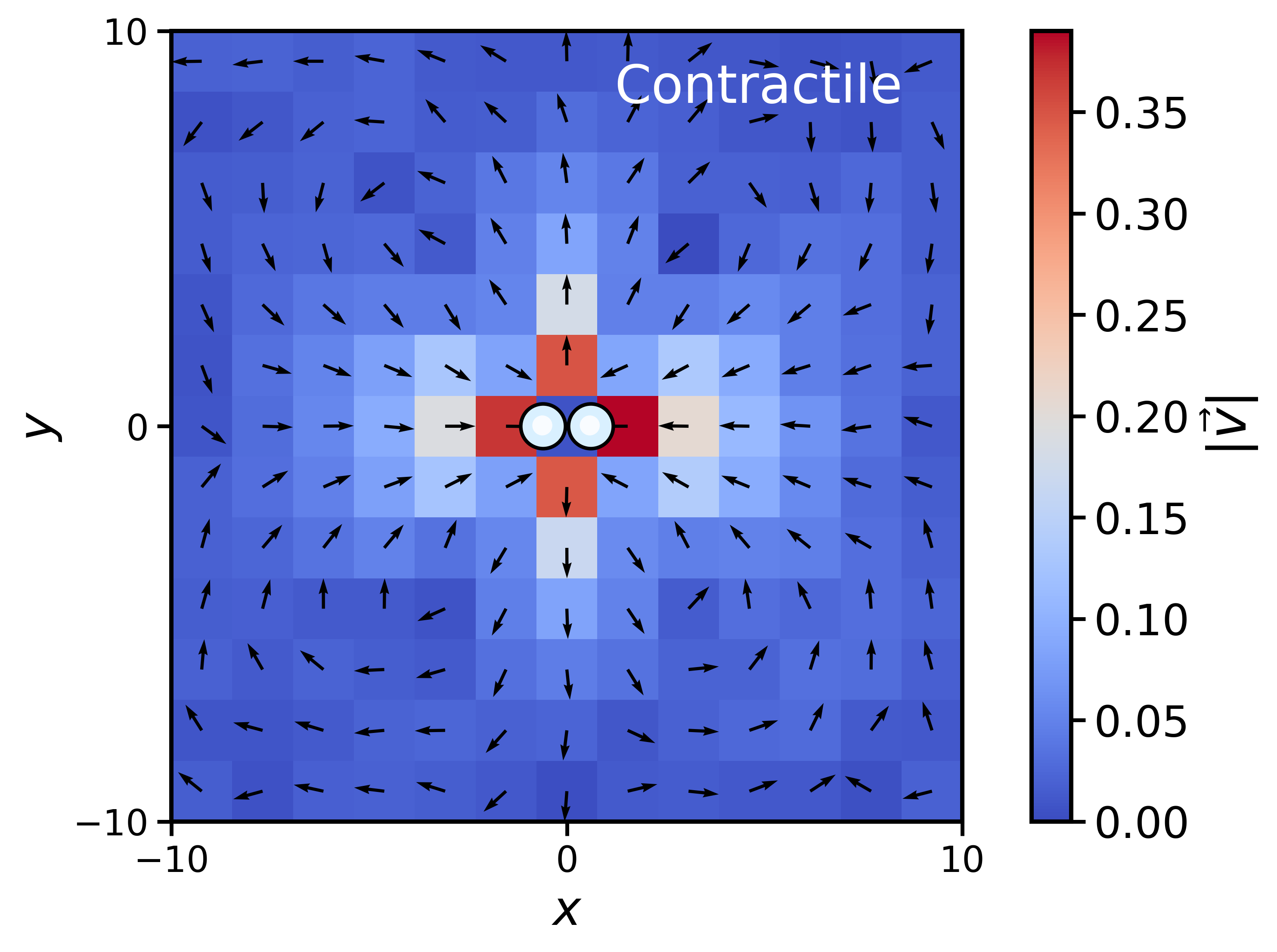}
        \label{fig:velocity_field_cont}
    \end{subfigure}
    \captionsetup{position=top, justification=raggedright, singlelinecheck=false}
    \caption{\justifying
    A two-dimensional slice of the velocity fields generated by an active force dipole, in three dimensions, for \textbf{(a)} extensile and \textbf{(b)} contractile activity. The arrows indicate the in-plane velocity direction, and the color map represents the magnitude of the two-dimensional velocity field.}
    \label{fig:velocity_field}
\end{figure*}

\begin{figure*}[!ht]
    \justifying
    \begin{subfigure}[t]{0.45\textwidth}
        \captionsetup{position=top, justification=raggedright, singlelinecheck=false}
        \caption{}
        \includegraphics[width=\linewidth]{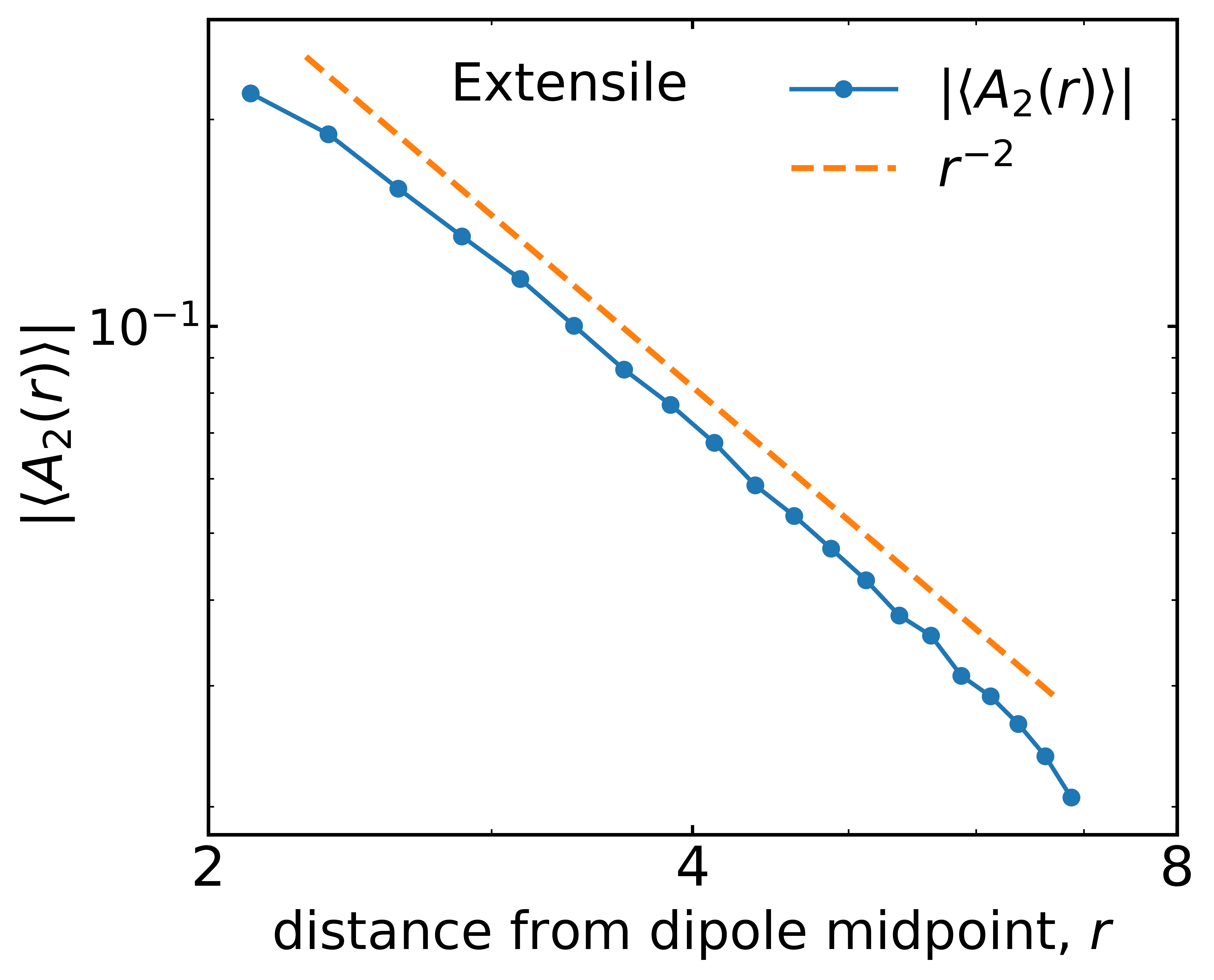}
        \label{fig:extensile_scaling}
    \end{subfigure}
    \hfill
    \begin{subfigure}[t]{0.45\textwidth}
        \captionsetup{position=top, justification=raggedright, singlelinecheck=false}
        \caption{}
        \includegraphics[width=\linewidth]{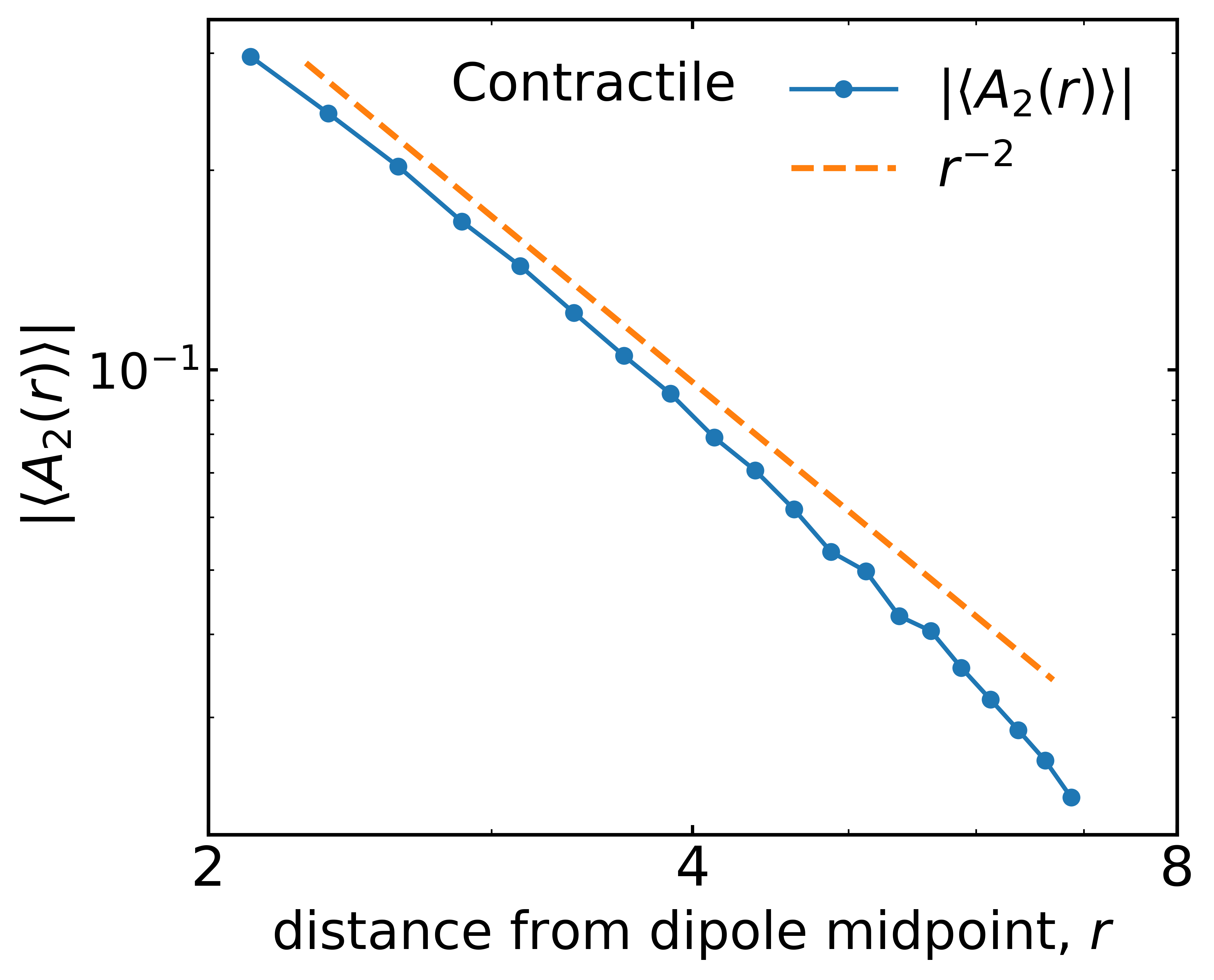}
        \label{fig:contractile_scaling}
    \end{subfigure}
    \captionsetup{position=top, justification=raggedright, singlelinecheck=false}
    \caption{\justifying
    Radial decay of the solvent velocity generated by an active force dipole for \textbf{(a)} extensile and \textbf{(b)} contractile activity. The radial velocity field is decomposed as $v_r(r,\theta)=A_2(r)P_2(\cos\theta)$, where $\theta$ is measured relative to the dipole axis and $A_2(r)$ is the dipolar radial velocity amplitude. The dashed lines indicate the expected $r^{-2}$ decay and are shown as guides to the eye.}
    \label{fig:velocity_scaling}
\end{figure*}

\begin{figure*}[!ht]
    \justifying
    \begin{subfigure}[t]{0.48\textwidth}
        \captionsetup{position=top, justification=raggedright, singlelinecheck=false}
        \caption{}
        \includegraphics[width=\linewidth]{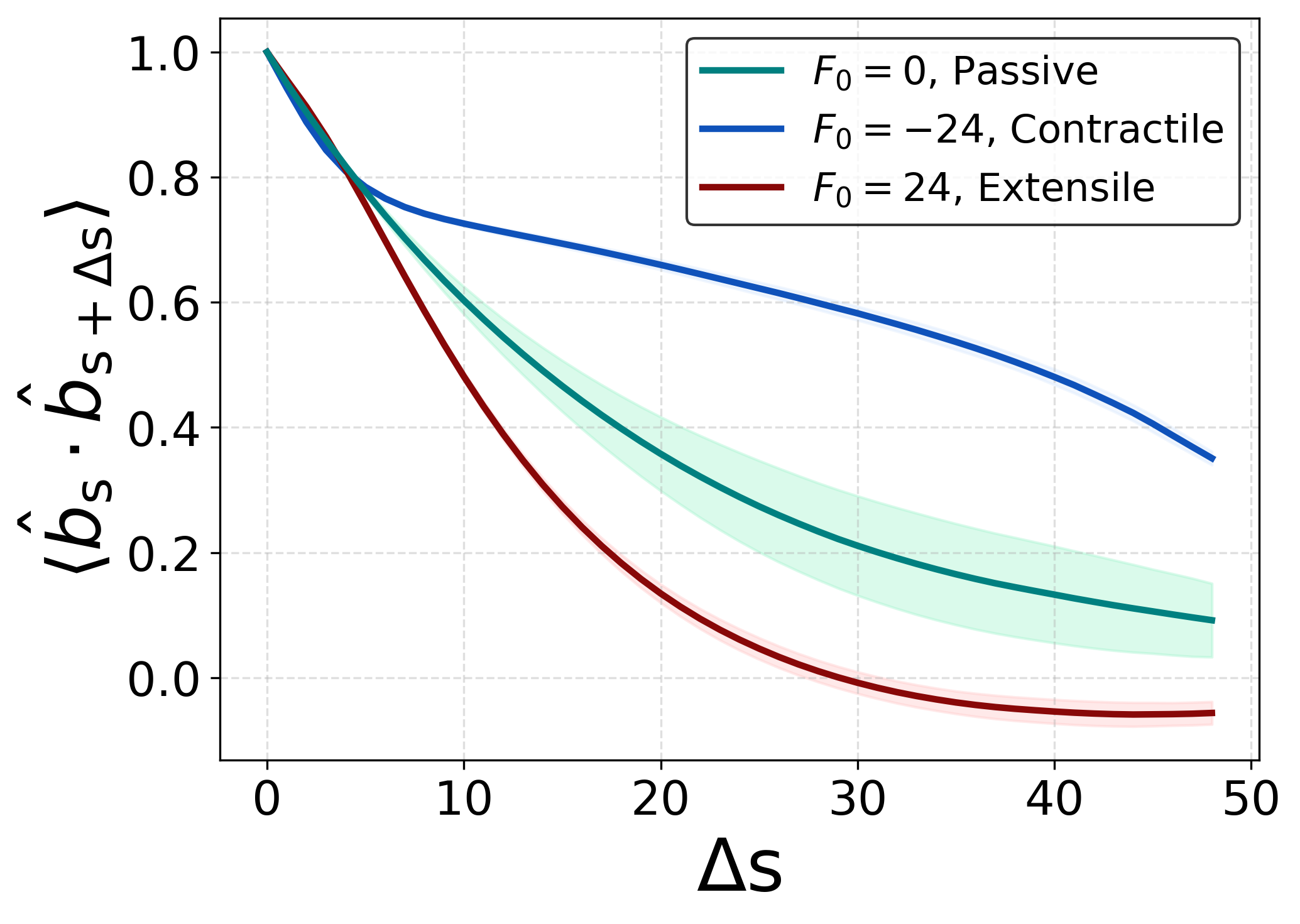}
        \label{fig:bond_corr_single_poly}
    \end{subfigure}
    \hfill
    \begin{subfigure}[t]{0.48\textwidth}
        \captionsetup{position=top, justification=raggedright, singlelinecheck=false}
        \caption{}
        \includegraphics[width=\linewidth]{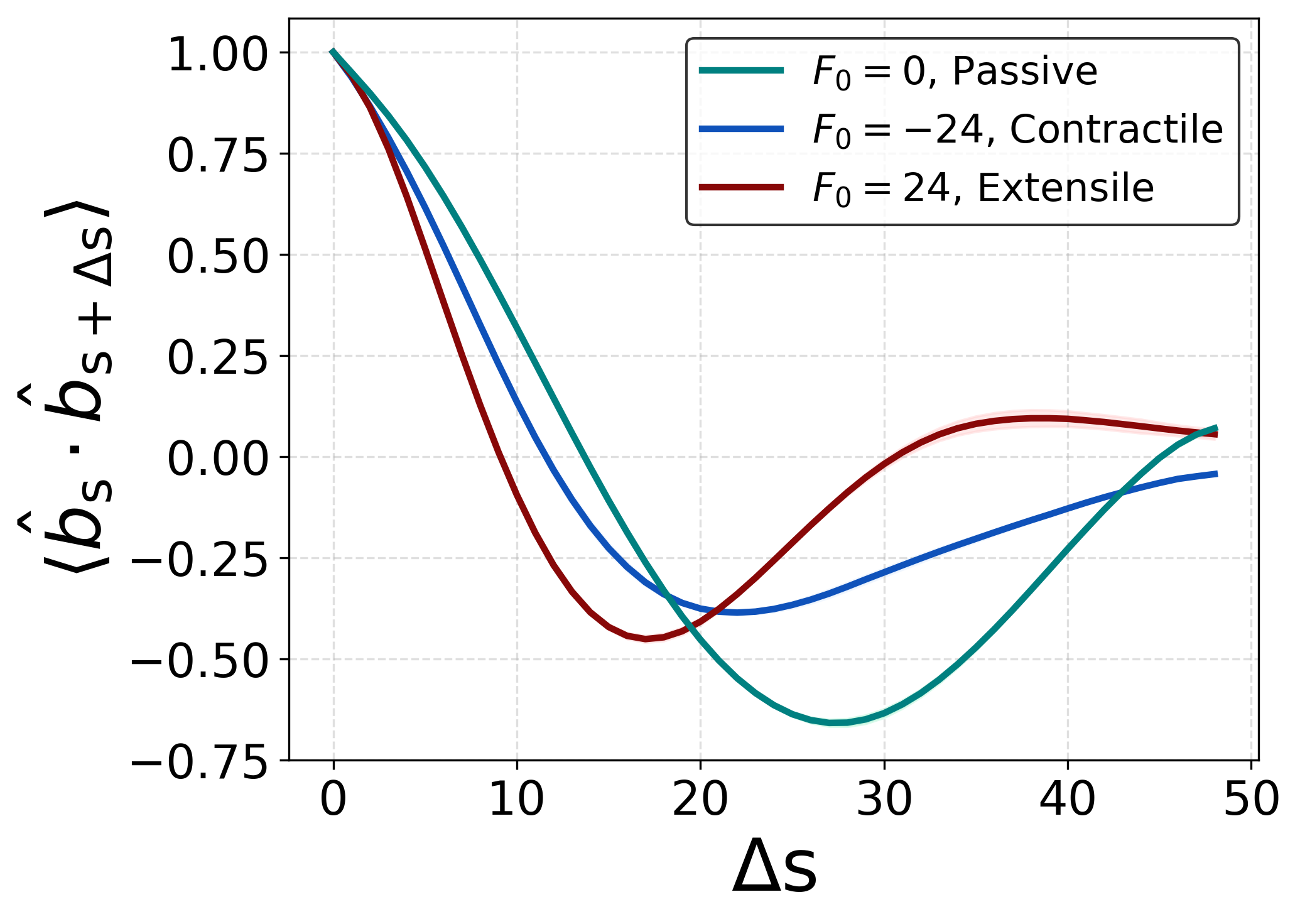}
        \label{fig:bond_corr_single_poly_conf}
    \end{subfigure}
    \captionsetup{position=top, justification=raggedright, singlelinecheck=false}
    \caption{
    Bond orientational correlation along the contour of a polymer comprising $N=50$ monomers for passive ($F_0=0$), contractile ($F_0=-24$), and extensile ($F_0=24$) activity; \textbf{(a)} when the  polymer is without confinement, and  \textbf{(b)} when the polymer is confined within a droplet of radius $R\sim6.6$. Shaded regions represent the standard deviation.}
    \label{fig:bond_corr_compare}
\end{figure*}

\begin{figure*}[!ht]
    \centering
    \begin{subfigure}[t]{0.52\textwidth}
        \captionsetup{position=top, justification=raggedright, singlelinecheck=false}
        \includegraphics[width=\linewidth]{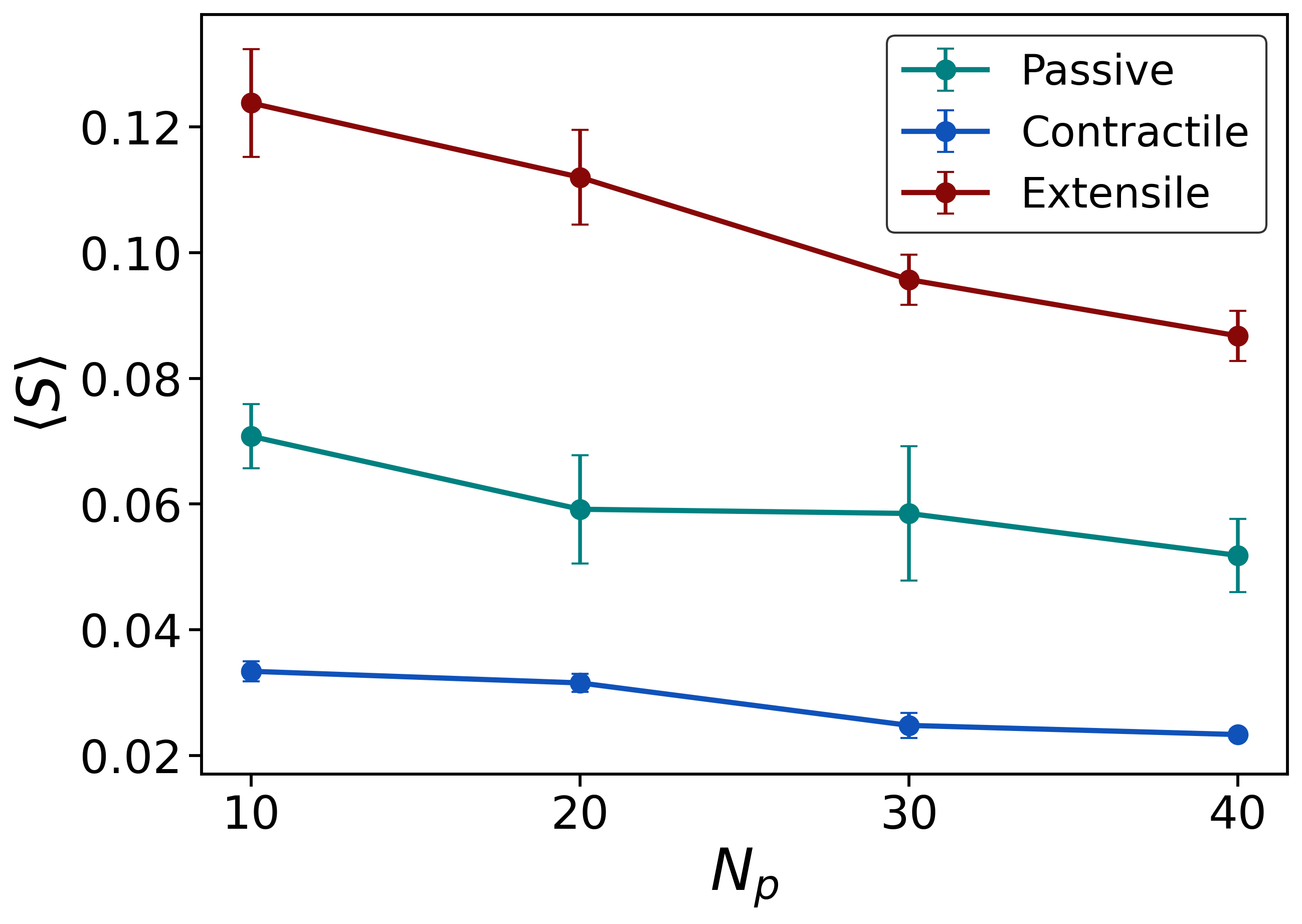}
    \end{subfigure}
    \captionsetup{position=top, justification=raggedright, singlelinecheck=false}
    \caption{\justifying
    Global nematic order parameter, $\langle S\rangle$, as a function of $N_p$ for droplets containing passive, contractile, and extensile polymers. For the active systems, the total activity is fixed at $N_p|F_0|=240$. Error bars represent standard deviations over independent simulations and are shown only when larger than the data symbols.}
    \label{fig:global_nematic}
\end{figure*}

\begin{figure*}[!ht]
    \centering
    \begin{subfigure}[t]{0.52\textwidth}
        \captionsetup{position=top, justification=raggedright, singlelinecheck=false}
        \includegraphics[width=\linewidth]{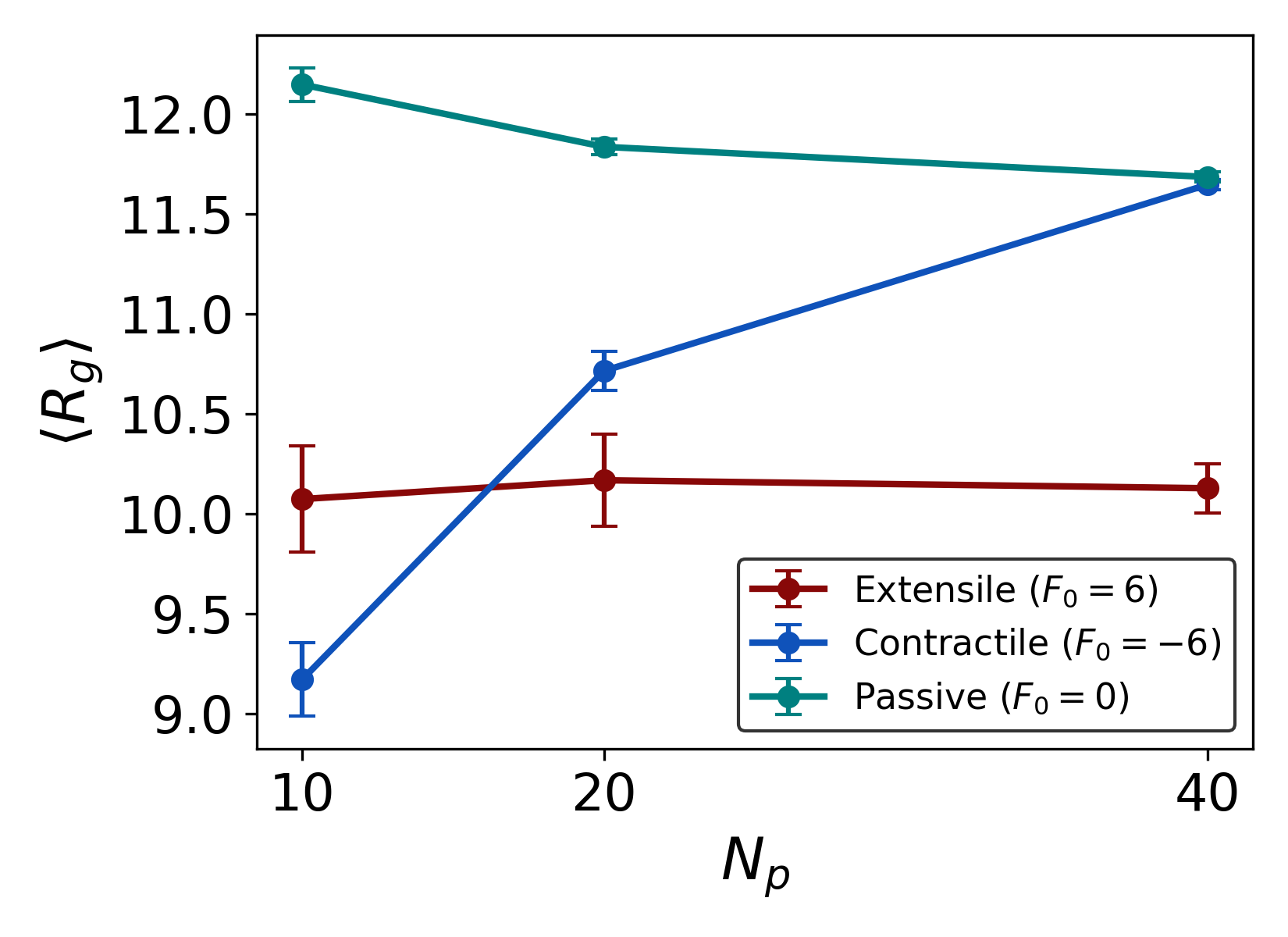}
    \end{subfigure}
    \captionsetup{position=top, justification=raggedright, singlelinecheck=false}
    \caption{\justifying
    Mean radius of gyration of the polymer cluster, $\langle R_g\rangle$, as a function of $N_p$ for passive polymers and for extensile and contractile polymers at fixed dipole strength $|F_0|=6$. Error bars represent standard deviations over independent simulations and are shown only when larger than the data symbols.}
    \label{fig:Rg_vs_Np_fix_F}
\end{figure*}

\begin{figure*}[!ht]
    \justifying
    \begin{subfigure}[t]{0.45\textwidth}
        \captionsetup{position=top, justification=raggedright, singlelinecheck=false}
        \caption{}
        \includegraphics[width=\linewidth]{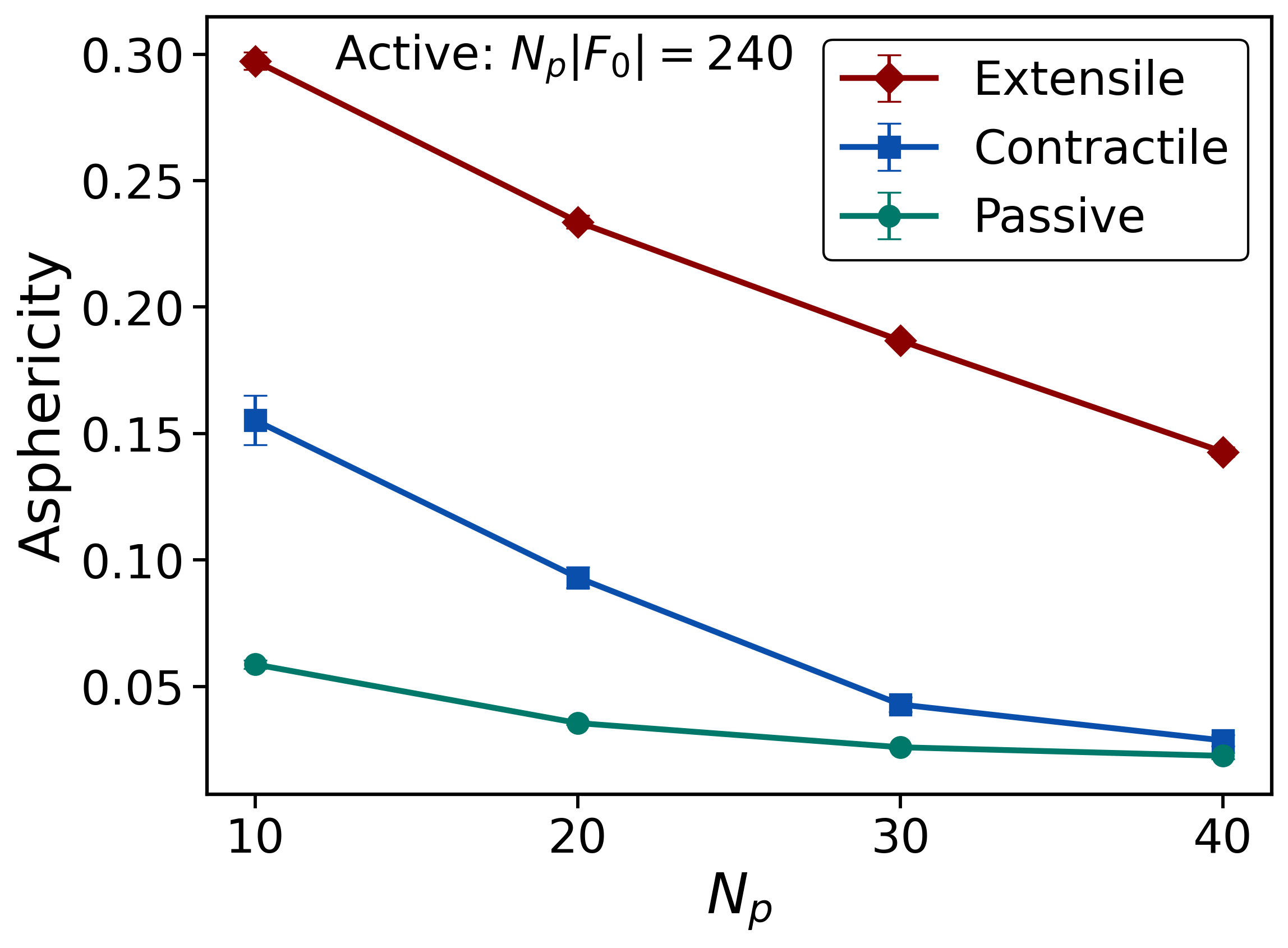}
        \label{fig:Asphericity_a_2}
    \end{subfigure}
    \hfill
    \begin{subfigure}[t]{0.45\textwidth}
        \captionsetup{position=top, justification=raggedright, singlelinecheck=false}
        \caption{}
        \includegraphics[width=\linewidth]{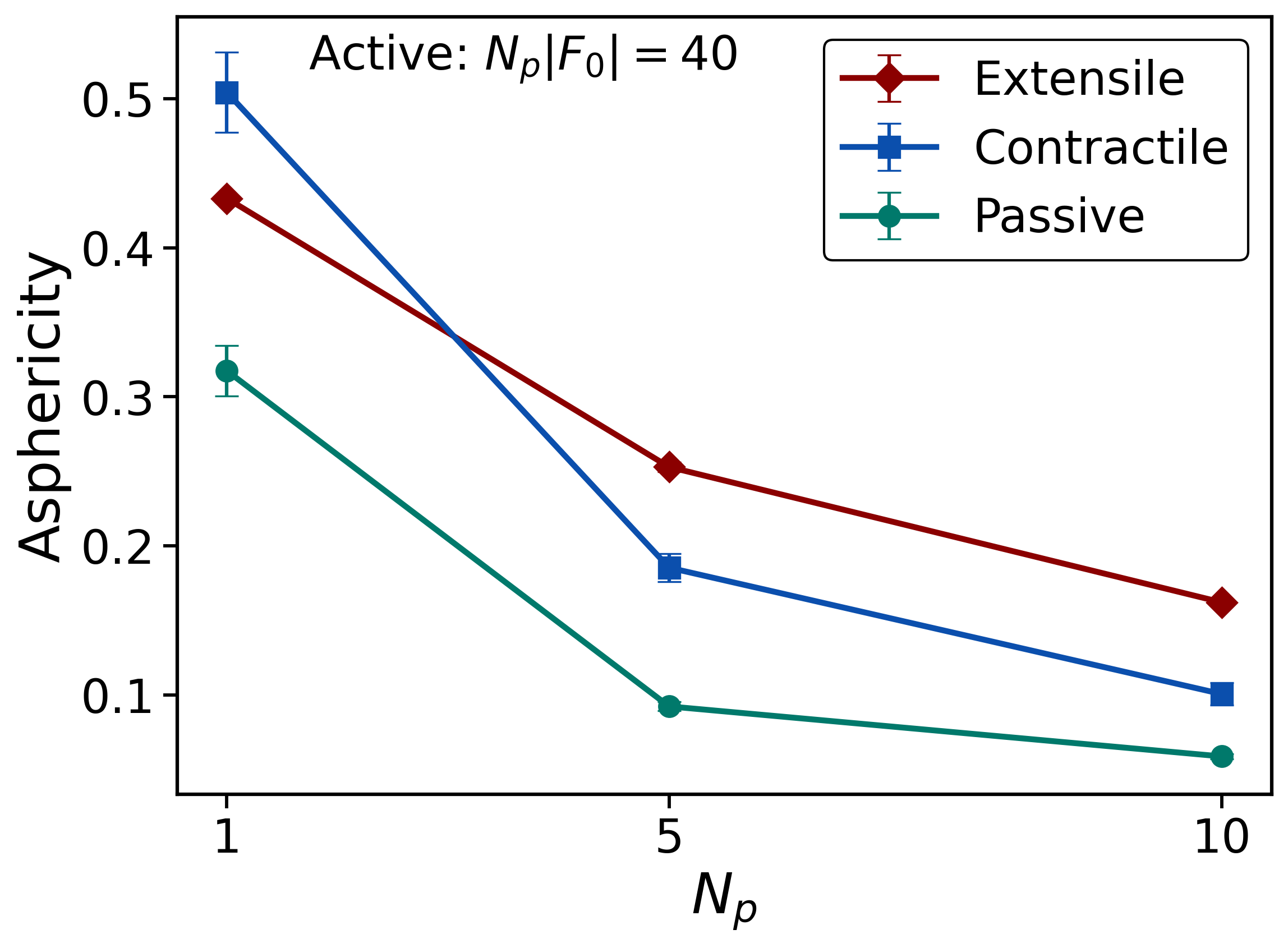}
        \label{fig:Asphericity_a_1}
    \end{subfigure}
    \captionsetup{position=top, justification=raggedright, singlelinecheck=false}
    \caption{\justifying
    Asphericity of the polymer cluster as a function of $N_p$ for passive, contractile, and extensile systems. For the active systems, the total activity is fixed at \textbf{(a)} $N_p|F_0|=240$ and \textbf{(b)} $N_p|F_0|=40$. Error bars represent standard deviations over independent simulations and are shown only when larger than the data symbols.}
    \label{fig:Asphericity}
\end{figure*}

\begin{figure*}[!ht]
    \justifying
    \begin{subfigure}[t]{0.32\textwidth}
        \captionsetup{position=top, justification=raggedright, singlelinecheck=false}
        \caption{}
        \includegraphics[width=\linewidth]{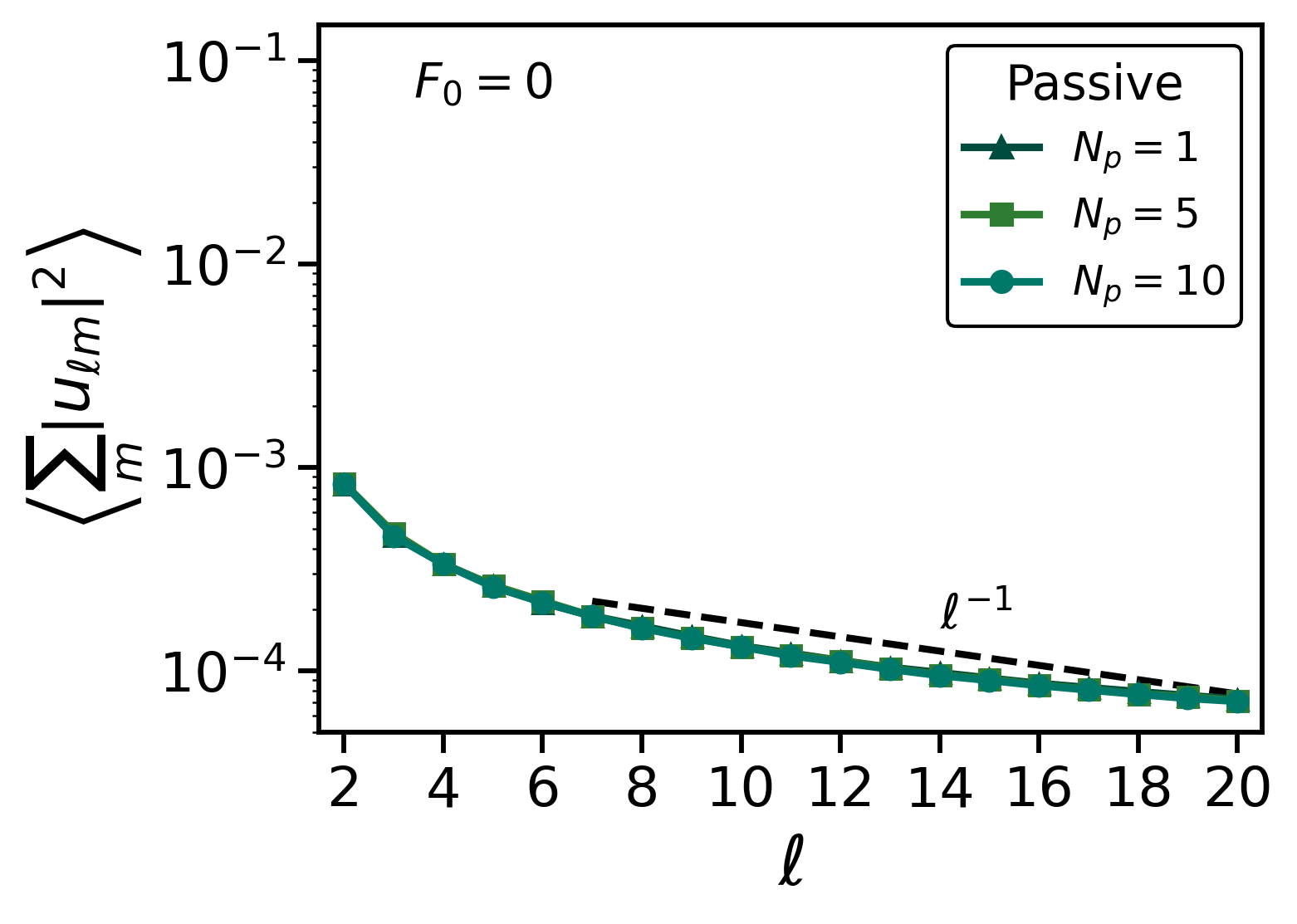}
        \label{fig:spectrum_passive_a_1}
    \end{subfigure}
    \hfill
    \begin{subfigure}[t]{0.32\textwidth}
        \captionsetup{position=top, justification=raggedright, singlelinecheck=false}
        \caption{}
        \includegraphics[width=\linewidth]{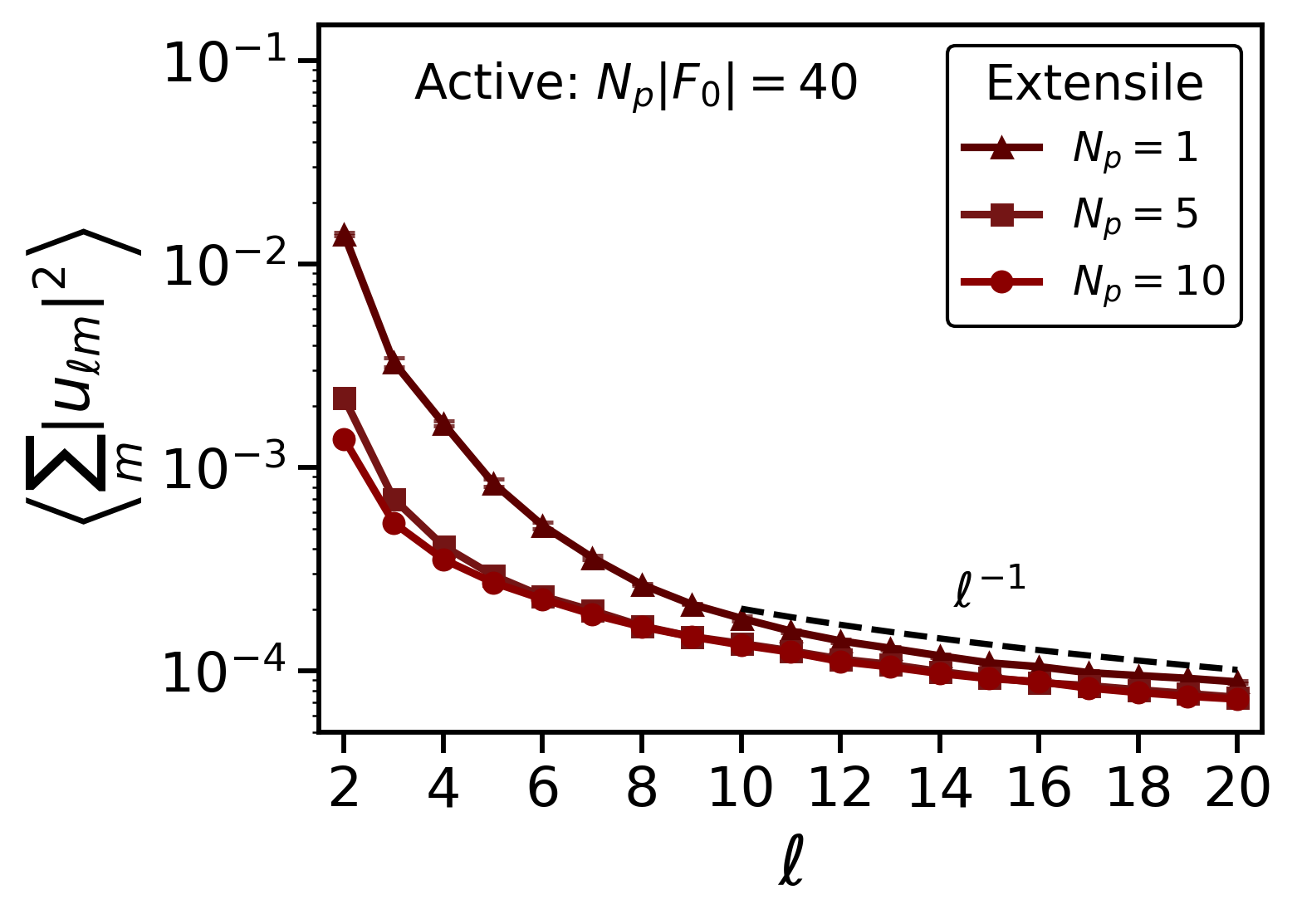}
        \label{fig:spectrum_extensile_a_1}
    \end{subfigure}
    \hfill
    \begin{subfigure}[t]{0.32\textwidth}
        \captionsetup{position=top, justification=raggedright, singlelinecheck=false}
        \caption{}
        \includegraphics[width=\linewidth]{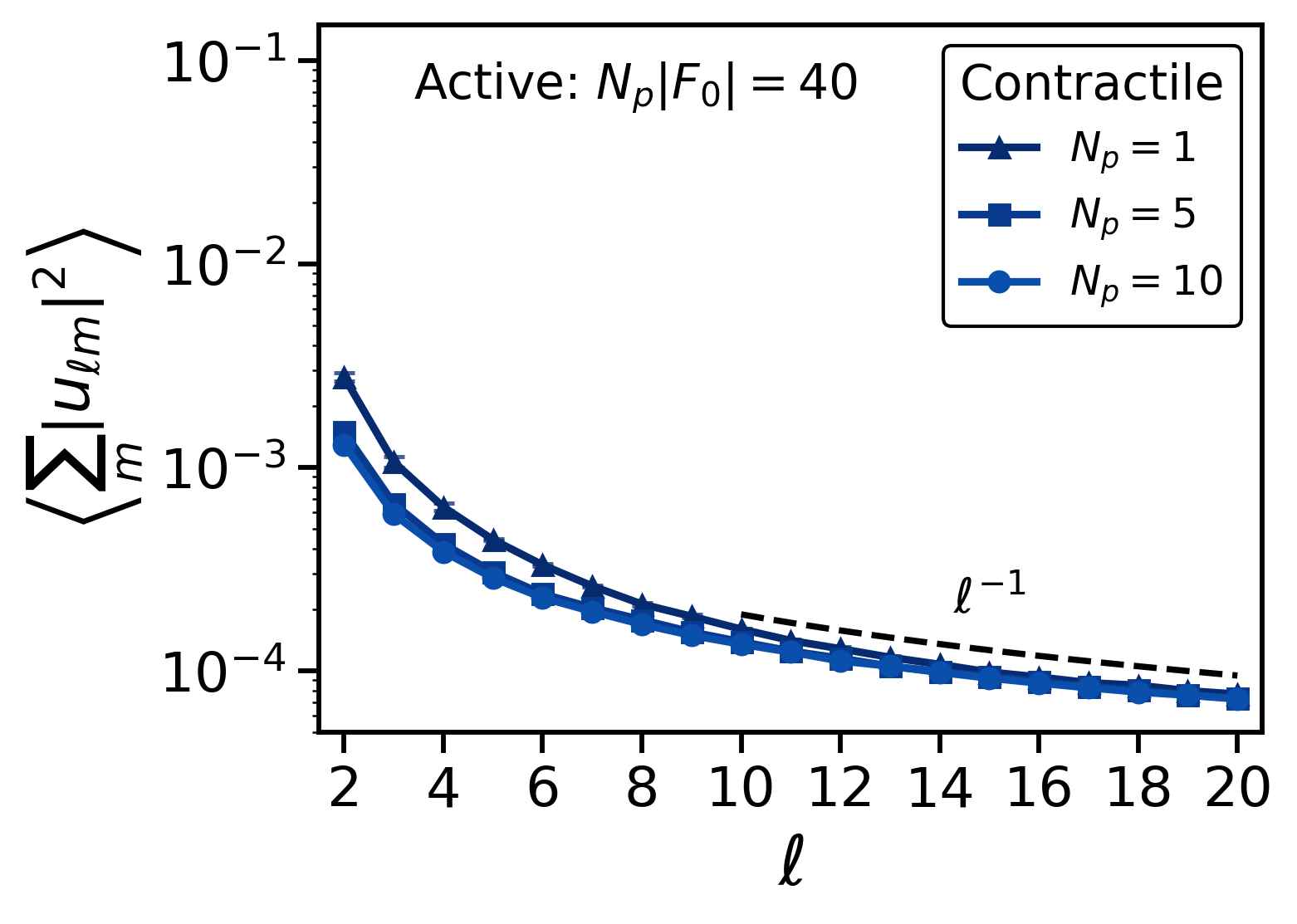}
        \label{fig:spectrum_contractile_a_1}
    \end{subfigure}
    \begin{subfigure}[t]{0.32\textwidth}
        \captionsetup{position=top, justification=raggedright, singlelinecheck=false}
        \caption{}
        \includegraphics[width=\linewidth]{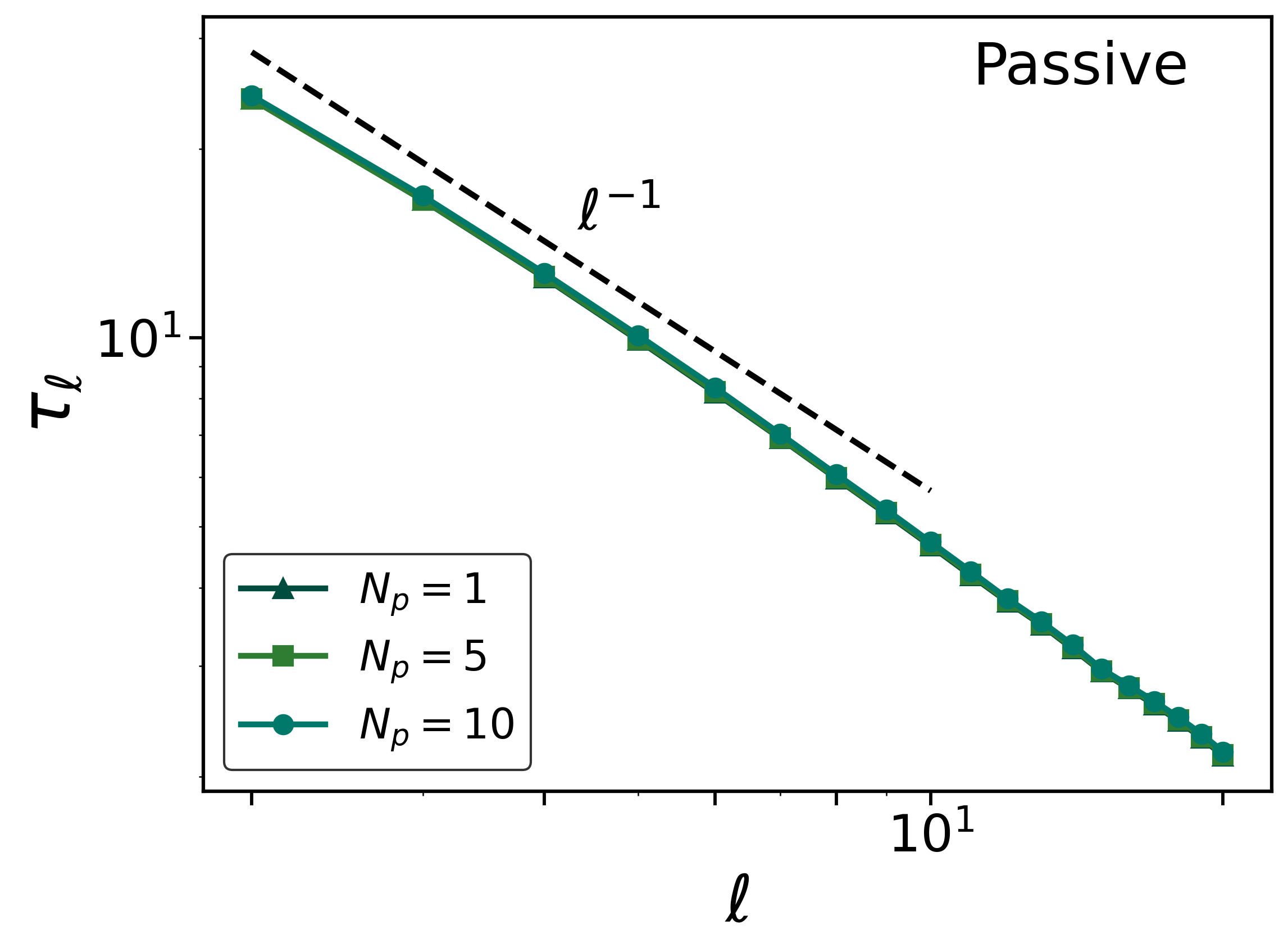}
        \label{fig:corr_time_pass_a_1}
    \end{subfigure}
    \hfill
    \begin{subfigure}[t]{0.32\textwidth}
        \captionsetup{position=top, justification=raggedright, singlelinecheck=false}
        \caption{}
        \includegraphics[width=\linewidth]{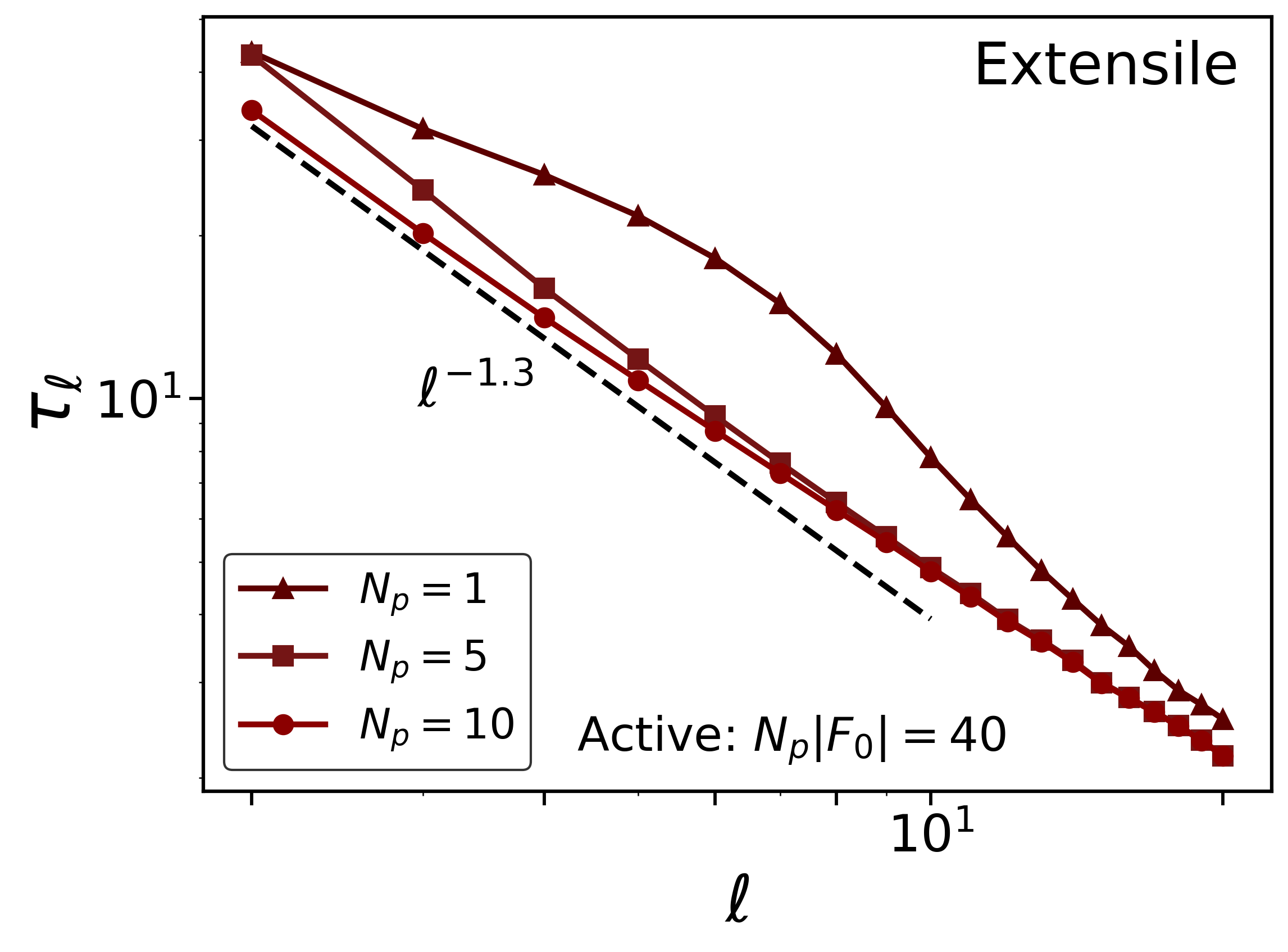}
        \label{fig:corr_time_ext_a_1}
    \end{subfigure}
    \hfill
    \begin{subfigure}[t]{0.32\textwidth}
        \captionsetup{position=top, justification=raggedright, singlelinecheck=false}
        \caption{}
        \includegraphics[width=\linewidth]{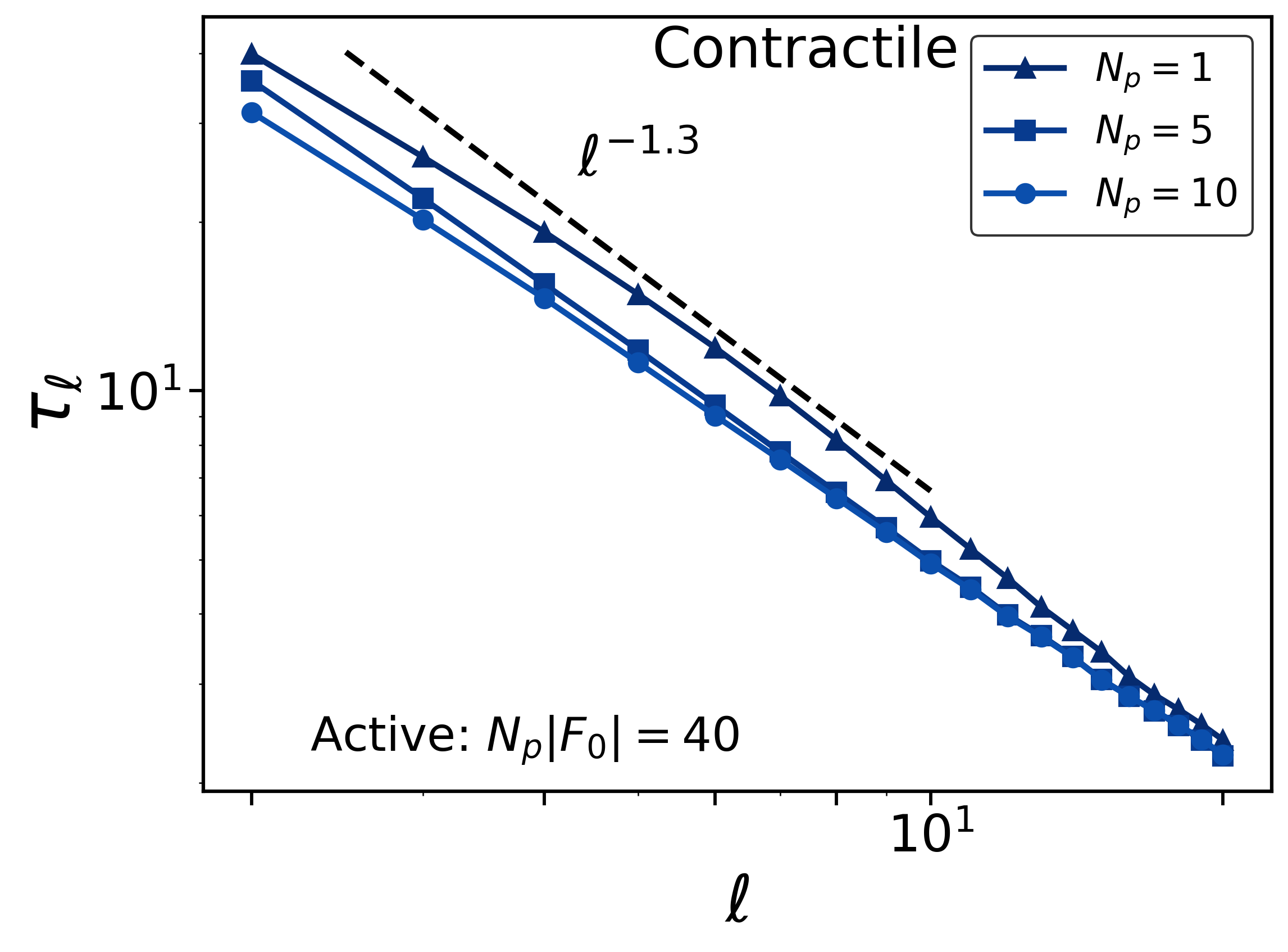}
        \label{fig:corr_time_cont_a_1}
    \end{subfigure}
    \captionsetup{position=top, justification=raggedright, singlelinecheck=false}
    \caption{\justifying
    Interfacial shape-fluctuation spectra as a function of the angular mode $\ell$ for droplets containing \textbf{(a)} passive, \textbf{(b)} extensile, and \textbf{(c)} contractile polymers at different polymer numbers, $N_p$.
    Shape-mode relaxation time, $\tau_{\ell}$, extracted from the first crossing of $C_{\ell}(\Delta t)=e^{-1}$, for droplets containing \textbf{(d)} passive, \textbf{(e)} extensile, and \textbf{(f)} contractile polymers. For the active systems, the total activity is $N_p|F_0|=40$.}
    \label{fig:shape_fluctuation_a_1}
\end{figure*}

\begin{figure*}[!ht]
    \justifying
    \begin{subfigure}[t]{0.48\textwidth}
        \captionsetup{position=top, justification=raggedright, singlelinecheck=false}
        \caption{}
        \includegraphics[width=\linewidth]{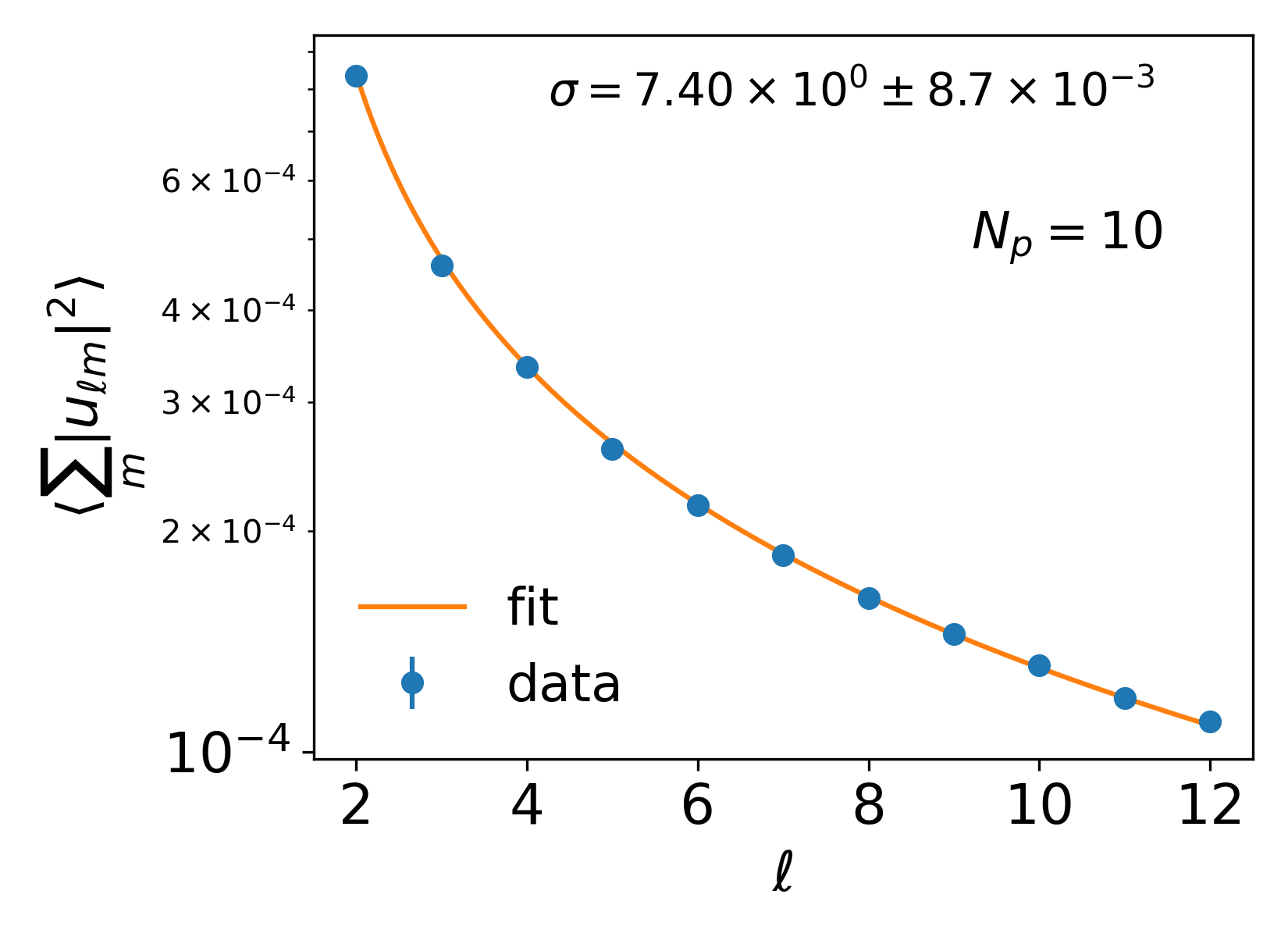}
        \label{fig:pass_10_fit}
    \end{subfigure}
    \hfill
    \begin{subfigure}[t]{0.48\textwidth}
        \captionsetup{position=top, justification=raggedright, singlelinecheck=false}
        \caption{}
        \includegraphics[width=\linewidth]{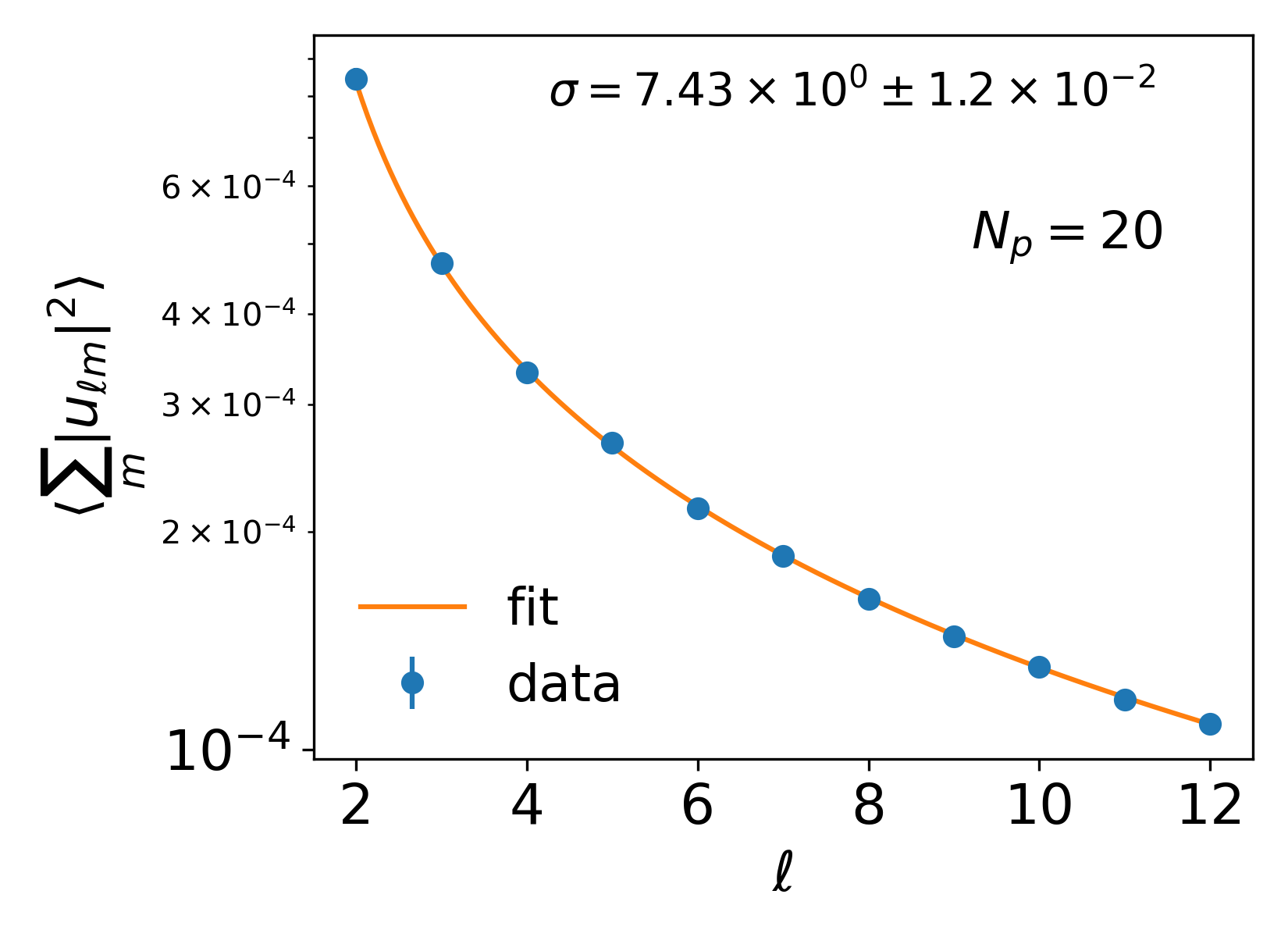}
        \label{fig:pass_20_fit}
    \end{subfigure}
    \begin{subfigure}[t]{0.48\textwidth}
        \captionsetup{position=top, justification=raggedright, singlelinecheck=false}
        \caption{}
        \includegraphics[width=\linewidth]{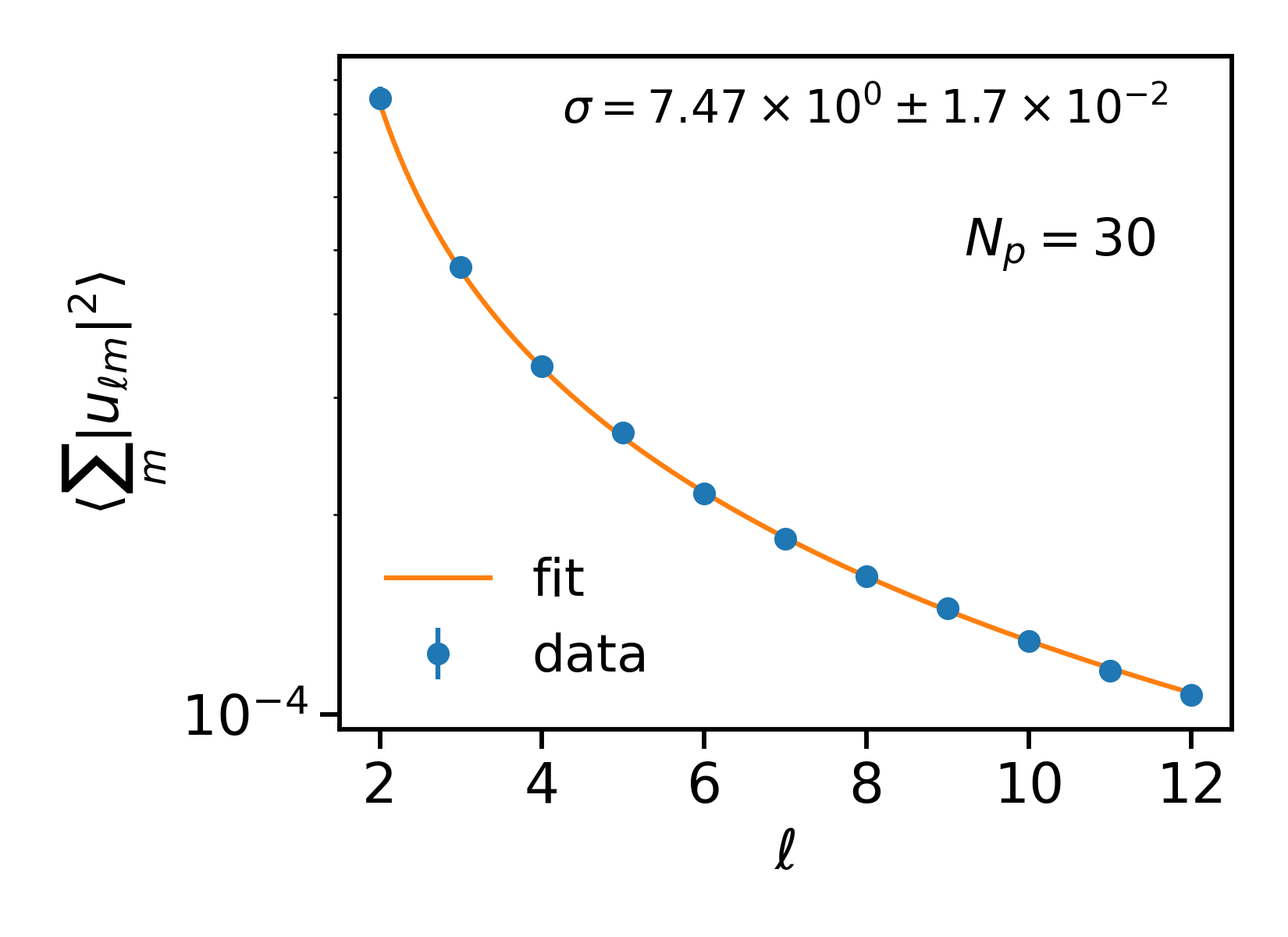}
        \label{fig:pass_30_fit}
    \end{subfigure}
    \hfill
    \begin{subfigure}[t]{0.48\textwidth}
        \captionsetup{position=top, justification=raggedright, singlelinecheck=false}
        \caption{}
        \includegraphics[width=\linewidth]{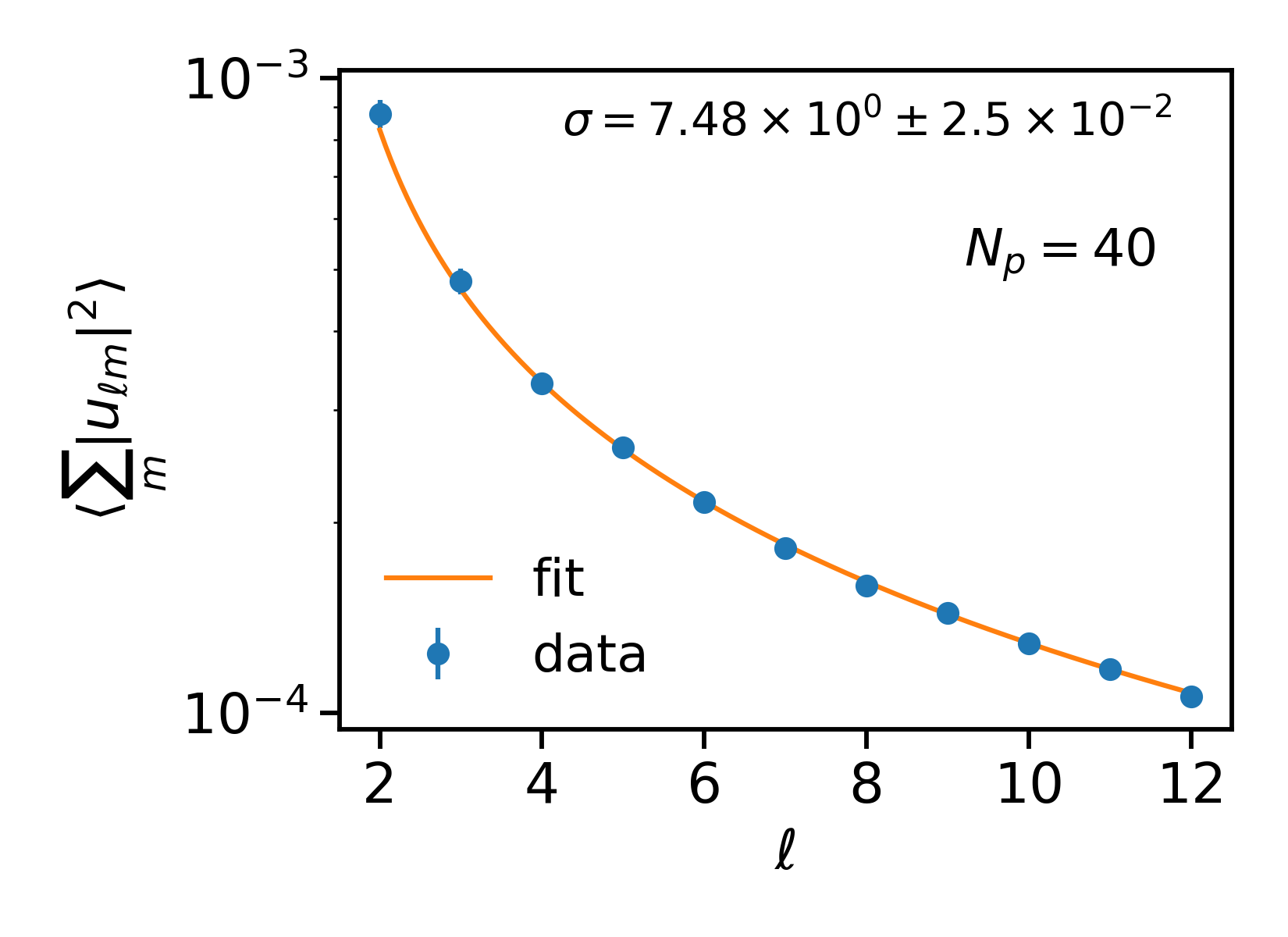}
        \label{fig:pass_40_fit}
    \end{subfigure}
    \captionsetup{position=top, justification=raggedright, singlelinecheck=false}
    \caption{
    Fluctuation spectra of passive polymer-containing droplets for \textbf{(a)} $N_p=10$, \textbf{(b)} $N_p=20$, \textbf{(c)} $N_p=30$, and \textbf{(d)} $N_p=40$. Symbols represent the simulation data, and solid lines show fits to the equilibrium capillary-wave spectrum given by Eq.~\ref{eq:equilibrium_fluctuation_spectrum}.}
    \label{fig:passive_spectrum_fit}
\end{figure*}

\begin{figure*}[!ht]
    \justifying
    \begin{subfigure}[t]{0.48\textwidth}
        \captionsetup{position=top, justification=raggedright, singlelinecheck=false}
        \caption{}
        \includegraphics[width=\linewidth]{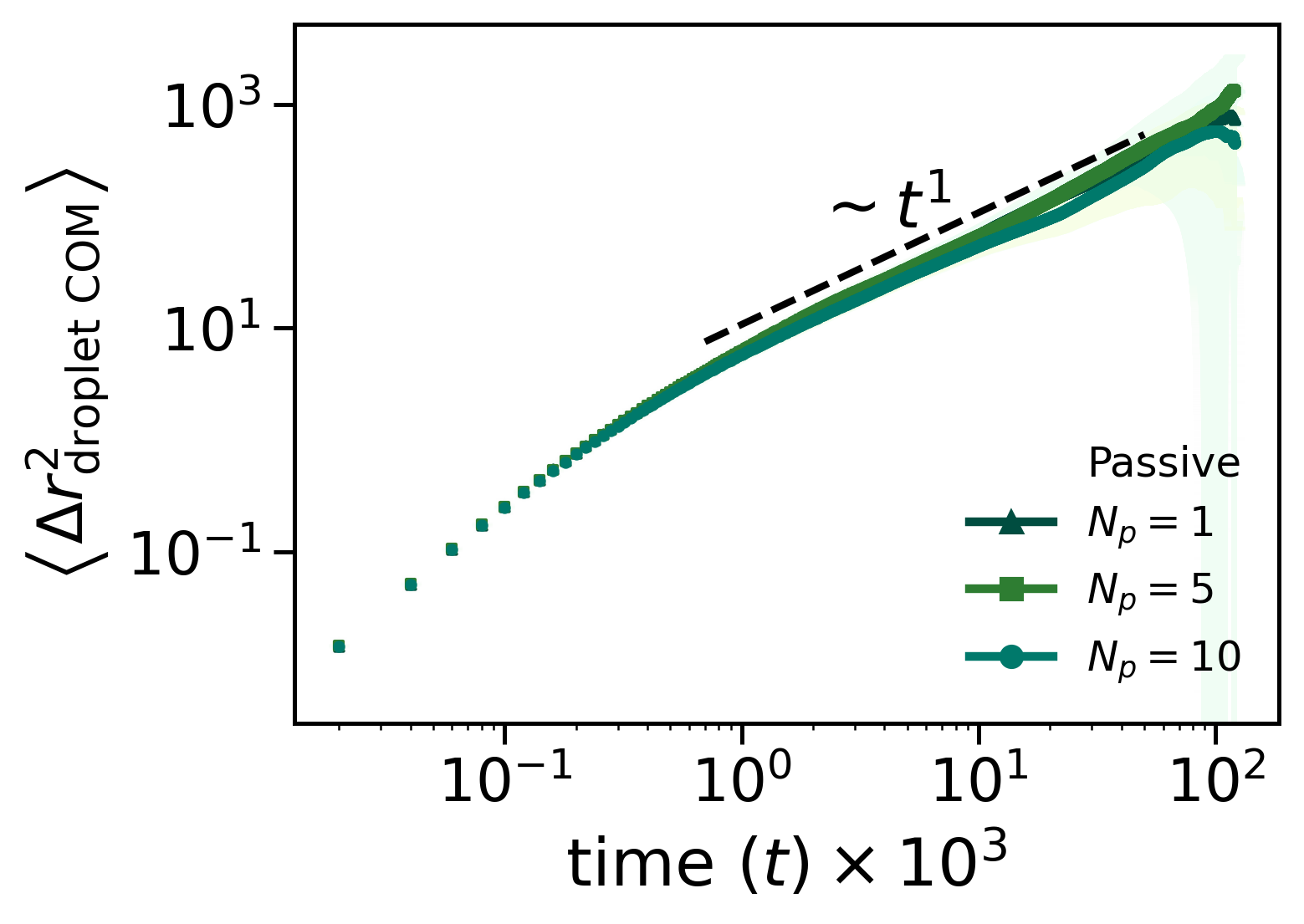}
        \label{fig:MSD_passive_a1}
    \end{subfigure}
    \hfill
    \begin{subfigure}[t]{0.48\textwidth}
        \captionsetup{position=top, justification=raggedright, singlelinecheck=false}
        \caption{}
        \includegraphics[width=\linewidth]{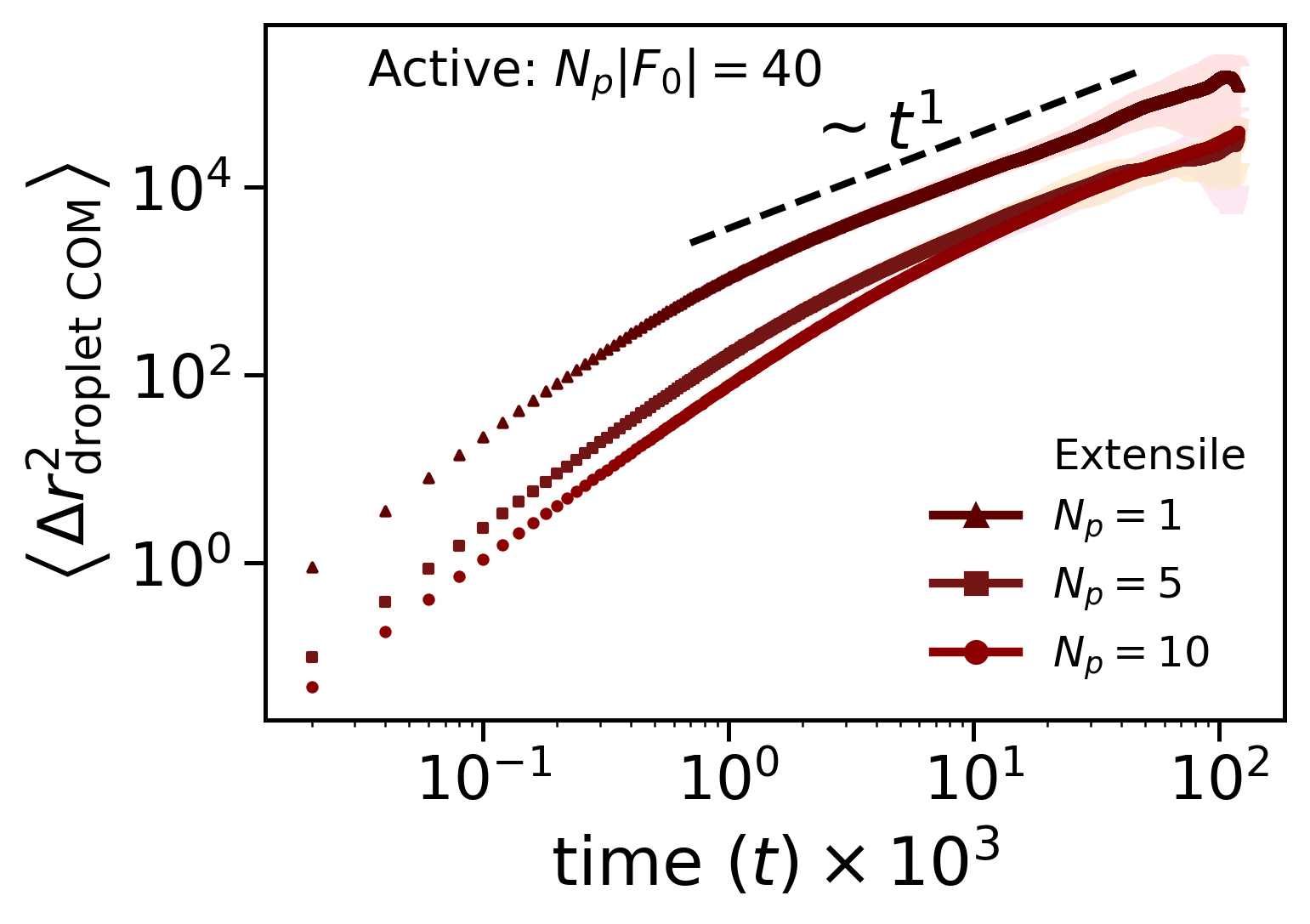}
        \label{fig:MSD_extensile_a1}
    \end{subfigure}
    \begin{subfigure}[t]{0.48\textwidth}
        \captionsetup{position=top, justification=raggedright, singlelinecheck=false}
        \caption{}
        \includegraphics[width=\linewidth]{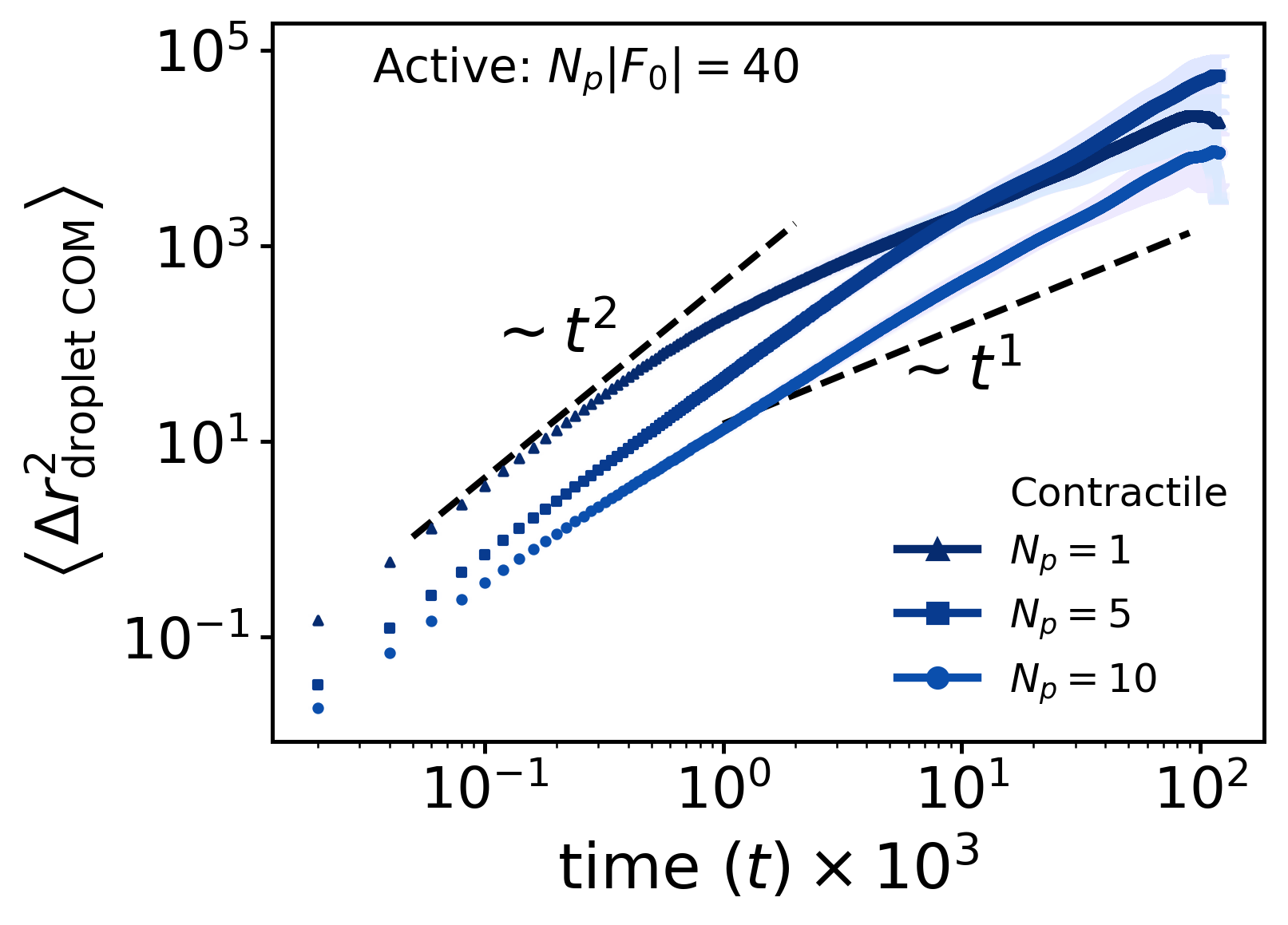}
        \label{fig:MSD_contractile_a1}
    \end{subfigure}
    \hfill
    \begin{subfigure}[t]{0.48\textwidth}
        \captionsetup{position=top, justification=raggedright, singlelinecheck=false}
        \caption{}
        \includegraphics[width=\linewidth]{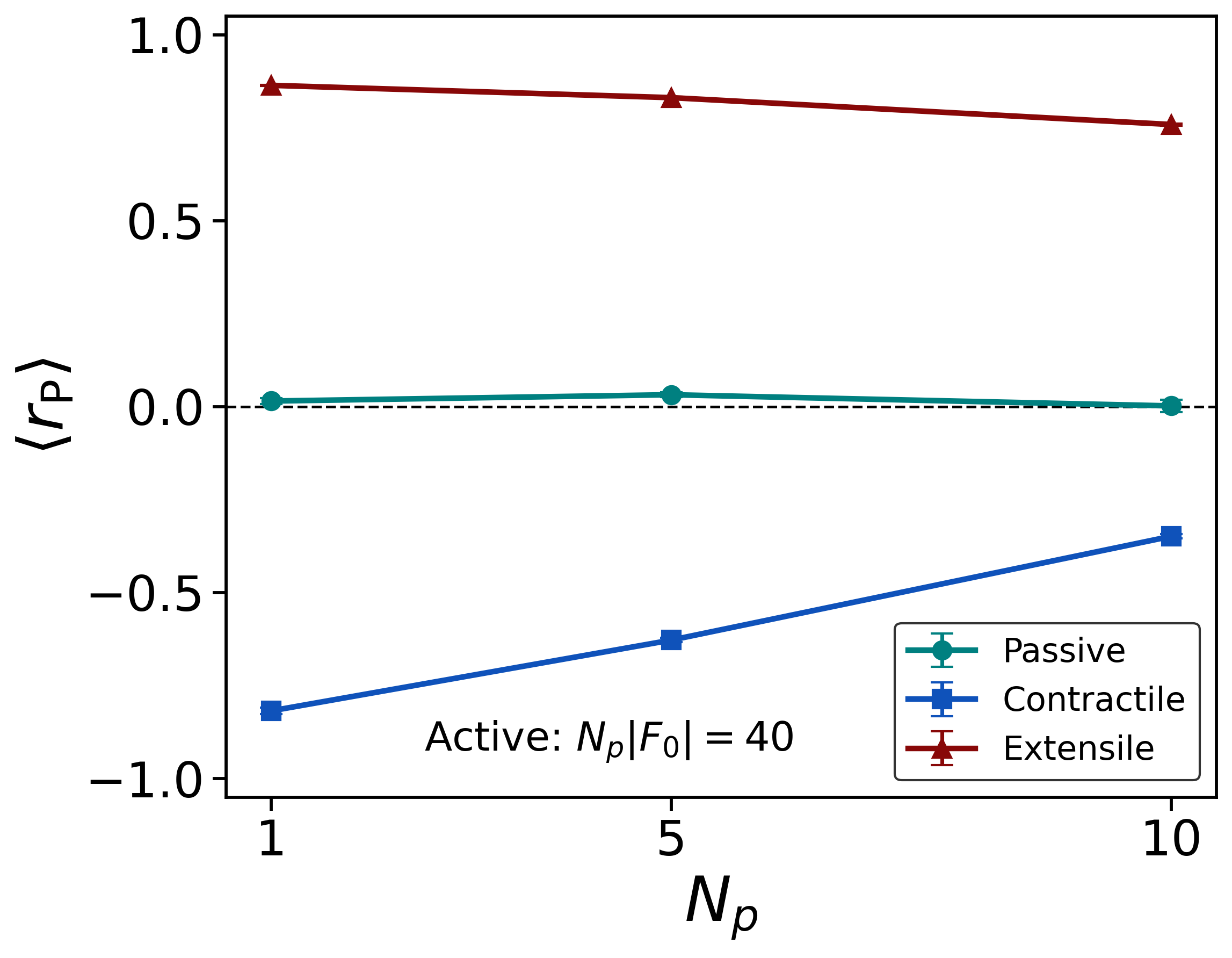}
        \label{fig:pearson_corr_a_1}
    \end{subfigure}
    \caption{\justifying
    Mean-square displacement of the droplet center of mass for different polymer numbers, $N_p$, for droplets containing \textbf{(a)} passive, \textbf{(b)} extensile, and \textbf{(c)} contractile polymers. \textbf{(d)} Pearson correlation coefficient, $r_p$, between the polymer center-of-mass offset relative to the droplet center, $\Delta\vec{R}_{\mathrm{COM}} =\vec{R}_{\mathrm{poly,COM}} -\vec{R}_{\mathrm{droplet,COM}}$, and the instantaneous droplet center-of-mass velocity, as a function of $N_p$. For the active systems, the total activity is $N_p|F_0|=40$. Shaded regions in (a--c) represent standard deviations.}
    \label{fig:droplet_motion_a_1}
\end{figure*}

\clearpage

\section{Supplementary Movies}

\textbf{Supplementary Movie 1: Internal organization and dynamics of confined polymers.} Representative dynamics of passive, extensile, and contractile polymers
confined within a deformable droplet for $N_p=10$, $20$, and $40$. For the active systems, the total activity is fixed at $N_p|F_0|=240$.

\textbf{Supplementary Movie 2: Activity-dependent fluctuations of the
droplet interface.} Dynamics of the droplet interface for passive, extensile, and
contractile systems with $N_p=10$, $20$, and $40$. For the active systems, the total activity is fixed at $N_p|F_0|=240$.

\textbf{Supplementary Movie 3: Activity-dependent droplet motility.} Droplet motion and center-of-mass trajectories for passive, extensile, and contractile systems with $N_p=10$ and $40$. For the active systems, the total activity is fixed at
$N_p|F_0|=240$.

\bibliographystyle{apsrev4-2} 
\bibliography{references} 